\documentclass[aps,prl,showpacs,superscriptaddress,twocolumn,longbibliography]{revtex4-1}

\usepackage{amsfonts}
\usepackage{subfigure}
\usepackage{amsmath}
\usepackage{txfonts}
\usepackage{amssymb}
\usepackage{amsbsy} 
\usepackage{epsfig}
\usepackage{graphicx}
\usepackage{epstopdf}

\def\be{\begin{equation}} \def\ee{\end{equation}}
\def\bea{\begin{eqnarray}} \def\eea{\end{eqnarray}}

\def\nn{\nonumber}

\def\be{{\bf e}}

\newcommand{\bra}[1]{\langle#1|}
\newcommand{\ket}[1]{|#1\rangle}

\def\ra{\rangle}

\def\rw{\rightarrow}

\begin{document}

\title{Edge states and topological invariants of non-Hermitian systems}



\author{Shunyu Yao}
 \affiliation{ Institute for
Advanced Study, Tsinghua University, Beijing,  100084, China }

\author{Zhong Wang} \altaffiliation{ wangzhongemail@gmail.com }
\affiliation{ Institute for
Advanced Study, Tsinghua University, Beijing,  100084, China }

\affiliation{Collaborative Innovation Center of Quantum Matter, Beijing, 100871, China }


\begin{abstract}

The bulk-boundary correspondence is among the central issues of non-Hermitian topological states. We show that a previously overlooked ``non-Hermitian skin effect'' necessitates redefinition of topological invariants in a generalized Brillouin zone. The resultant phase diagrams dramatically differ from the usual Bloch theory. Specifically, we obtain the phase diagram of non-Hermitian Su-Schrieffer-Heeger model, whose  topological zero modes are determined by the non-Bloch winding number instead of the Bloch-Hamiltonian-based topological number. Our work settles the issue of the breakdown of conventional bulk-boundary correspondence and introduces the non-Bloch bulk-boundary correspondence.

\end{abstract}

\maketitle

\emph{Introduction.--}Topological materials are characterized by robust boundary states immune to perturbations\cite{hasan2010,qi2011,Chiu2016rmp, bernevig2013topological,Bansil2016}. According to the principle of bulk-boundary correspondence, the existence of boundary states is dictated by the bulk topological invariants, which,  in the band-theory framework, are defined in terms of the Bloch Hamiltonian. The Hamiltonian is often assumed to be Hermitian. In many physical systems, however, non-Hermitian Hamiltonians are more appropriate\cite{bender1998real,bender2007making}. For example, they are widely used in describing open systems\cite{rotter2009non,malzard2015open, carmichael1993,zhen2015spawning, diehl2011topology,cao2015microcavities,choi2010coalescence,san2016majorana,lee2014heralded,lee2014entanglement}, wave systems with gain and loss\cite{makris2008beam,longhi2009bloch,klaiman2008branch,regensburger2012parity,bittner2012,
ruter2010observation,lin2011unidirectional,feng2013experimental,guo2009complex, liertzer2012pumpinduced,peng2014lossinduced,fleury2015invisible,chang2014PT,hodaei2017enhanced, hodaei2014PT,feng2014singlemode,gao2015billiard,xu2016topological, ashida2017parity,kawabata2017retrieval,chen2017exceptional,ding2016multiple,downing2017} (e.g. photonic and acoustic \cite{ozawa2018rmp,Lu2014review,el2018non,longhi2018}), and solid-state systems where electron-electron interactions or disorders introduce a non-Hermitian self energy into the effective Hamiltonian of quasiparticle\cite{kozii2017,papa2018bulk,shen2018quantum}.
With these physical motivations, there have recently been growing efforts, both theoretically\cite{esaki2011, lee2016anomalous,leykam2017,menke2017,lieu2018ssh,shen2017topological,yin2018ssh,li2017kitaev, rudner2009topological,liang2013topological, hu2011absence,gong2017zeno,gong2010geometrical,rudner2016survival,
kawabata2018PT,ni2018exceptional,zyuzin2018flat,cerjan2018weyl, zhou2017dynamical,gonzalez2017,klett2017sshkitaev,klett2018ssh,yuce2016majorana, yuce2015topological,xu2017weyl,hu2017exceptional,wang2015spontaneous, ke2017topological,rivolta2017,gong2018nonhermitian,harari2018topological} and experimentally\cite{zeuner2015bulk,zhan2017detecting,xiao2017observation, weimann2017topologically,parto2017SSHexperiment,zhao2017Topological,zhou2017observation}, to investigate topological phenomena of non-Hermitian Hamiltonians.

Among the key issues is the fate of bulk-boundary correspondence in non-Hermitian systems. Recently, numerical results in a one-dimensional (1D) model show that open-boundary spectra look notably different from periodic-boundary ones, which seems to indicate a complete breakdown of bulk-boundary correspondence\cite{lee2016anomalous,xiong2017}. In view of this breakdown, a possible scenario is that the topological edge states depend on all sample details, without any general rule telling their existence or absence. Here, we ask the following questions:  Is there a generalized bulk-boundary correspondence? Are there bulk topological invariants responsible for the topological edge states? Affirmative answers are obtained in this paper.

We start from solving a 1D model. Interestingly, all the eigenstates of an open chain are found to be localized near the boundary (dubbed ``non-Hermtian skin effect''), in contrast to the extended Bloch waves in Hermitian cases. In the simplest situations, this effect can be understood in terms of an imaginary gauge field\cite{Longhi2017,Hatano1996}. We show that the non-Hermitian skin effect has dramatic consequences in establishing a ``non-Bloch bulk-boundary correspondence'' in which the topological boundary modes are determined by ``non-Bloch topological invariants''.

Previous non-Hermitian topological invariants\cite{esaki2011,rudner2009topological,leykam2017,lee2016anomalous, shen2017topological,menke2017,lieu2018ssh,yin2018ssh,li2017kitaev} are formulated in terms of the Bloch Hamiltonian. The crucial non-Bloch-wave nature of eigenstates (non-Hermitian skin effect) is untouched, therefore, the number of topological edge modes is not generally related to these topological invariants. In view of the non-Hermitian skin effect, we introduce a non-Bloch topological invariant, which faithfully determines the number of topological edge modes. It embodies the non-Bloch bulk-boundary correspondence of non-Hermitian systems.

\begin{figure}
\includegraphics[width=8.0cm, height=1.7cm]{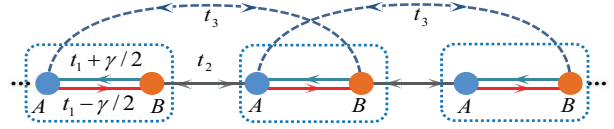}
\caption{Non-Hermitian SSH model. The
dotted box indicates the unit cell. }\label{sketch}
\end{figure}

\emph{Model.--}The non-Hermitian Su-Schrieffer-Heeger (SSH) model\cite{su1980}\footnote{Related models have been studied, for example, in Refs. \cite{zhu2014PT,yin2018ssh,lieu2018ssh}.} is pictorially shown in Fig.\ref{sketch}. Related models are relevant to quite a few experiments\cite{weimann2017topologically,zeuner2015bulk,poli2015selective}.
The Bloch Hamiltonian is
\begin{equation}
\begin{aligned}
H(k)=d_x\sigma_x+ (d_y+i\frac{\gamma}{2} )\sigma_y,
\end{aligned} \label{nHSSH}
\end{equation} where $d_x=t_1+(t_2+t_3)\cos k$, $d_y=(t_2-t_3)\sin k$, and $\sigma_{x,y}$ are the Pauli matrices. A mathematically equivalent model was studied in Ref. \cite{lee2016anomalous}, where $\sigma_y$ was replaced by $\sigma_z$; as such, the physical interpretation was not SSH. The model has a chiral symmetry\cite{Chiu2016rmp} $\sigma_z^{-1} H(k)\sigma_z=-H(k)$, which ensures that the eigenvalues appear in $(E,-E)$ pairs: $E_\pm(k)=\pm\sqrt{d_x^2+(d_y+i\gamma/2)^2}$. Let us first take $t_3=0$ for simplicity (nonzero $t_3$ will be included later). The energy gap closes at the exceptional points $(d_x,d_y)=(\pm\gamma/2,0)$, which requires $t_1= t_2 \pm\gamma/2$ ($k=\pi$) or $t_1= -t_2 \pm\gamma/2$ ($k=0$).

\begin{figure}
\subfigure{\includegraphics[width=7.6cm, height=4.3cm]{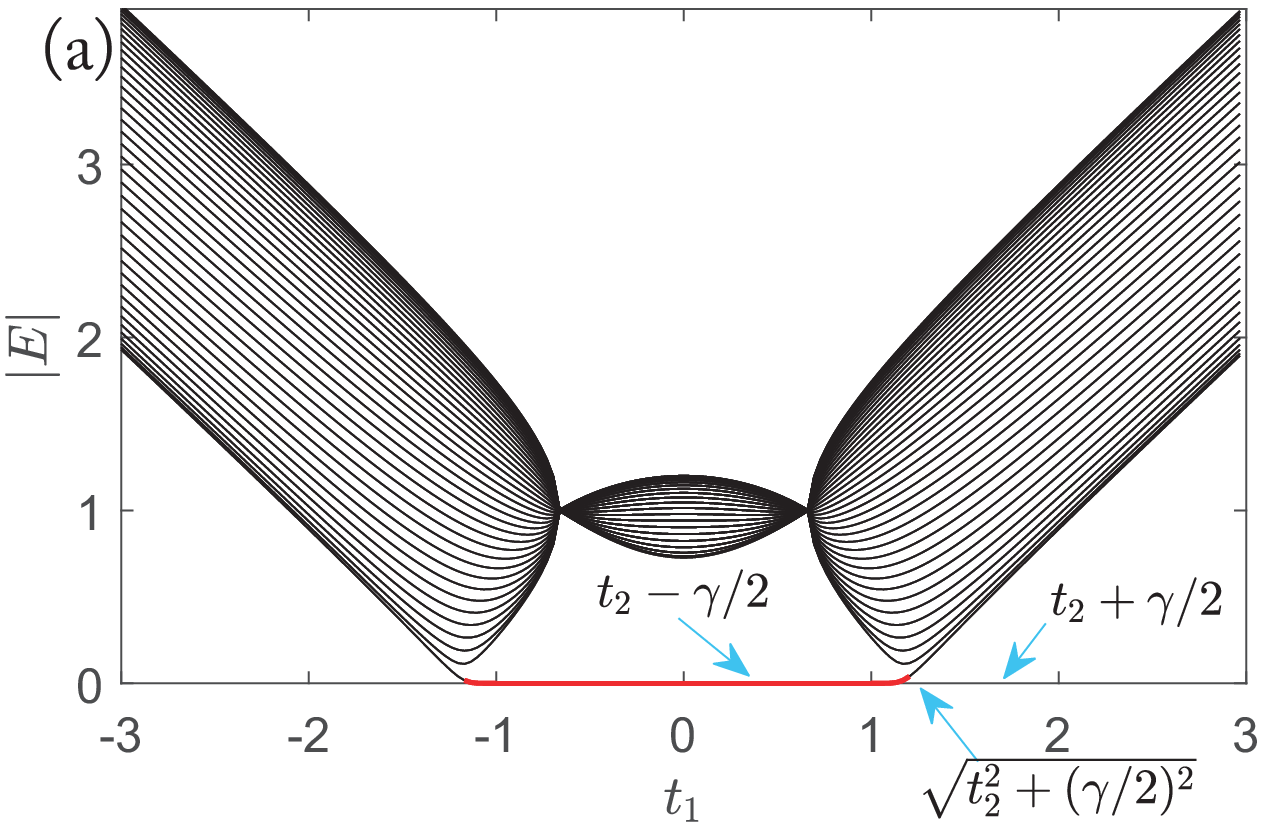}}
\subfigure{\includegraphics[width=3.7cm, height=2.8cm]{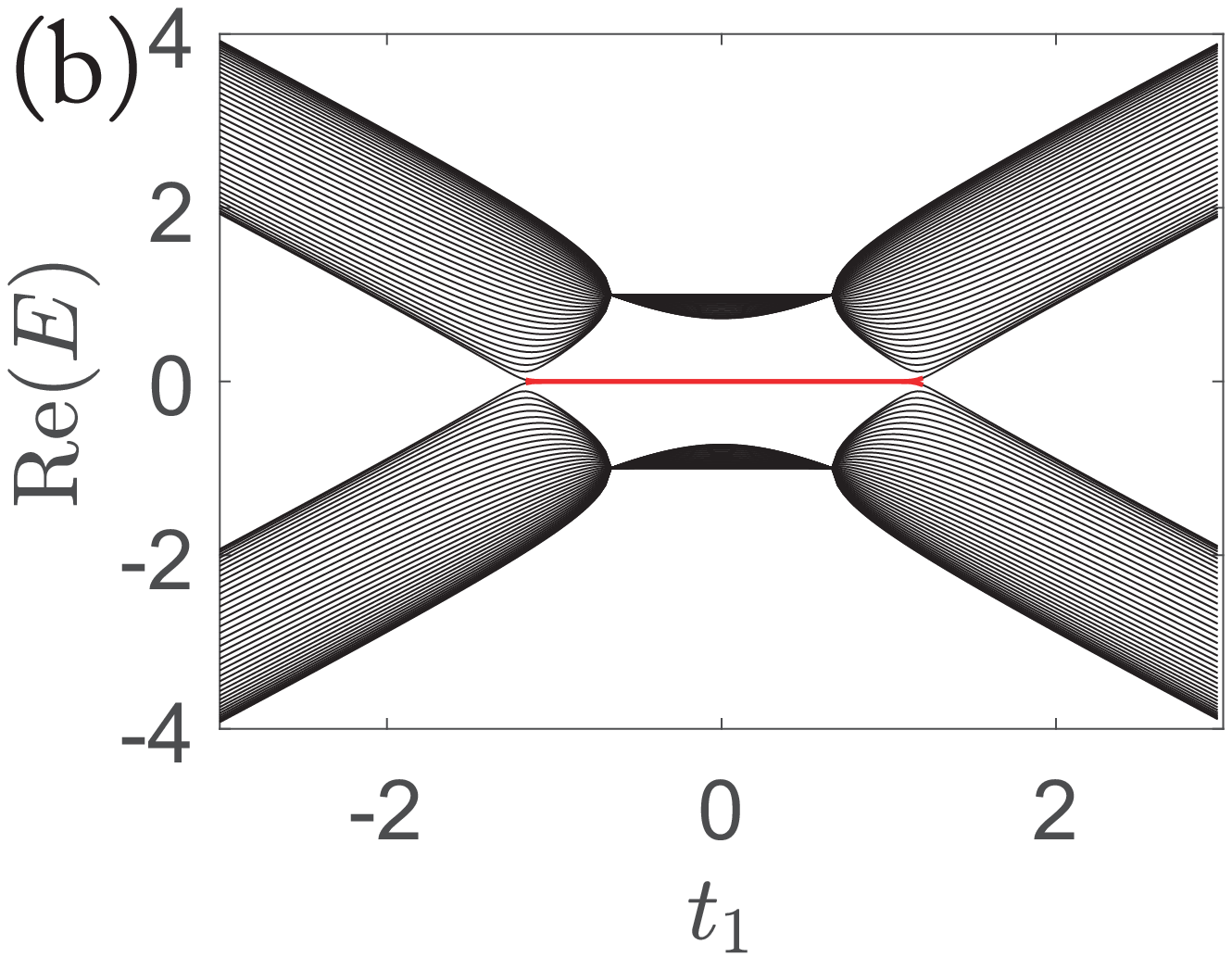}}
\subfigure{\includegraphics[width=3.7cm, height=2.8cm]{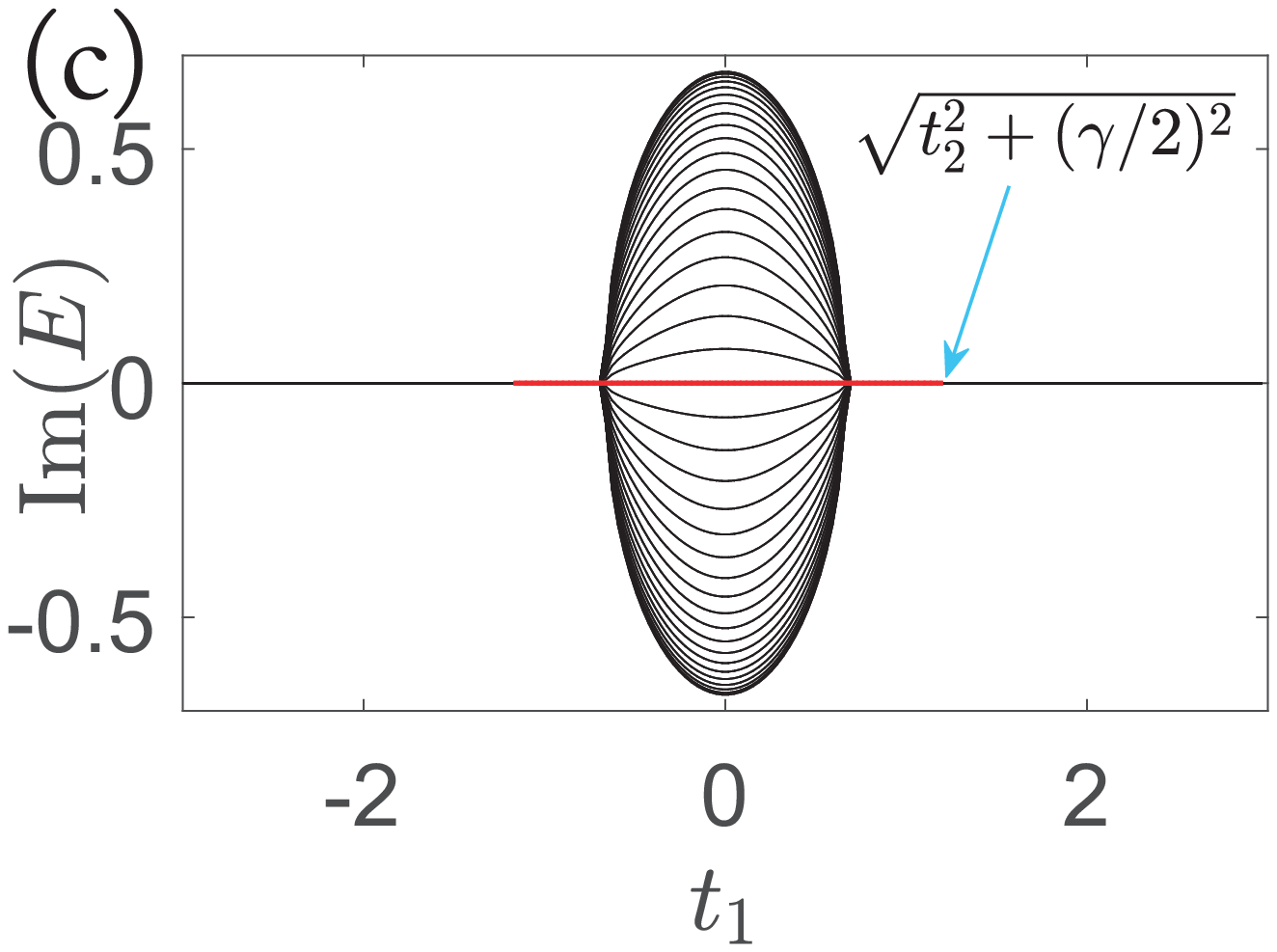}}
\subfigure{\includegraphics[width=7.6cm, height=4.3cm]{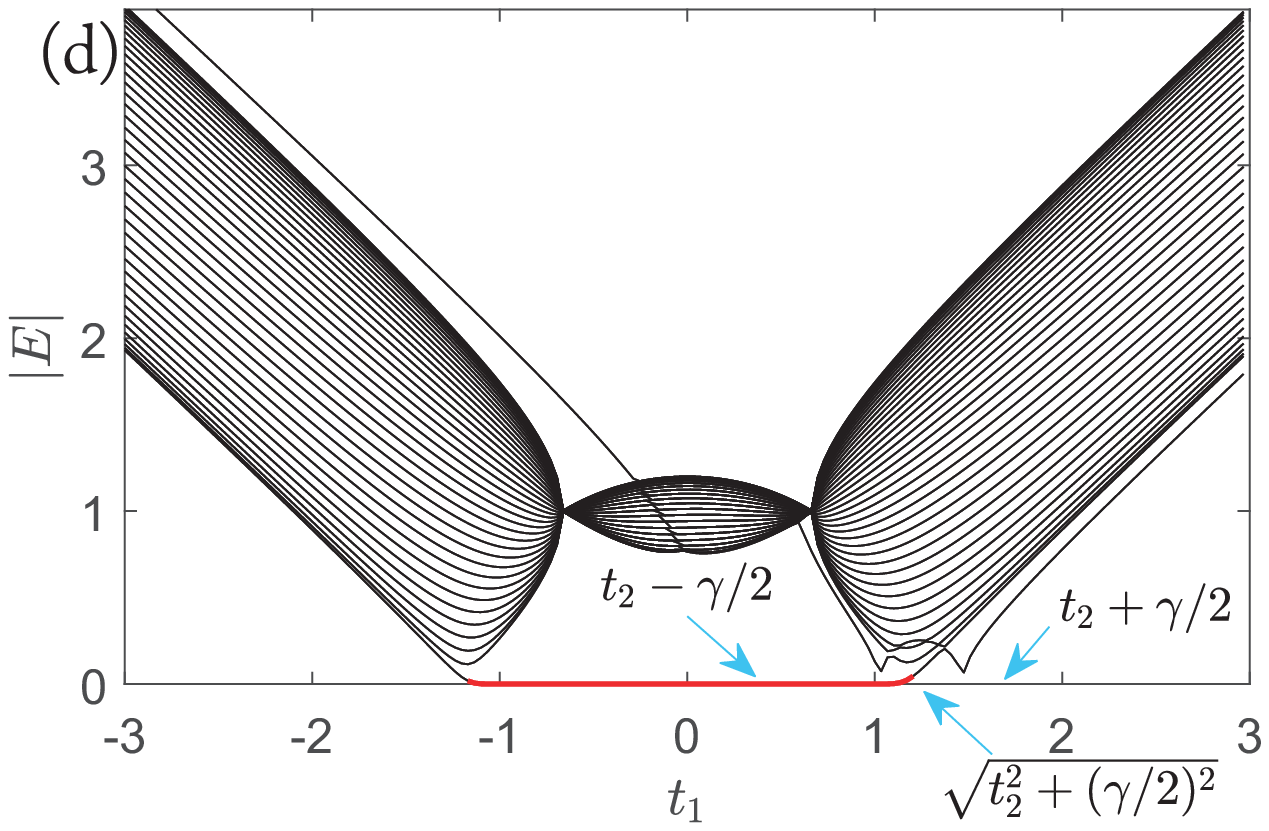}}
\caption{Numerical spectra of of an open chain with length $L=40$ (unit cell). $t_2=1$, $\gamma=4/3$; $t_1$ varies in $[-3,3]$.  (a) $|E|$ as functions of $t_1$. The zero-mode line is shown in red (twofold degenerate, ignoring an indiscernible split). The true transition point ($\sqrt{t_2^2+(\gamma/2)^2}\approx 1.20$) and the $H(k)$-gap-closing points ($t_2\pm\gamma/2$) are indicated by arrows.  (b,c) The real and imaginary parts of $E$.  (d) The same as (a) except that the value of $t_1$ at the leftmost bond is replaced by $t_1-0.8$, which generates additional nonzero modes, but the zero modes are unaffected.    }\label{spectsmall}
\end{figure}

The open-boundary spectrum is noticeably different from that of periodic boundary\cite{lee2016anomalous}\footnote{We note that the numerical precision of Ref.\cite{lee2016anomalous} is improvable. According to our exact results, the zero-mode line in their Fig. 3(a) should span the entire $[-1/\sqrt{2},1/\sqrt{2}]$ interval, instead of the two disconnected lines there.}, which can be seen in the numerical spectra of real-space Hamiltonian of an open chain [Fig.\ref{spectsmall}]. The zero modes are robust to perturbation [Fig.\ref{spectsmall}(d)], which indicates their topological origin. A transition point is located at $t_1\approx 1.20$, which is a quite unremarkable point from the perspective of $H(k)$ whose spectrum is gapped there ($|E_\pm(k)|\neq 0$). As such, the topology of $H(k)$ cannot determine the zero modes, which challenges the familiar Hermitian wisdom. The question arises: What topological invariant predicts the zero modes?

\emph{Shortcut solution.--}To gain insights, we analytically solve an open chain. The wavefunction is written as  $\ket{\psi}=(\psi_{1,A},\psi_{1,B},\psi_{2,A},\psi_{2,B},\cdots,\psi_{L,A},\psi_{L,B})^T$. We first present a shortcut, which is applicable only to the $t_3=0$ case. The real-space eigen-equation $H\ket{\psi}=E\ket{\psi}$ is equivalent to $\bar{H}\ket{\bar{\psi}}=E\ket{\bar{\psi}}$ with $\ket{\bar{\psi}}=S^{-1}\ket{\psi}$ and \bea \bar{H}= S^{-1}HS. \eea  We can judiciously choose $S$ in this similarity transformation. Let us take $S$ to be a diagonal matrix whose diagonal elements are $\{1,r,r,r^2,r^2,\cdots, r^{L-1},r^{L-1}, r^L\}$, then in $\bar{H}$ we have $r^{\pm 1}(t_1\pm \gamma/2)$ in the place of $t_1\pm \gamma/2$ (Fig.\ref{sketch}).
If we take $r= \sqrt{|\frac{t_1-\gamma/2}{t_1+\gamma/2}|}$, $\bar{H}$ becomes the standard SSH model for $|t_1|>|\gamma/2|$, with intracell and intercell hoppings \bea \bar{t}_1=\sqrt{(t_1-\gamma/2)(t_1+\gamma/2)},\quad \bar{t}_2=t_2.\eea The $k$-space expression is \bea \bar{H}(k)=(\bar{t}_1+\bar{t}_2\cos k)\sigma_x +\bar{t}_2\sin k\sigma_y. \eea The transition points are $\bar{t}_1=\bar{t}_2$, namely \bea t_1=\pm\sqrt{t_2^2+(\gamma/2)^2}. \label{shortcut}\eea For the parameters in Fig.\ref{spectsmall}, Eq.(\ref{shortcut}) gives $t_1\approx\pm 1.20$. Note that any $H(k)$-based topological invariants\cite{esaki2011,rudner2009topological,leykam2017,lee2016anomalous, shen2017topological,menke2017,lieu2018ssh,yin2018ssh,li2017kitaev}  can jump only at $t_1=\pm t_2\pm\gamma/2$, where the gap of $H(k)$ closes.

A bulk eigenstate $\ket{\bar{\psi}_l}$ of Hermitian $\bar{H}$ is extended, therefore, $H$'s eigenstate $\ket{\psi_l}=S\ket{\bar{\psi}_l}$ is exponentially localized at an end of the chain when $\gamma\neq 0$. It implies that the usual Bloch phase factor $e^{ik}$ is replaced by $\beta\equiv re^{ik}$ in the open-boundary system (i.e., the wavevector acquires an imaginary part: $k\rw k-i\ln r$).  Although this intuitive picture is based on the shortcut solution, we believe that the exponential-decay behavior of eigenstates (``non-Hermitian skin effect'') is a general feature of non-Hermitian bands.

\emph{Generalizable solution.--}The intuitive shortcut solution has limitations; e.g., it is inapplicable when $t_3\neq 0$. Here, we re-derive the solution in a more generalizable way (still focusing on $t_3=0$ for simplicity). The real-space eigen-equation leads to
$t_2\psi_{n-1,B}+(t_1+\frac{\gamma}{2})\psi_{n,B}=E\psi_{n,A}$ and
$(t_1-\frac{\gamma}{2})\psi_{n,A}+t_2\psi_{n+1,A}=E\psi_{n,B}$ in the bulk of chain. We take the ansatz that $\ket{\psi}=\sum_j\ket{\phi^{(j)}}$, where each $\ket{\phi^{(j)}}$ takes the exponential form (omitting the $j$ index temporarily): $(\phi_{n,A},\phi_{n,B}) = \beta^{n}(\phi_A,\phi_B)$,
which satisfies
\bea \label{ansatz}
[(t_1+\frac{\gamma}{2})+t_2\beta^{-1}]\phi_{B}   =E\phi_{A}, \,\,
[(t_1-\frac{\gamma}{2})+t_2\beta]\phi_{A}   =E\phi_{B}.
\eea
Therefore, we have
\bea \label{bulkeigen}
[(t_1-\frac{\gamma}{2})+t_2\beta][(t_1+\frac{\gamma}{2})+t_2\beta^{-1}]=E^2,
\eea
which has two solutions, namely $\beta_{1,2}(E)= \frac{E^2+\gamma^2/4-t_1^2-t_2^2\pm\sqrt{ (E^2+\gamma^2/4-t_1^2-t_2^2)^2 -4t_2^2(t_1^2-\gamma^2/4)}}{2t_2(t_1+\gamma/2)}$,
where $+(-)$ corresponds to $\beta_1(\beta_2)$.
In the $E\rw 0$ limit, we have
\begin{equation}
\begin{aligned}
\beta_{1,2}^{E\to 0}=-\frac{t_1-\gamma/2}{t_2},\, -\frac{t_2}{t_1+\gamma/2}.
\end{aligned} \label{betazero}
\end{equation}
They can also be seen from Eq.(\ref{ansatz}). These two solutions correspond to $\phi_B=0$ and $\phi_A=0$, respectively.

\begin{figure*}[htb]
\subfigure{\includegraphics[width=6.0cm, height=3.7cm]{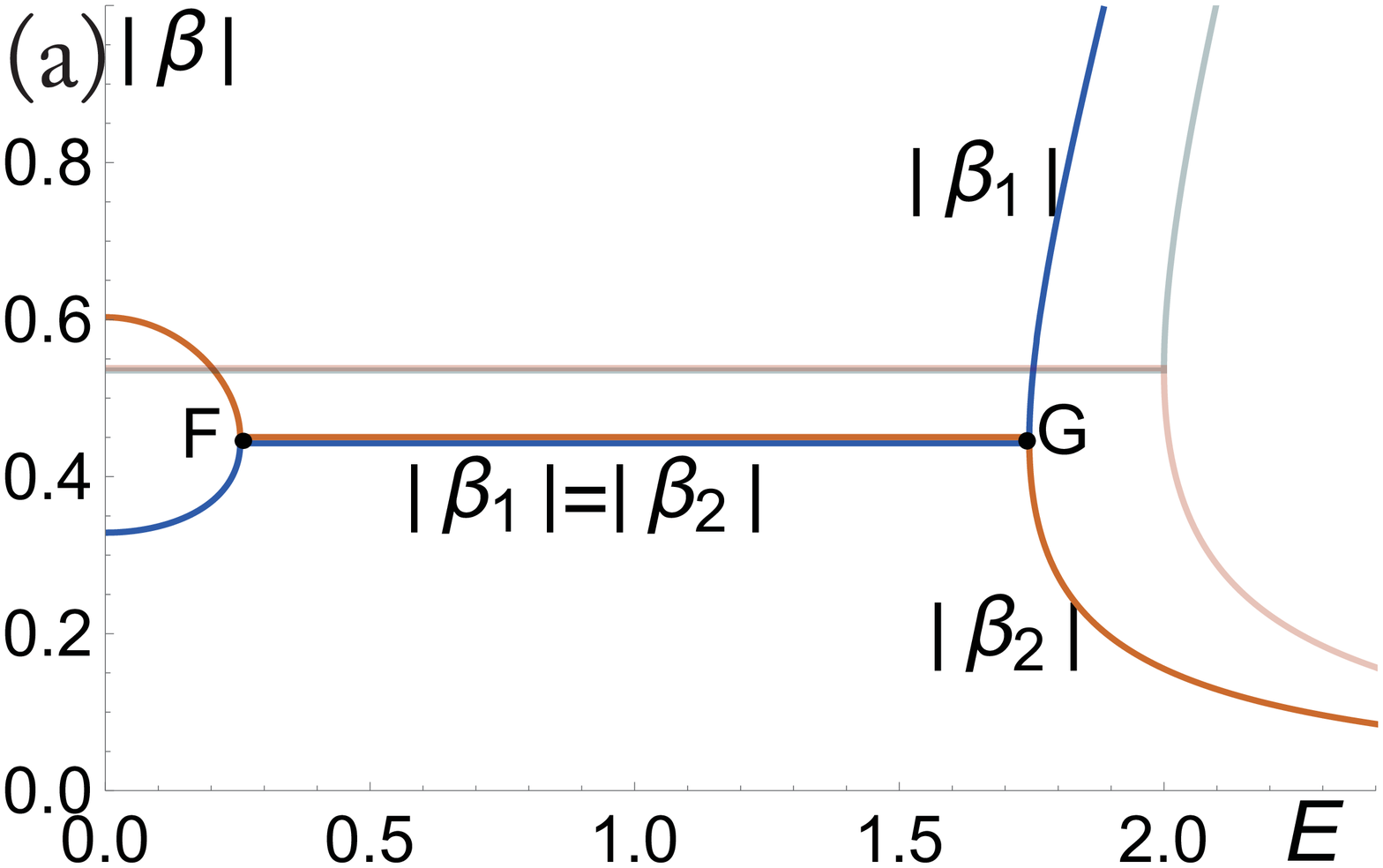}}
\subfigure{\includegraphics[width=4cm, height=4cm]{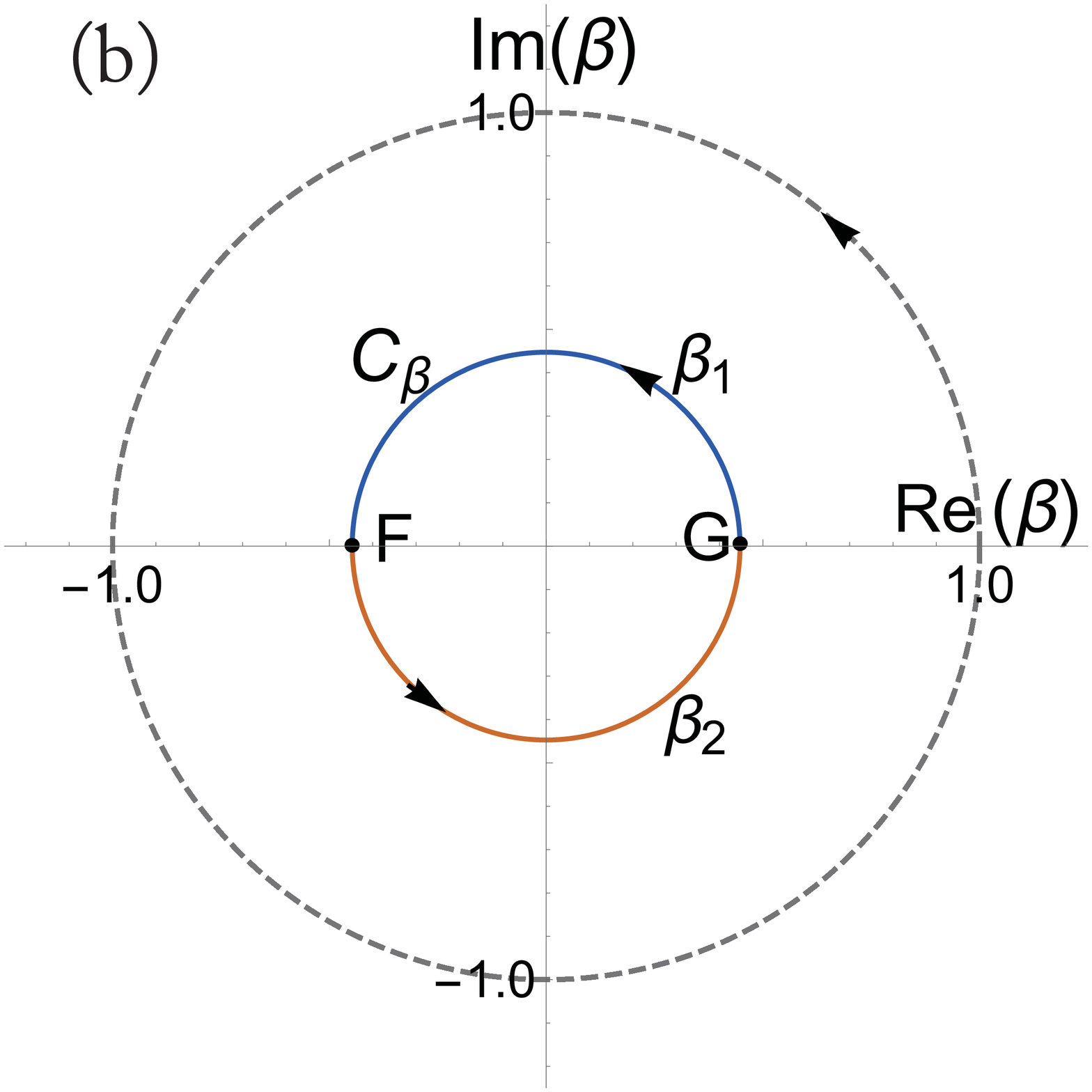}}
\subfigure{\includegraphics[width=5cm, height=3.7cm]{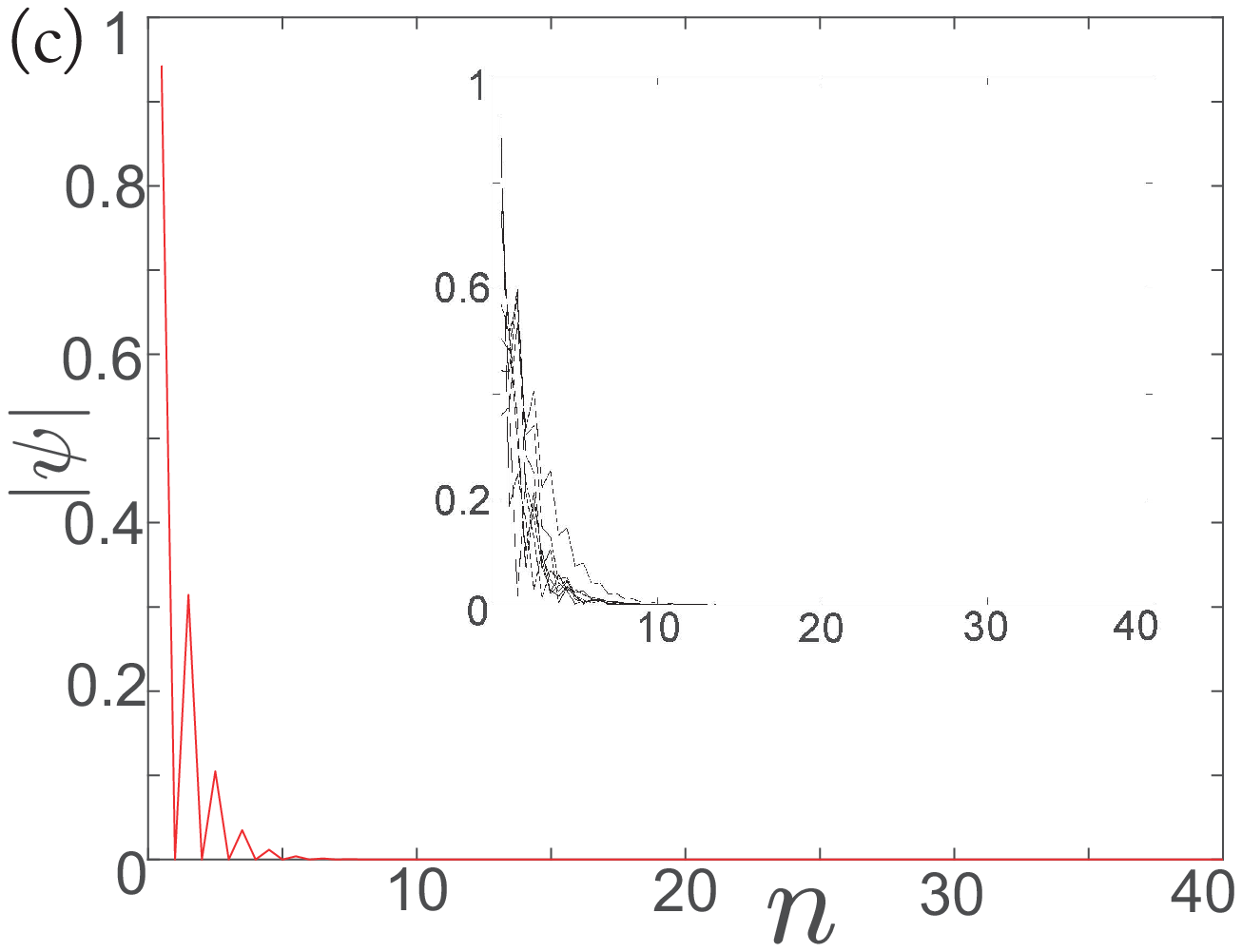}}
\caption{(a) $|\beta_j|$-$E$ curves from Eq.(\ref{bulkeigen}). $t_1=1$ (dark color) and $\sqrt{t_2^2+(\gamma/2)^2}\approx 1.20$ (light color).  (b)  Complex-valued $\beta_j$'s form a closed loop $C_\beta$, which is a circle for the present model [by Eq.(\ref{bulkbeta})]. The shown one is for $t_1=1$. $C_\beta$ can be viewed as a deformed Brillouin zone that generalizes the usual one. In Hermitian cases, $C_\beta$ is a unit circle (dashed line). (c) Profile of a zero mode (main figure) and eight randomly chosen bulk eigenstates (inset), illustrating the ``non-Hermitian skin effect'' found in the analytic solution, namely, all the bulk eigenstates are localized near the boundary. $t_1=1$. Common parameters: $t_2=1,\gamma=4/3$. }\label{absbeta}
\end{figure*}

Restoring the $j$ index in $\ket{\phi^{(j)}}$, we have
\begin{equation} \label{abratio}
\begin{aligned}
\phi_A^{(j)}=\frac{E}{ t_1-\gamma/2 +t_2\beta_j}\phi_B^{(j)},\quad \phi_B^{(j)}=\frac{E}{t_1+\gamma/2 +t_2\beta_j^{-1}}\phi_A^{(j)}.
\end{aligned}
\end{equation} These two equations are equivalent because of Eq.(\ref{bulkeigen}).
The general solution is written as a linear combination:
\begin{equation}
\label{}
\begin{aligned}
\begin{pmatrix}
\psi_{n,A} \\
\psi_{n,B}\\
\end{pmatrix}
=\beta_1^{n}
\begin{pmatrix}
\phi_A^{(1)} \\
\phi_B^{(1)}\\
\end{pmatrix}
+\beta_2^{n}
\begin{pmatrix}
\phi_A^{(2)} \\
\phi_B^{(2)}\\
\end{pmatrix}
\end{aligned},
\end{equation}
which should satisfy the boundary condition
\bea \label{boundary}
 (t_1+\frac{\gamma}{2})\psi_{1,B}-E\psi_{1,A}=0,\,\,
 (t_1-\frac{\gamma}{2})\psi_{L,A}-E\psi_{L,B}=0.
\eea
Together with Eq.(\ref{abratio}), they lead to \bea
\beta_1^{L+1}(t_1-\gamma/2+t_2\beta_2)=\beta_2^{L+1}(t_1-\gamma/2+t_2\beta_1). \label{L} \eea  We are concerned about the spectrum for a long chain, which necessitates $|\beta_1|=|\beta_2|$ for the bulk eigenstates. If not, suppose that $|\beta_1|<|\beta_2|$, we would be able to discard the tiny $\beta_1^{L+1}$ term in Eq.(\ref{L}), and the equation becomes $\beta_2=0$ or $t_1-\gamma/2+t_2\beta_1=0$ (without the appearance of $L$). As a bulk-band property,  $|\beta_1(E)|=|\beta_2(E)|$ remains valid in the presence of perturbations near the edges [e.g., Fig.\ref{spectsmall}(d)], and essentially determines the bulk-band energies\cite{supplemental}.  Combined with $\beta_1\beta_2= \frac{t_1-\gamma/2}{t_1+\gamma/2}$ coming from Eq.(\ref{bulkeigen}), $|\beta_1|=|\beta_2|$ leads to
\begin{equation} \label{bulkbeta}
\begin{aligned}
|\beta_{j}|=r \equiv\sqrt{|\frac{t_1-\gamma/2}{t_1+\gamma/2}|}
\end{aligned}
\end{equation} for bulk eigenstates (i.e., eigenstates in the continuum spectrum). The same $r$ has just been used in the shortcut solution.

We emphasize that $r<1$ indicates that all the eigenstates are localized at the left end of the chain [see Fig.\ref{absbeta}(c) for illustration]\footnote{Recently we noticed Ref.\cite{alvarez2017}, in which similar localization is found numerically; however, in contrast to our viewpoint, it is suggested there that the localization lessens the relevance of zero modes and destroys bulk-boundary correspondence. Also note that the zero-mode interval in their Fig.1 differs from our exact solutions. }\footnote{We emphasize that the bulk energy spectra remain insensitive to a small perturbation at the ends of a long chain. In fact, the eigenstates of $H^\dag$ (namely left eigenstates) have opposite exponential decay and the outcome of a perturbation depends on the product of right and left eigenstates.}. In Hermitian systems, the orthogonality of eigenstates excludes this ``non-Hermitian skin effect'.

There are various ways to re-derive the transition points in Eq.(\ref{shortcut}). To introduce one of them, we first plot in Fig.\ref{absbeta}(a) the $|\beta|$-$E$ curve solved from Eq.(\ref{bulkeigen}) for $t_1=t_2=1,\gamma=4/3$. The spectrum is real for this set of parameters, therefore, no imaginary part of $E$ is needed (This reality is related to $PT$ symmetry\cite{bender1998real,bender2007making}). The expected $|\beta_1|=|\beta_2|=r$ relation is found on the line $FG$ (Fig.\ref{absbeta}(a))), which is associated with bulk spectra. As $t_1$ is increased from $1$, $F$ moves towards left, and finally hits the $|\beta|$ axis ($E=0$ axis). Apparently, it occurs when $|\beta_1^{E\rw 0}|=|\beta_2^{E\rw 0}|=r$. Inserting Eq.(\ref{betazero}) into this equation, we have \bea t_1=\pm\sqrt{t_2^2+(\gamma/2)^2}  \quad\text{or}\quad\pm\sqrt{-t_2^2+(\gamma/2)^2}. \label{tp} \eea At these points, the open-boundary continuum spectra touch zero energy, enabling topological transitions.

A simpler way to re-derive Eq.(\ref{shortcut}) is to calculate the open-boundary spectra. According to Eq.(\ref{bulkbeta}), we can take $\beta=re^{ik}$ ($k\in[0,2\pi]$) in Eq.(\ref{bulkeigen}) to obtain the spectra:
\bea E^2(k) = &&t_1^2+t_2^2 -\gamma^2/4 +t_2 \sqrt{|t_1^2-\gamma^2/4|}[\text{sgn}(t_1+\gamma/2)e^{ik} \nn \\ &&+\text{sgn}(t_1-\gamma/2)e^{-ik}], \label{spectra} \eea which recovers the spectrum of SSH model when $\gamma=0$. The spectra are real when $|t_1|>|\gamma|/2$. Eq.(\ref{tp}) can be readily re-derived as the gap-closing condition of Eq.(\ref{spectra}) ($|E(k)|=0$).

Before proceeding, we comment on a subtle issue in the standard method of finding zero modes. For concreteness, let us consider the present model, and focus on zero modes at the left end of a long chain. One can see that $\ket{\psi^\text{zero}}$ with $(\psi^\text{zero}_{n,A},\psi^\text{zero}_{n,B})=(\beta_1^{E\rw 0})^n (1, 0)$ appears as a zero-energy eigenstate (see Eq.(\ref{betazero}) for $\beta_1^{E\rw 0}$). In the standard approach, the normalizable condition $|\beta_1^{E\rw 0}|<1$ is imposed, and the transition points satisfy $|\beta_1^{E\rw 0}|=1$, which predicts $t_1=t_2+\gamma/2$ as a transition point, being consistent with the gap closing of $H(k)$. Such an apparent but misleading consistency has hidden the true transition points and topological invariants in quite a few previous studies of non-Hermitian models. The implicit assumption was that the bulk eigenstates are extended Bloch waves with $|\beta|=1$, into which the zero modes merge at transitions. In reality, the bulk eigenstates have $|\beta|=r$ (eigenstate skin effect); therefore, the true merging-into-bulk condition is \bea |\beta_1^{E\rw 0}|=r, \eea   which correctly produces $t_1=\sqrt{t_2^2+(\gamma/2)^2}$. This is a manifestation of the non-Bloch bulk-boundary correspondence.

\emph{Non-Bloch topological invariant.--}The bulk-boundary correspondence is fulfilled if we can find a bulk topological invariant that determines the edge modes. Previous constructions take $H(k)$ as the starting point\cite{esaki2011,rudner2009topological,leykam2017,lee2016anomalous, shen2017topological,menke2017,lieu2018ssh,yin2018ssh,li2017kitaev}, which should be revised in view of the non-Hermitian skin effect. The usual Bloch waves carry a pure phase factor $e^{ik}$, whose role is now played by $\beta$. In addition to the phase factor, $\beta$ has a modulus $|\beta|\neq 1$ in general [e.g., Eq.(\ref{bulkbeta})]. Therefore, we start from the ``non-Bloch Hamiltonian'' obtained from $H(k)$ by the replacement $e^{ik}\rw\beta, e^{-ik}\rw\beta^{-1}$:
\begin{equation} \label{}
\begin{aligned}
H(\beta)= (t_1-\frac{\gamma}{2} + \beta t_2 )\sigma_{-}+ (t_1+\frac{\gamma}{2}+\beta^{-1} t_2)\sigma_{+},
\end{aligned}
\end{equation}
where $\sigma_{\pm}=(\sigma_x\pm i\sigma_y)/2$. We have taken $t_3=0$ for simplicity. As explained in both the shortcut and generalizable solutions,  $\beta$ takes values in a non-unit circle $|\beta|=r$ (In other words, $k$ acquires an imaginary part $-i\ln r$). It is notable that the open-boundary spectra in Eq.(\ref{spectra}) are given by $H(\beta)$ instead of $H(k)$. The right and left eigenvectors are defined by  \bea H(\beta)\ket{u_\text{R}}=E(\beta) \ket{u_\text{R}},\quad H^\dag(\beta)\ket{u_\text{L}}=E^*(\beta)\ket{u_\text{L}}. \eea Chiral symmetry ensures that $\ket{\tilde{u}_\text{R}}\equiv \sigma_z \ket{u_\text{R}}$ and $\ket{\tilde{u}_\text{L}}\equiv\sigma_z \ket{u_\text{L}}$ is also right and left eigenvector, with eigenvalues $-E$ and $-E^*$, respectively. In fact, one can diagonalize the matrix as $H(\beta)=TJT^{-1}$ with $J=\begin{pmatrix}
 E &   \\
    & -E
\end{pmatrix}$, then each column of $T$ and $(T^{-1})^\dag$ is a right and left eigenvector, respectively, and the normalization condition $\bra{u_\text{L}}u_\text{R}\ra=\bra{\tilde{u}_\text{L}}\tilde{u}_\text{R}\ra=1, \bra{u_\text{L}}\tilde{u}_\text{R}\ra=\bra{\tilde{u}_\text{L}}u_\text{R}\ra=0$ is guaranteed. As a generalization of the usual ``$Q$ matrix''\cite{Chiu2016rmp}, we define
\bea Q(\beta)=\ket{\tilde{u}_\text{R}(\beta)}\bra{\tilde{u}_\text{L}(\beta)} - \ket{u_\text{R}(\beta)}\bra{u_\text{L}(\beta)}, \label{Q} \eea which is off-diagonal due to the chiral symmetry $\sigma_z^{-1}Q\sigma_z=-Q$, namely $Q=\begin{pmatrix}
  & q  \\
 q^{-1}  &
\end{pmatrix}$.
Now we introduce the non-Bloch winding number:
\begin{equation}
\label{winding}
\begin{aligned}
&W =\frac{i}{2\pi}\int_{C_\beta}  q^{-1}dq.
\end{aligned}
\end{equation} Crucially, it is defined on the ``generalized Brillouin zone'' $C_\beta$ [Fig.\ref{absbeta}(b)]. It is useful to mention that the conventional formulations using $H(k)$ may sometimes produce correct phase diagrams, if $C_\beta$ happens to be a unit circle\footnote{For example, we find that it is the case for the model numerically studied in Ref. \cite{lieu2018ssh}.}.

\begin{figure}
\includegraphics[width=8.0cm, height=4.5cm]{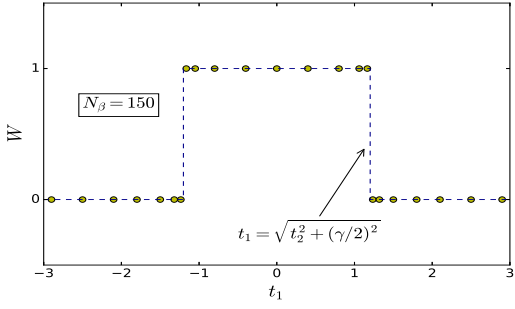}
\caption{Numerical result of topological invariant. $N_\beta$ is the number of grid point on $C_\beta$. $t_2=1,\gamma=4/3$. }\label{invariant}
\end{figure}

The numerical results for $t_3=0$ is shown in Fig.\ref{invariant}, which is consistent with the analytical spectra obtained above. Quantitatively, $2W$ counts the total number of robust zero modes at the left and right ends. For example, corresponding to Fig.\ref{spectsmall}, there are two zero modes for $t_1\in [-\sqrt{t_2^2+(\gamma/2)^2},\sqrt{t_2^2+(\gamma/2)^2}]$, and none elsewhere. The analytic solution shows that, for $[t_2-\gamma/2,\sqrt{t_2^2+(\gamma/2)^2}]$, both modes live at the left end; for $[-t_2+\gamma/2,t_2-\gamma/2]$, one for each end; and for $[-\sqrt{t_2^2+(\gamma/2)^2}, -t_2+\gamma/2]$, both at the right end. Thus, the $H(k)$-gap closing points $\pm (t_2-\gamma/2)$ are where zero modes migrate from one end to the other, conserving the total mode number. In fact, one can see $|\beta_{j=1\,\text{or}\,2}^{E\rw 0}|=1$ at $\pm (t_2-\gamma/2)$, indicating the penetration into bulk.

\begin{figure}
{\includegraphics[width=5.0cm, height=4.1cm]{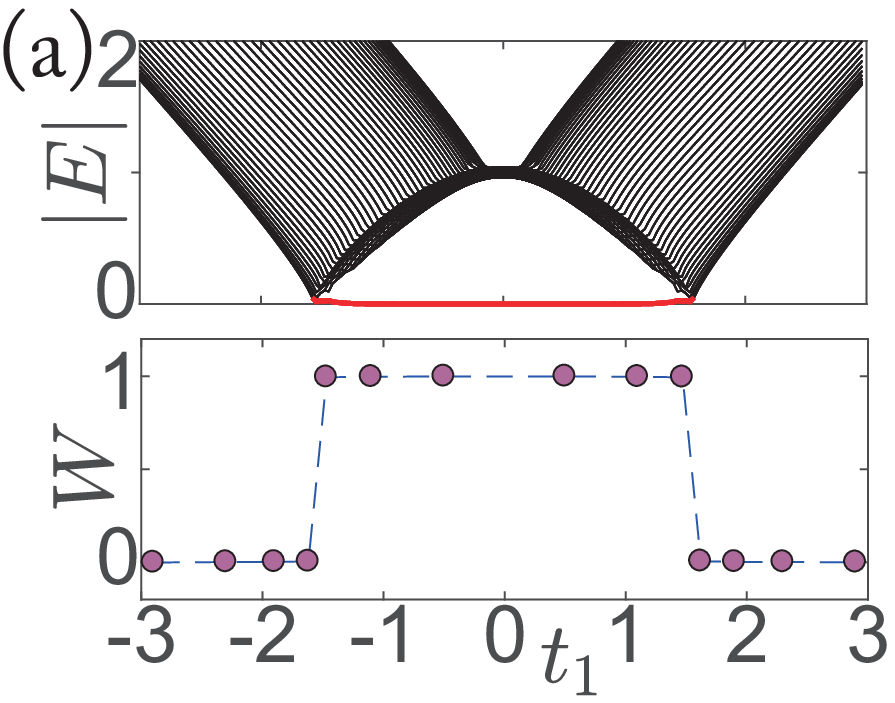}}
{\includegraphics[width=4.6cm, height=4.4cm]{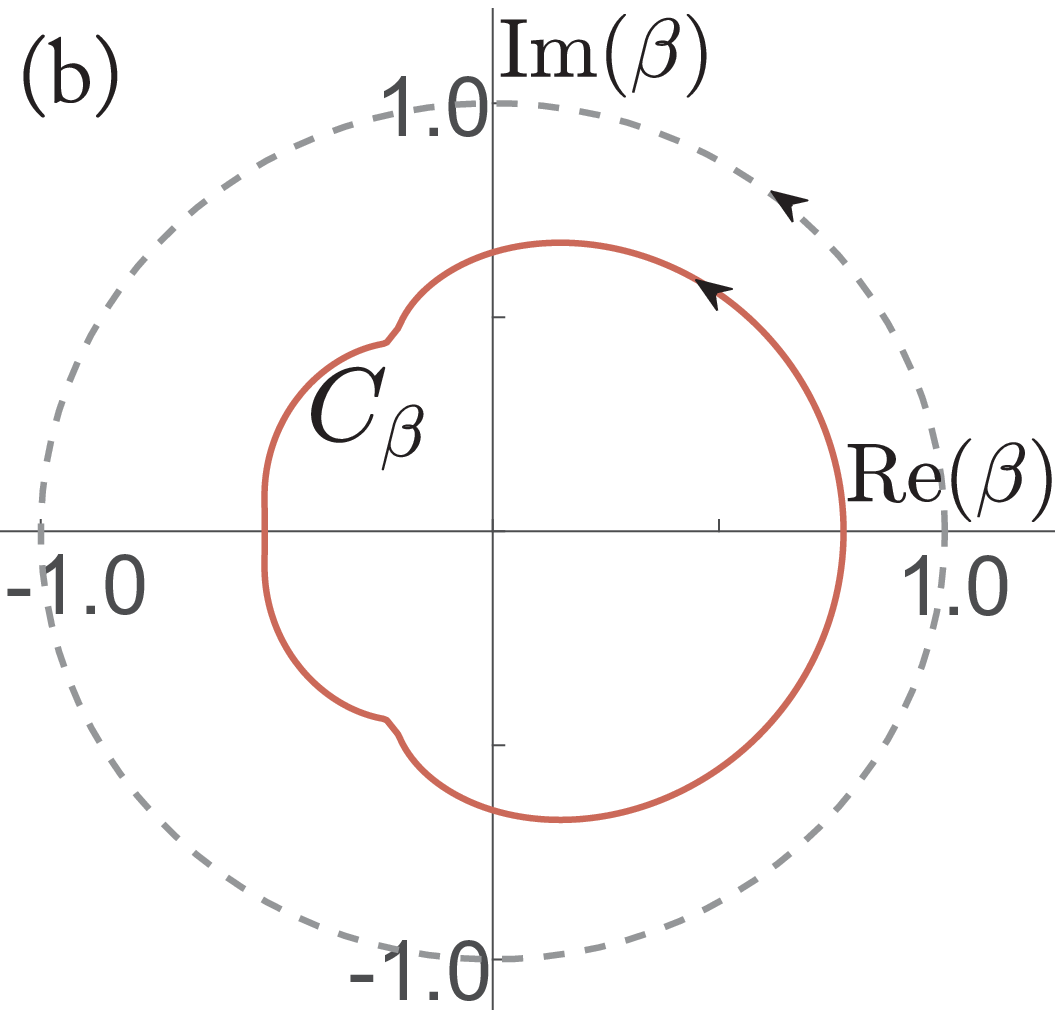}}
\caption{ The nonzero $t_3$ case. (a) Upper panel: Spectrum of an open chain; $t_2=1, \gamma=4/3, t_3=1/5$; $L=100$. Lower panel: topological invariant calculated using 200 grid points on $C_\beta$. The transition points are $t_1\approx\pm 1.56$. (b) $C_\beta$ for $t_1=1.1$.  }\label{t3}
\end{figure}

To provide a more generic exemplification, we take a nonzero $t_3$. Now we find\cite{supplemental} that $C_\beta$ is no longer a circle (bulk eigenstates with different energies have different $|\beta|$), yet $2W$ correctly predicts the total zero-mode number (Fig.\ref{t3}).

Finally, we remarked that Eq.(\ref{winding}) can be generalized to multi-band systems. Each pair of bands (labeled by $l$) possesses a $C^{(l)}_{\beta}$ curve, and the $Q$ matrix [Eq.(\ref{Q})] becomes $Q^{(l)}$, each one defining a winding number $W^{(l)}$ (with matrix trace). The topological invariant is $W=\sum_l W^{(l)}$.

\emph{Conclusions.--}Through the analytic solution of non-Hermitian SSH model, we explained why the usual bulk-boundary correspondence breaks down, and how the non-Bloch bulk-boundary correspondence takes its place. Two of the key concepts are the non-Hermitian skin effect and generalized Brillouin zone. We formulate the generalized bulk-boundary correspondence by introducing a precise topological invariant that faithfully predicts the topological edge modes. The physics presented here can be generalized to a rich variety of non-Hermitian systems, which will be left for future studies.

\emph{Acknowledgements.--}This work is supported by NSFC under Grant No. 11674189.

\bibliography{dirac}

\begin{thebibliography}{98}%
\makeatletter
\providecommand \@ifxundefined [1]{%
 \@ifx{#1\undefined}
}%
\providecommand \@ifnum [1]{%
 \ifnum #1\expandafter \@firstoftwo
 \else \expandafter \@secondoftwo
 \fi
}%
\providecommand \@ifx [1]{%
 \ifx #1\expandafter \@firstoftwo
 \else \expandafter \@secondoftwo
 \fi
}%
\providecommand \natexlab [1]{#1}%
\providecommand \enquote  [1]{``#1''}%
\providecommand \bibnamefont  [1]{#1}%
\providecommand \bibfnamefont [1]{#1}%
\providecommand \citenamefont [1]{#1}%
\providecommand \href@noop [0]{\@secondoftwo}%
\providecommand \href [0]{\begingroup \@sanitize@url \@href}%
\providecommand \@href[1]{\@@startlink{#1}\@@href}%
\providecommand \@@href[1]{\endgroup#1\@@endlink}%
\providecommand \@sanitize@url [0]{\catcode `\\12\catcode `\$12\catcode
  `\&12\catcode `\#12\catcode `\^12\catcode `\_12\catcode `\%12\relax}%
\providecommand \@@startlink[1]{}%
\providecommand \@@endlink[0]{}%
\providecommand \url  [0]{\begingroup\@sanitize@url \@url }%
\providecommand \@url [1]{\endgroup\@href {#1}{\urlprefix }}%
\providecommand \urlprefix  [0]{URL }%
\providecommand \Eprint [0]{\href }%
\providecommand \doibase [0]{http://dx.doi.org/}%
\providecommand \selectlanguage [0]{\@gobble}%
\providecommand \bibinfo  [0]{\@secondoftwo}%
\providecommand \bibfield  [0]{\@secondoftwo}%
\providecommand \translation [1]{[#1]}%
\providecommand \BibitemOpen [0]{}%
\providecommand \bibitemStop [0]{}%
\providecommand \bibitemNoStop [0]{.\EOS\space}%
\providecommand \EOS [0]{\spacefactor3000\relax}%
\providecommand \BibitemShut  [1]{\csname bibitem#1\endcsname}%
\let\auto@bib@innerbib\@empty
\bibitem [{\citenamefont {Hasan}\ and\ \citenamefont {Kane}(2010)}]{hasan2010}%
  \BibitemOpen
  \bibfield  {author} {\bibinfo {author} {\bibfnamefont {M.~Z.}\ \bibnamefont
  {Hasan}}\ and\ \bibinfo {author} {\bibfnamefont {C.~L.}\ \bibnamefont
  {Kane}},\ }\bibfield  {title} {\enquote {\bibinfo {title}
  {\textit{Colloquium} : Topological insulators},}\ }\href {\doibase
  10.1103/RevModPhys.82.3045} {\bibfield  {journal} {\bibinfo  {journal} {Rev.
  Mod. Phys.}\ }\textbf {\bibinfo {volume} {82}},\ \bibinfo {pages}
  {3045--3067} (\bibinfo {year} {2010})}\BibitemShut {NoStop}%
\bibitem [{\citenamefont {Qi}\ and\ \citenamefont {Zhang}(2011)}]{qi2011}%
  \BibitemOpen
  \bibfield  {author} {\bibinfo {author} {\bibfnamefont {Xiao-Liang}\
  \bibnamefont {Qi}}\ and\ \bibinfo {author} {\bibfnamefont {Shou-Cheng}\
  \bibnamefont {Zhang}},\ }\bibfield  {title} {\enquote {\bibinfo {title}
  {Topological insulators and superconductors},}\ }\href {\doibase
  10.1103/RevModPhys.83.1057} {\bibfield  {journal} {\bibinfo  {journal} {Rev.
  Mod. Phys.}\ }\textbf {\bibinfo {volume} {83}},\ \bibinfo {pages}
  {1057--1110} (\bibinfo {year} {2011})}\BibitemShut {NoStop}%
\bibitem [{\citenamefont {Chiu}\ \emph {et~al.}(2016)\citenamefont {Chiu},
  \citenamefont {Teo}, \citenamefont {Schnyder},\ and\ \citenamefont
  {Ryu}}]{Chiu2016rmp}%
  \BibitemOpen
  \bibfield  {author} {\bibinfo {author} {\bibfnamefont {Ching-Kai}\
  \bibnamefont {Chiu}}, \bibinfo {author} {\bibfnamefont {Jeffrey C.~Y.}\
  \bibnamefont {Teo}}, \bibinfo {author} {\bibfnamefont {Andreas~P.}\
  \bibnamefont {Schnyder}}, \ and\ \bibinfo {author} {\bibfnamefont {Shinsei}\
  \bibnamefont {Ryu}},\ }\bibfield  {title} {\enquote {\bibinfo {title}
  {Classification of topological quantum matter with symmetries},}\ }\href
  {\doibase 10.1103/RevModPhys.88.035005} {\bibfield  {journal} {\bibinfo
  {journal} {Rev. Mod. Phys.}\ }\textbf {\bibinfo {volume} {88}},\ \bibinfo
  {pages} {035005} (\bibinfo {year} {2016})}\BibitemShut {NoStop}%
\bibitem [{\citenamefont {Bernevig}\ and\ \citenamefont
  {Hughes}(2013)}]{bernevig2013topological}%
  \BibitemOpen
  \bibfield  {author} {\bibinfo {author} {\bibfnamefont {B~Andrei}\
  \bibnamefont {Bernevig}}\ and\ \bibinfo {author} {\bibfnamefont {Taylor~L}\
  \bibnamefont {Hughes}},\ }\href@noop {} {\emph {\bibinfo {title} {Topological
  insulators and topological superconductors}}}\ (\bibinfo  {publisher}
  {Princeton University Press, Princeton, NJ},\ \bibinfo {year}
  {2013})\BibitemShut {NoStop}%
\bibitem [{\citenamefont {Bansil}\ \emph {et~al.}(2016)\citenamefont {Bansil},
  \citenamefont {Lin},\ and\ \citenamefont {Das}}]{Bansil2016}%
  \BibitemOpen
  \bibfield  {author} {\bibinfo {author} {\bibfnamefont {A.}~\bibnamefont
  {Bansil}}, \bibinfo {author} {\bibfnamefont {Hsin}\ \bibnamefont {Lin}}, \
  and\ \bibinfo {author} {\bibfnamefont {Tanmoy}\ \bibnamefont {Das}},\
  }\bibfield  {title} {\enquote {\bibinfo {title} {\textit{Colloquium} :
  Topological band theory},}\ }\href {\doibase 10.1103/RevModPhys.88.021004}
  {\bibfield  {journal} {\bibinfo  {journal} {Rev. Mod. Phys.}\ }\textbf
  {\bibinfo {volume} {88}},\ \bibinfo {pages} {021004} (\bibinfo {year}
  {2016})}\BibitemShut {NoStop}%
\bibitem [{\citenamefont {Bender}\ and\ \citenamefont
  {Boettcher}(1998)}]{bender1998real}%
  \BibitemOpen
  \bibfield  {author} {\bibinfo {author} {\bibfnamefont {Carl~M}\ \bibnamefont
  {Bender}}\ and\ \bibinfo {author} {\bibfnamefont {Stefan}\ \bibnamefont
  {Boettcher}},\ }\bibfield  {title} {\enquote {\bibinfo {title} {Real spectra
  in non-hermitian hamiltonians having p t symmetry},}\ }\href@noop {}
  {\bibfield  {journal} {\bibinfo  {journal} {Physical Review Letters}\
  }\textbf {\bibinfo {volume} {80}},\ \bibinfo {pages} {5243} (\bibinfo {year}
  {1998})}\BibitemShut {NoStop}%
\bibitem [{\citenamefont {Bender}(2007)}]{bender2007making}%
  \BibitemOpen
  \bibfield  {author} {\bibinfo {author} {\bibfnamefont {Carl~M}\ \bibnamefont
  {Bender}},\ }\bibfield  {title} {\enquote {\bibinfo {title} {Making sense of
  non-hermitian hamiltonians},}\ }\href@noop {} {\bibfield  {journal} {\bibinfo
   {journal} {Reports on Progress in Physics}\ }\textbf {\bibinfo {volume}
  {70}},\ \bibinfo {pages} {947} (\bibinfo {year} {2007})}\BibitemShut
  {NoStop}%
\bibitem [{\citenamefont {Rotter}(2009)}]{rotter2009non}%
  \BibitemOpen
  \bibfield  {author} {\bibinfo {author} {\bibfnamefont {Ingrid}\ \bibnamefont
  {Rotter}},\ }\bibfield  {title} {\enquote {\bibinfo {title} {A non-hermitian
  hamilton operator and the physics of open quantum systems},}\ }\href@noop {}
  {\bibfield  {journal} {\bibinfo  {journal} {Journal of Physics A:
  Mathematical and Theoretical}\ }\textbf {\bibinfo {volume} {42}},\ \bibinfo
  {pages} {153001} (\bibinfo {year} {2009})}\BibitemShut {NoStop}%
\bibitem [{\citenamefont {Malzard}\ \emph {et~al.}(2015)\citenamefont
  {Malzard}, \citenamefont {Poli},\ and\ \citenamefont
  {Schomerus}}]{malzard2015open}%
  \BibitemOpen
  \bibfield  {author} {\bibinfo {author} {\bibfnamefont {Simon}\ \bibnamefont
  {Malzard}}, \bibinfo {author} {\bibfnamefont {Charles}\ \bibnamefont {Poli}},
  \ and\ \bibinfo {author} {\bibfnamefont {Henning}\ \bibnamefont
  {Schomerus}},\ }\bibfield  {title} {\enquote {\bibinfo {title} {Topologically
  protected defect states in open photonic systems with non-hermitian
  charge-conjugation and parity-time symmetry},}\ }\href {\doibase
  10.1103/PhysRevLett.115.200402} {\bibfield  {journal} {\bibinfo  {journal}
  {Phys. Rev. Lett.}\ }\textbf {\bibinfo {volume} {115}},\ \bibinfo {pages}
  {200402} (\bibinfo {year} {2015})}\BibitemShut {NoStop}%
\bibitem [{\citenamefont {Carmichael}(1993)}]{carmichael1993}%
  \BibitemOpen
  \bibfield  {author} {\bibinfo {author} {\bibfnamefont {H.~J.}\ \bibnamefont
  {Carmichael}},\ }\bibfield  {title} {\enquote {\bibinfo {title} {Quantum
  trajectory theory for cascaded open systems},}\ }\href {\doibase
  10.1103/PhysRevLett.70.2273} {\bibfield  {journal} {\bibinfo  {journal}
  {Phys. Rev. Lett.}\ }\textbf {\bibinfo {volume} {70}},\ \bibinfo {pages}
  {2273--2276} (\bibinfo {year} {1993})}\BibitemShut {NoStop}%
\bibitem [{\citenamefont {Zhen}\ \emph {et~al.}(2015)\citenamefont {Zhen},
  \citenamefont {Hsu}, \citenamefont {Igarashi}, \citenamefont {Lu},
  \citenamefont {Kaminer}, \citenamefont {Pick}, \citenamefont {Chua},
  \citenamefont {Joannopoulos},\ and\ \citenamefont
  {Solja{\v{c}}i{\'c}}}]{zhen2015spawning}%
  \BibitemOpen
  \bibfield  {author} {\bibinfo {author} {\bibfnamefont {Bo}~\bibnamefont
  {Zhen}}, \bibinfo {author} {\bibfnamefont {Chia~Wei}\ \bibnamefont {Hsu}},
  \bibinfo {author} {\bibfnamefont {Yuichi}\ \bibnamefont {Igarashi}}, \bibinfo
  {author} {\bibfnamefont {Ling}\ \bibnamefont {Lu}}, \bibinfo {author}
  {\bibfnamefont {Ido}\ \bibnamefont {Kaminer}}, \bibinfo {author}
  {\bibfnamefont {Adi}\ \bibnamefont {Pick}}, \bibinfo {author} {\bibfnamefont
  {Song-Liang}\ \bibnamefont {Chua}}, \bibinfo {author} {\bibfnamefont
  {John~D}\ \bibnamefont {Joannopoulos}}, \ and\ \bibinfo {author}
  {\bibfnamefont {Marin}\ \bibnamefont {Solja{\v{c}}i{\'c}}},\ }\bibfield
  {title} {\enquote {\bibinfo {title} {Spawning rings of exceptional points out
  of dirac cones},}\ }\href@noop {} {\bibfield  {journal} {\bibinfo  {journal}
  {Nature}\ }\textbf {\bibinfo {volume} {525}},\ \bibinfo {pages} {354}
  (\bibinfo {year} {2015})}\BibitemShut {NoStop}%
\bibitem [{\citenamefont {Diehl}\ \emph {et~al.}(2011)\citenamefont {Diehl},
  \citenamefont {Rico}, \citenamefont {Baranov},\ and\ \citenamefont
  {Zoller}}]{diehl2011topology}%
  \BibitemOpen
  \bibfield  {author} {\bibinfo {author} {\bibfnamefont {Sebastian}\
  \bibnamefont {Diehl}}, \bibinfo {author} {\bibfnamefont {Enrique}\
  \bibnamefont {Rico}}, \bibinfo {author} {\bibfnamefont {Mikhail~A}\
  \bibnamefont {Baranov}}, \ and\ \bibinfo {author} {\bibfnamefont {Peter}\
  \bibnamefont {Zoller}},\ }\bibfield  {title} {\enquote {\bibinfo {title}
  {Topology by dissipation in atomic quantum wires},}\ }\href@noop {}
  {\bibfield  {journal} {\bibinfo  {journal} {Nature Physics}\ }\textbf
  {\bibinfo {volume} {7}},\ \bibinfo {pages} {971--977} (\bibinfo {year}
  {2011})}\BibitemShut {NoStop}%
\bibitem [{\citenamefont {Cao}\ and\ \citenamefont
  {Wiersig}(2015)}]{cao2015microcavities}%
  \BibitemOpen
  \bibfield  {author} {\bibinfo {author} {\bibfnamefont {Hui}\ \bibnamefont
  {Cao}}\ and\ \bibinfo {author} {\bibfnamefont {Jan}\ \bibnamefont
  {Wiersig}},\ }\bibfield  {title} {\enquote {\bibinfo {title} {Dielectric
  microcavities: Model systems for wave chaos and non-hermitian physics},}\
  }\href {\doibase 10.1103/RevModPhys.87.61} {\bibfield  {journal} {\bibinfo
  {journal} {Rev. Mod. Phys.}\ }\textbf {\bibinfo {volume} {87}},\ \bibinfo
  {pages} {61--111} (\bibinfo {year} {2015})}\BibitemShut {NoStop}%
\bibitem [{\citenamefont {Choi}\ \emph {et~al.}(2010)\citenamefont {Choi},
  \citenamefont {Kang}, \citenamefont {Lim}, \citenamefont {Kim}, \citenamefont
  {Kim}, \citenamefont {Lee},\ and\ \citenamefont {An}}]{choi2010coalescence}%
  \BibitemOpen
  \bibfield  {author} {\bibinfo {author} {\bibfnamefont {Youngwoon}\
  \bibnamefont {Choi}}, \bibinfo {author} {\bibfnamefont {Sungsam}\
  \bibnamefont {Kang}}, \bibinfo {author} {\bibfnamefont {Sooin}\ \bibnamefont
  {Lim}}, \bibinfo {author} {\bibfnamefont {Wookrae}\ \bibnamefont {Kim}},
  \bibinfo {author} {\bibfnamefont {Jung-Ryul}\ \bibnamefont {Kim}}, \bibinfo
  {author} {\bibfnamefont {Jai-Hyung}\ \bibnamefont {Lee}}, \ and\ \bibinfo
  {author} {\bibfnamefont {Kyungwon}\ \bibnamefont {An}},\ }\bibfield  {title}
  {\enquote {\bibinfo {title} {Quasieigenstate coalescence in an atom-cavity
  quantum composite},}\ }\href {\doibase 10.1103/PhysRevLett.104.153601}
  {\bibfield  {journal} {\bibinfo  {journal} {Phys. Rev. Lett.}\ }\textbf
  {\bibinfo {volume} {104}},\ \bibinfo {pages} {153601} (\bibinfo {year}
  {2010})}\BibitemShut {NoStop}%
\bibitem [{\citenamefont {San-Jose}\ \emph {et~al.}(2016)\citenamefont
  {San-Jose}, \citenamefont {Cayao}, \citenamefont {Prada},\ and\ \citenamefont
  {Aguado}}]{san2016majorana}%
  \BibitemOpen
  \bibfield  {author} {\bibinfo {author} {\bibfnamefont {Pablo}\ \bibnamefont
  {San-Jose}}, \bibinfo {author} {\bibfnamefont {Jorge}\ \bibnamefont {Cayao}},
  \bibinfo {author} {\bibfnamefont {Elsa}\ \bibnamefont {Prada}}, \ and\
  \bibinfo {author} {\bibfnamefont {Ram{\'o}n}\ \bibnamefont {Aguado}},\
  }\bibfield  {title} {\enquote {\bibinfo {title} {Majorana bound states from
  exceptional points in non-topological superconductors},}\ }\href@noop {}
  {\bibfield  {journal} {\bibinfo  {journal} {Scientific reports}\ }\textbf
  {\bibinfo {volume} {6}},\ \bibinfo {pages} {21427} (\bibinfo {year}
  {2016})}\BibitemShut {NoStop}%
\bibitem [{\citenamefont {Lee}\ and\ \citenamefont
  {Chan}(2014)}]{lee2014heralded}%
  \BibitemOpen
  \bibfield  {author} {\bibinfo {author} {\bibfnamefont {Tony~E.}\ \bibnamefont
  {Lee}}\ and\ \bibinfo {author} {\bibfnamefont {Ching-Kit}\ \bibnamefont
  {Chan}},\ }\bibfield  {title} {\enquote {\bibinfo {title} {Heralded magnetism
  in non-hermitian atomic systems},}\ }\href {\doibase
  10.1103/PhysRevX.4.041001} {\bibfield  {journal} {\bibinfo  {journal} {Phys.
  Rev. X}\ }\textbf {\bibinfo {volume} {4}},\ \bibinfo {pages} {041001}
  (\bibinfo {year} {2014})}\BibitemShut {NoStop}%
\bibitem [{\citenamefont {Lee}\ \emph {et~al.}(2014)\citenamefont {Lee},
  \citenamefont {Reiter},\ and\ \citenamefont
  {Moiseyev}}]{lee2014entanglement}%
  \BibitemOpen
  \bibfield  {author} {\bibinfo {author} {\bibfnamefont {Tony~E.}\ \bibnamefont
  {Lee}}, \bibinfo {author} {\bibfnamefont {Florentin}\ \bibnamefont {Reiter}},
  \ and\ \bibinfo {author} {\bibfnamefont {Nimrod}\ \bibnamefont {Moiseyev}},\
  }\bibfield  {title} {\enquote {\bibinfo {title} {Entanglement and spin
  squeezing in non-hermitian phase transitions},}\ }\href {\doibase
  10.1103/PhysRevLett.113.250401} {\bibfield  {journal} {\bibinfo  {journal}
  {Phys. Rev. Lett.}\ }\textbf {\bibinfo {volume} {113}},\ \bibinfo {pages}
  {250401} (\bibinfo {year} {2014})}\BibitemShut {NoStop}%
\bibitem [{\citenamefont {Makris}\ \emph {et~al.}(2008)\citenamefont {Makris},
  \citenamefont {El-Ganainy}, \citenamefont {Christodoulides},\ and\
  \citenamefont {Musslimani}}]{makris2008beam}%
  \BibitemOpen
  \bibfield  {author} {\bibinfo {author} {\bibfnamefont {K.~G.}\ \bibnamefont
  {Makris}}, \bibinfo {author} {\bibfnamefont {R.}~\bibnamefont {El-Ganainy}},
  \bibinfo {author} {\bibfnamefont {D.~N.}\ \bibnamefont {Christodoulides}}, \
  and\ \bibinfo {author} {\bibfnamefont {Z.~H.}\ \bibnamefont {Musslimani}},\
  }\bibfield  {title} {\enquote {\bibinfo {title} {Beam dynamics in
  $\mathcal{P}\mathcal{T}$ symmetric optical lattices},}\ }\href {\doibase
  10.1103/PhysRevLett.100.103904} {\bibfield  {journal} {\bibinfo  {journal}
  {Phys. Rev. Lett.}\ }\textbf {\bibinfo {volume} {100}},\ \bibinfo {pages}
  {103904} (\bibinfo {year} {2008})}\BibitemShut {NoStop}%
\bibitem [{\citenamefont {Longhi}(2009)}]{longhi2009bloch}%
  \BibitemOpen
  \bibfield  {author} {\bibinfo {author} {\bibfnamefont {S.}~\bibnamefont
  {Longhi}},\ }\bibfield  {title} {\enquote {\bibinfo {title} {Bloch
  oscillations in complex crystals with $\mathcal{P}\mathcal{T}$ symmetry},}\
  }\href {\doibase 10.1103/PhysRevLett.103.123601} {\bibfield  {journal}
  {\bibinfo  {journal} {Phys. Rev. Lett.}\ }\textbf {\bibinfo {volume} {103}},\
  \bibinfo {pages} {123601} (\bibinfo {year} {2009})}\BibitemShut {NoStop}%
\bibitem [{\citenamefont {Klaiman}\ \emph {et~al.}(2008)\citenamefont
  {Klaiman}, \citenamefont {G\"unther},\ and\ \citenamefont
  {Moiseyev}}]{klaiman2008branch}%
  \BibitemOpen
  \bibfield  {author} {\bibinfo {author} {\bibfnamefont {Shachar}\ \bibnamefont
  {Klaiman}}, \bibinfo {author} {\bibfnamefont {Uwe}\ \bibnamefont
  {G\"unther}}, \ and\ \bibinfo {author} {\bibfnamefont {Nimrod}\ \bibnamefont
  {Moiseyev}},\ }\bibfield  {title} {\enquote {\bibinfo {title} {Visualization
  of branch points in $\mathcal{P}\mathcal{T}$-symmetric waveguides},}\ }\href
  {\doibase 10.1103/PhysRevLett.101.080402} {\bibfield  {journal} {\bibinfo
  {journal} {Phys. Rev. Lett.}\ }\textbf {\bibinfo {volume} {101}},\ \bibinfo
  {pages} {080402} (\bibinfo {year} {2008})}\BibitemShut {NoStop}%
\bibitem [{\citenamefont {Regensburger}\ \emph {et~al.}(2012)\citenamefont
  {Regensburger}, \citenamefont {Bersch}, \citenamefont {Miri}, \citenamefont
  {Onishchukov}, \citenamefont {Christodoulides},\ and\ \citenamefont
  {Peschel}}]{regensburger2012parity}%
  \BibitemOpen
  \bibfield  {author} {\bibinfo {author} {\bibfnamefont {Alois}\ \bibnamefont
  {Regensburger}}, \bibinfo {author} {\bibfnamefont {Christoph}\ \bibnamefont
  {Bersch}}, \bibinfo {author} {\bibfnamefont {Mohammad-Ali}\ \bibnamefont
  {Miri}}, \bibinfo {author} {\bibfnamefont {Georgy}\ \bibnamefont
  {Onishchukov}}, \bibinfo {author} {\bibfnamefont {Demetrios~N}\ \bibnamefont
  {Christodoulides}}, \ and\ \bibinfo {author} {\bibfnamefont {Ulf}\
  \bibnamefont {Peschel}},\ }\bibfield  {title} {\enquote {\bibinfo {title}
  {Parity--time synthetic photonic lattices},}\ }\href@noop {} {\bibfield
  {journal} {\bibinfo  {journal} {Nature}\ }\textbf {\bibinfo {volume} {488}},\
  \bibinfo {pages} {167} (\bibinfo {year} {2012})}\BibitemShut {NoStop}%
\bibitem [{\citenamefont {Bittner}\ \emph {et~al.}(2012)\citenamefont
  {Bittner}, \citenamefont {Dietz}, \citenamefont {G\"unther}, \citenamefont
  {Harney}, \citenamefont {Miski-Oglu}, \citenamefont {Richter},\ and\
  \citenamefont {Sch\"afer}}]{bittner2012}%
  \BibitemOpen
  \bibfield  {author} {\bibinfo {author} {\bibfnamefont {S.}~\bibnamefont
  {Bittner}}, \bibinfo {author} {\bibfnamefont {B.}~\bibnamefont {Dietz}},
  \bibinfo {author} {\bibfnamefont {U.}~\bibnamefont {G\"unther}}, \bibinfo
  {author} {\bibfnamefont {H.~L.}\ \bibnamefont {Harney}}, \bibinfo {author}
  {\bibfnamefont {M.}~\bibnamefont {Miski-Oglu}}, \bibinfo {author}
  {\bibfnamefont {A.}~\bibnamefont {Richter}}, \ and\ \bibinfo {author}
  {\bibfnamefont {F.}~\bibnamefont {Sch\"afer}},\ }\bibfield  {title} {\enquote
  {\bibinfo {title} {Pt-symmetry and spontaneous symmetry breaking in a
  microwave billiard},}\ }\href {\doibase 10.1103/PhysRevLett.108.024101}
  {\bibfield  {journal} {\bibinfo  {journal} {Phys. Rev. Lett.}\ }\textbf
  {\bibinfo {volume} {108}},\ \bibinfo {pages} {024101} (\bibinfo {year}
  {2012})}\BibitemShut {NoStop}%
\bibitem [{\citenamefont {R{\"u}ter}\ \emph {et~al.}(2010)\citenamefont
  {R{\"u}ter}, \citenamefont {Makris}, \citenamefont {El-Ganainy},
  \citenamefont {Christodoulides}, \citenamefont {Segev},\ and\ \citenamefont
  {Kip}}]{ruter2010observation}%
  \BibitemOpen
  \bibfield  {author} {\bibinfo {author} {\bibfnamefont {Christian~E}\
  \bibnamefont {R{\"u}ter}}, \bibinfo {author} {\bibfnamefont {Konstantinos~G}\
  \bibnamefont {Makris}}, \bibinfo {author} {\bibfnamefont {Ramy}\ \bibnamefont
  {El-Ganainy}}, \bibinfo {author} {\bibfnamefont {Demetrios~N}\ \bibnamefont
  {Christodoulides}}, \bibinfo {author} {\bibfnamefont {Mordechai}\
  \bibnamefont {Segev}}, \ and\ \bibinfo {author} {\bibfnamefont {Detlef}\
  \bibnamefont {Kip}},\ }\bibfield  {title} {\enquote {\bibinfo {title}
  {Observation of parity--time symmetry in optics},}\ }\href@noop {} {\bibfield
   {journal} {\bibinfo  {journal} {Nature physics}\ }\textbf {\bibinfo {volume}
  {6}},\ \bibinfo {pages} {192} (\bibinfo {year} {2010})}\BibitemShut {NoStop}%
\bibitem [{\citenamefont {Lin}\ \emph {et~al.}(2011)\citenamefont {Lin},
  \citenamefont {Ramezani}, \citenamefont {Eichelkraut}, \citenamefont
  {Kottos}, \citenamefont {Cao},\ and\ \citenamefont
  {Christodoulides}}]{lin2011unidirectional}%
  \BibitemOpen
  \bibfield  {author} {\bibinfo {author} {\bibfnamefont {Zin}\ \bibnamefont
  {Lin}}, \bibinfo {author} {\bibfnamefont {Hamidreza}\ \bibnamefont
  {Ramezani}}, \bibinfo {author} {\bibfnamefont {Toni}\ \bibnamefont
  {Eichelkraut}}, \bibinfo {author} {\bibfnamefont {Tsampikos}\ \bibnamefont
  {Kottos}}, \bibinfo {author} {\bibfnamefont {Hui}\ \bibnamefont {Cao}}, \
  and\ \bibinfo {author} {\bibfnamefont {Demetrios~N.}\ \bibnamefont
  {Christodoulides}},\ }\bibfield  {title} {\enquote {\bibinfo {title}
  {Unidirectional invisibility induced by $\mathcal{P}\mathcal{T}$-symmetric
  periodic structures},}\ }\href {\doibase 10.1103/PhysRevLett.106.213901}
  {\bibfield  {journal} {\bibinfo  {journal} {Phys. Rev. Lett.}\ }\textbf
  {\bibinfo {volume} {106}},\ \bibinfo {pages} {213901} (\bibinfo {year}
  {2011})}\BibitemShut {NoStop}%
\bibitem [{\citenamefont {Feng}\ \emph {et~al.}(2013)\citenamefont {Feng},
  \citenamefont {Xu}, \citenamefont {Fegadolli}, \citenamefont {Lu},
  \citenamefont {Oliveira}, \citenamefont {Almeida}, \citenamefont {Chen},\
  and\ \citenamefont {Scherer}}]{feng2013experimental}%
  \BibitemOpen
  \bibfield  {author} {\bibinfo {author} {\bibfnamefont {Liang}\ \bibnamefont
  {Feng}}, \bibinfo {author} {\bibfnamefont {Ye-Long}\ \bibnamefont {Xu}},
  \bibinfo {author} {\bibfnamefont {William~S}\ \bibnamefont {Fegadolli}},
  \bibinfo {author} {\bibfnamefont {Ming-Hui}\ \bibnamefont {Lu}}, \bibinfo
  {author} {\bibfnamefont {Jos{\'e}~EB}\ \bibnamefont {Oliveira}}, \bibinfo
  {author} {\bibfnamefont {Vilson~R}\ \bibnamefont {Almeida}}, \bibinfo
  {author} {\bibfnamefont {Yan-Feng}\ \bibnamefont {Chen}}, \ and\ \bibinfo
  {author} {\bibfnamefont {Axel}\ \bibnamefont {Scherer}},\ }\bibfield  {title}
  {\enquote {\bibinfo {title} {Experimental demonstration of a unidirectional
  reflectionless parity-time metamaterial at optical frequencies},}\
  }\href@noop {} {\bibfield  {journal} {\bibinfo  {journal} {Nature materials}\
  }\textbf {\bibinfo {volume} {12}},\ \bibinfo {pages} {108} (\bibinfo {year}
  {2013})}\BibitemShut {NoStop}%
\bibitem [{\citenamefont {Guo}\ \emph {et~al.}(2009)\citenamefont {Guo},
  \citenamefont {Salamo}, \citenamefont {Duchesne}, \citenamefont {Morandotti},
  \citenamefont {Volatier-Ravat}, \citenamefont {Aimez}, \citenamefont
  {Siviloglou},\ and\ \citenamefont {Christodoulides}}]{guo2009complex}%
  \BibitemOpen
  \bibfield  {author} {\bibinfo {author} {\bibfnamefont {A.}~\bibnamefont
  {Guo}}, \bibinfo {author} {\bibfnamefont {G.~J.}\ \bibnamefont {Salamo}},
  \bibinfo {author} {\bibfnamefont {D.}~\bibnamefont {Duchesne}}, \bibinfo
  {author} {\bibfnamefont {R.}~\bibnamefont {Morandotti}}, \bibinfo {author}
  {\bibfnamefont {M.}~\bibnamefont {Volatier-Ravat}}, \bibinfo {author}
  {\bibfnamefont {V.}~\bibnamefont {Aimez}}, \bibinfo {author} {\bibfnamefont
  {G.~A.}\ \bibnamefont {Siviloglou}}, \ and\ \bibinfo {author} {\bibfnamefont
  {D.~N.}\ \bibnamefont {Christodoulides}},\ }\bibfield  {title} {\enquote
  {\bibinfo {title} {Observation of $\mathcal{P}\mathcal{T}$-symmetry breaking
  in complex optical potentials},}\ }\href {\doibase
  10.1103/PhysRevLett.103.093902} {\bibfield  {journal} {\bibinfo  {journal}
  {Phys. Rev. Lett.}\ }\textbf {\bibinfo {volume} {103}},\ \bibinfo {pages}
  {093902} (\bibinfo {year} {2009})}\BibitemShut {NoStop}%
\bibitem [{\citenamefont {Liertzer}\ \emph {et~al.}(2012)\citenamefont
  {Liertzer}, \citenamefont {Ge}, \citenamefont {Cerjan}, \citenamefont
  {Stone}, \citenamefont {T\"ureci},\ and\ \citenamefont
  {Rotter}}]{liertzer2012pumpinduced}%
  \BibitemOpen
  \bibfield  {author} {\bibinfo {author} {\bibfnamefont {M.}~\bibnamefont
  {Liertzer}}, \bibinfo {author} {\bibfnamefont {Li}~\bibnamefont {Ge}},
  \bibinfo {author} {\bibfnamefont {A.}~\bibnamefont {Cerjan}}, \bibinfo
  {author} {\bibfnamefont {A.~D.}\ \bibnamefont {Stone}}, \bibinfo {author}
  {\bibfnamefont {H.~E.}\ \bibnamefont {T\"ureci}}, \ and\ \bibinfo {author}
  {\bibfnamefont {S.}~\bibnamefont {Rotter}},\ }\bibfield  {title} {\enquote
  {\bibinfo {title} {Pump-induced exceptional points in lasers},}\ }\href
  {\doibase 10.1103/PhysRevLett.108.173901} {\bibfield  {journal} {\bibinfo
  {journal} {Phys. Rev. Lett.}\ }\textbf {\bibinfo {volume} {108}},\ \bibinfo
  {pages} {173901} (\bibinfo {year} {2012})}\BibitemShut {NoStop}%
\bibitem [{\citenamefont {Peng}\ \emph {et~al.}(2014)\citenamefont {Peng},
  \citenamefont {{\"O}zdemir}, \citenamefont {Rotter}, \citenamefont {Yilmaz},
  \citenamefont {Liertzer}, \citenamefont {Monifi}, \citenamefont {Bender},
  \citenamefont {Nori},\ and\ \citenamefont {Yang}}]{peng2014lossinduced}%
  \BibitemOpen
  \bibfield  {author} {\bibinfo {author} {\bibfnamefont {B}~\bibnamefont
  {Peng}}, \bibinfo {author} {\bibfnamefont {{\c{S}}K}~\bibnamefont
  {{\"O}zdemir}}, \bibinfo {author} {\bibfnamefont {S}~\bibnamefont {Rotter}},
  \bibinfo {author} {\bibfnamefont {H}~\bibnamefont {Yilmaz}}, \bibinfo
  {author} {\bibfnamefont {M}~\bibnamefont {Liertzer}}, \bibinfo {author}
  {\bibfnamefont {F}~\bibnamefont {Monifi}}, \bibinfo {author} {\bibfnamefont
  {CM}~\bibnamefont {Bender}}, \bibinfo {author} {\bibfnamefont
  {F}~\bibnamefont {Nori}}, \ and\ \bibinfo {author} {\bibfnamefont
  {L}~\bibnamefont {Yang}},\ }\bibfield  {title} {\enquote {\bibinfo {title}
  {Loss-induced suppression and revival of lasing},}\ }\href@noop {} {\bibfield
   {journal} {\bibinfo  {journal} {Science}\ }\textbf {\bibinfo {volume}
  {346}},\ \bibinfo {pages} {328--332} (\bibinfo {year} {2014})}\BibitemShut
  {NoStop}%
\bibitem [{\citenamefont {Fleury}\ \emph {et~al.}(2015)\citenamefont {Fleury},
  \citenamefont {Sounas},\ and\ \citenamefont {Al{\`u}}}]{fleury2015invisible}%
  \BibitemOpen
  \bibfield  {author} {\bibinfo {author} {\bibfnamefont {Romain}\ \bibnamefont
  {Fleury}}, \bibinfo {author} {\bibfnamefont {Dimitrios}\ \bibnamefont
  {Sounas}}, \ and\ \bibinfo {author} {\bibfnamefont {Andrea}\ \bibnamefont
  {Al{\`u}}},\ }\bibfield  {title} {\enquote {\bibinfo {title} {An invisible
  acoustic sensor based on parity-time symmetry},}\ }\href@noop {} {\bibfield
  {journal} {\bibinfo  {journal} {Nature communications}\ }\textbf {\bibinfo
  {volume} {6}},\ \bibinfo {pages} {5905} (\bibinfo {year} {2015})}\BibitemShut
  {NoStop}%
\bibitem [{\citenamefont {Chang}\ \emph {et~al.}(2014)\citenamefont {Chang},
  \citenamefont {Jiang}, \citenamefont {Hua}, \citenamefont {Yang},
  \citenamefont {Wen}, \citenamefont {Jiang}, \citenamefont {Li}, \citenamefont
  {Wang},\ and\ \citenamefont {Xiao}}]{chang2014PT}%
  \BibitemOpen
  \bibfield  {author} {\bibinfo {author} {\bibfnamefont {Long}\ \bibnamefont
  {Chang}}, \bibinfo {author} {\bibfnamefont {Xiaoshun}\ \bibnamefont {Jiang}},
  \bibinfo {author} {\bibfnamefont {Shiyue}\ \bibnamefont {Hua}}, \bibinfo
  {author} {\bibfnamefont {Chao}\ \bibnamefont {Yang}}, \bibinfo {author}
  {\bibfnamefont {Jianming}\ \bibnamefont {Wen}}, \bibinfo {author}
  {\bibfnamefont {Liang}\ \bibnamefont {Jiang}}, \bibinfo {author}
  {\bibfnamefont {Guanyu}\ \bibnamefont {Li}}, \bibinfo {author} {\bibfnamefont
  {Guanzhong}\ \bibnamefont {Wang}}, \ and\ \bibinfo {author} {\bibfnamefont
  {Min}\ \bibnamefont {Xiao}},\ }\bibfield  {title} {\enquote {\bibinfo {title}
  {Parity--time symmetry and variable optical isolation in
  active--passive-coupled microresonators},}\ }\href@noop {} {\bibfield
  {journal} {\bibinfo  {journal} {Nature photonics}\ }\textbf {\bibinfo
  {volume} {8}},\ \bibinfo {pages} {524} (\bibinfo {year} {2014})}\BibitemShut
  {NoStop}%
\bibitem [{\citenamefont {Hodaei}\ \emph {et~al.}(2017)\citenamefont {Hodaei},
  \citenamefont {Hassan}, \citenamefont {Wittek}, \citenamefont
  {Garcia-Gracia}, \citenamefont {El-Ganainy}, \citenamefont
  {Christodoulides},\ and\ \citenamefont {Khajavikhan}}]{hodaei2017enhanced}%
  \BibitemOpen
  \bibfield  {author} {\bibinfo {author} {\bibfnamefont {Hossein}\ \bibnamefont
  {Hodaei}}, \bibinfo {author} {\bibfnamefont {Absar~U}\ \bibnamefont
  {Hassan}}, \bibinfo {author} {\bibfnamefont {Steffen}\ \bibnamefont
  {Wittek}}, \bibinfo {author} {\bibfnamefont {Hipolito}\ \bibnamefont
  {Garcia-Gracia}}, \bibinfo {author} {\bibfnamefont {Ramy}\ \bibnamefont
  {El-Ganainy}}, \bibinfo {author} {\bibfnamefont {Demetrios~N}\ \bibnamefont
  {Christodoulides}}, \ and\ \bibinfo {author} {\bibfnamefont {Mercedeh}\
  \bibnamefont {Khajavikhan}},\ }\bibfield  {title} {\enquote {\bibinfo {title}
  {Enhanced sensitivity at higher-order exceptional points},}\ }\href@noop {}
  {\bibfield  {journal} {\bibinfo  {journal} {Nature}\ }\textbf {\bibinfo
  {volume} {548}},\ \bibinfo {pages} {187} (\bibinfo {year}
  {2017})}\BibitemShut {NoStop}%
\bibitem [{\citenamefont {Hodaei}\ \emph {et~al.}(2014)\citenamefont {Hodaei},
  \citenamefont {Miri}, \citenamefont {Heinrich}, \citenamefont
  {Christodoulides},\ and\ \citenamefont {Khajavikhan}}]{hodaei2014PT}%
  \BibitemOpen
  \bibfield  {author} {\bibinfo {author} {\bibfnamefont {Hossein}\ \bibnamefont
  {Hodaei}}, \bibinfo {author} {\bibfnamefont {Mohammad-Ali}\ \bibnamefont
  {Miri}}, \bibinfo {author} {\bibfnamefont {Matthias}\ \bibnamefont
  {Heinrich}}, \bibinfo {author} {\bibfnamefont {Demetrios~N}\ \bibnamefont
  {Christodoulides}}, \ and\ \bibinfo {author} {\bibfnamefont {Mercedeh}\
  \bibnamefont {Khajavikhan}},\ }\bibfield  {title} {\enquote {\bibinfo {title}
  {Parity-time--symmetric microring lasers},}\ }\href@noop {} {\bibfield
  {journal} {\bibinfo  {journal} {Science}\ }\textbf {\bibinfo {volume}
  {346}},\ \bibinfo {pages} {975--978} (\bibinfo {year} {2014})}\BibitemShut
  {NoStop}%
\bibitem [{\citenamefont {Feng}\ \emph {et~al.}(2014)\citenamefont {Feng},
  \citenamefont {Wong}, \citenamefont {Ma}, \citenamefont {Wang},\ and\
  \citenamefont {Zhang}}]{feng2014singlemode}%
  \BibitemOpen
  \bibfield  {author} {\bibinfo {author} {\bibfnamefont {Liang}\ \bibnamefont
  {Feng}}, \bibinfo {author} {\bibfnamefont {Zi~Jing}\ \bibnamefont {Wong}},
  \bibinfo {author} {\bibfnamefont {Ren-Min}\ \bibnamefont {Ma}}, \bibinfo
  {author} {\bibfnamefont {Yuan}\ \bibnamefont {Wang}}, \ and\ \bibinfo
  {author} {\bibfnamefont {Xiang}\ \bibnamefont {Zhang}},\ }\bibfield  {title}
  {\enquote {\bibinfo {title} {Single-mode laser by parity-time symmetry
  breaking},}\ }\href@noop {} {\bibfield  {journal} {\bibinfo  {journal}
  {Science}\ }\textbf {\bibinfo {volume} {346}},\ \bibinfo {pages} {972--975}
  (\bibinfo {year} {2014})}\BibitemShut {NoStop}%
\bibitem [{\citenamefont {Gao}\ \emph {et~al.}(2015)\citenamefont {Gao},
  \citenamefont {Estrecho}, \citenamefont {Bliokh}, \citenamefont {Liew},
  \citenamefont {Fraser}, \citenamefont {Brodbeck}, \citenamefont {Kamp},
  \citenamefont {Schneider}, \citenamefont {H{\"o}fling}, \citenamefont
  {Yamamoto} \emph {et~al.}}]{gao2015billiard}%
  \BibitemOpen
  \bibfield  {author} {\bibinfo {author} {\bibfnamefont {Tiejun}\ \bibnamefont
  {Gao}}, \bibinfo {author} {\bibfnamefont {E}~\bibnamefont {Estrecho}},
  \bibinfo {author} {\bibfnamefont {KY}~\bibnamefont {Bliokh}}, \bibinfo
  {author} {\bibfnamefont {TCH}\ \bibnamefont {Liew}}, \bibinfo {author}
  {\bibfnamefont {MD}~\bibnamefont {Fraser}}, \bibinfo {author} {\bibfnamefont
  {Sebastian}\ \bibnamefont {Brodbeck}}, \bibinfo {author} {\bibfnamefont
  {Martin}\ \bibnamefont {Kamp}}, \bibinfo {author} {\bibfnamefont {Christian}\
  \bibnamefont {Schneider}}, \bibinfo {author} {\bibfnamefont {Sven}\
  \bibnamefont {H{\"o}fling}}, \bibinfo {author} {\bibfnamefont
  {Y}~\bibnamefont {Yamamoto}},  \emph {et~al.},\ }\bibfield  {title} {\enquote
  {\bibinfo {title} {Observation of non-hermitian degeneracies in a chaotic
  exciton-polariton billiard},}\ }\href@noop {} {\bibfield  {journal} {\bibinfo
   {journal} {Nature}\ }\textbf {\bibinfo {volume} {526}},\ \bibinfo {pages}
  {554} (\bibinfo {year} {2015})}\BibitemShut {NoStop}%
\bibitem [{\citenamefont {Xu}\ \emph {et~al.}(2016)\citenamefont {Xu},
  \citenamefont {Mason}, \citenamefont {Jiang},\ and\ \citenamefont
  {Harris}}]{xu2016topological}%
  \BibitemOpen
  \bibfield  {author} {\bibinfo {author} {\bibfnamefont {Haitan}\ \bibnamefont
  {Xu}}, \bibinfo {author} {\bibfnamefont {David}\ \bibnamefont {Mason}},
  \bibinfo {author} {\bibfnamefont {Luyao}\ \bibnamefont {Jiang}}, \ and\
  \bibinfo {author} {\bibfnamefont {JGE}\ \bibnamefont {Harris}},\ }\bibfield
  {title} {\enquote {\bibinfo {title} {Topological energy transfer in an
  optomechanical system with exceptional points},}\ }\href@noop {} {\bibfield
  {journal} {\bibinfo  {journal} {Nature}\ }\textbf {\bibinfo {volume} {537}},\
  \bibinfo {pages} {80} (\bibinfo {year} {2016})}\BibitemShut {NoStop}%
\bibitem [{\citenamefont {Ashida}\ \emph {et~al.}(2017)\citenamefont {Ashida},
  \citenamefont {Furukawa},\ and\ \citenamefont {Ueda}}]{ashida2017parity}%
  \BibitemOpen
  \bibfield  {author} {\bibinfo {author} {\bibfnamefont {Yuto}\ \bibnamefont
  {Ashida}}, \bibinfo {author} {\bibfnamefont {Shunsuke}\ \bibnamefont
  {Furukawa}}, \ and\ \bibinfo {author} {\bibfnamefont {Masahito}\ \bibnamefont
  {Ueda}},\ }\bibfield  {title} {\enquote {\bibinfo {title}
  {Parity-time-symmetric quantum critical phenomena},}\ }\href@noop {}
  {\bibfield  {journal} {\bibinfo  {journal} {Nature communications}\ }\textbf
  {\bibinfo {volume} {8}},\ \bibinfo {pages} {15791} (\bibinfo {year}
  {2017})}\BibitemShut {NoStop}%
\bibitem [{\citenamefont {Kawabata}\ \emph {et~al.}(2017)\citenamefont
  {Kawabata}, \citenamefont {Ashida},\ and\ \citenamefont
  {Ueda}}]{kawabata2017retrieval}%
  \BibitemOpen
  \bibfield  {author} {\bibinfo {author} {\bibfnamefont {Kohei}\ \bibnamefont
  {Kawabata}}, \bibinfo {author} {\bibfnamefont {Yuto}\ \bibnamefont {Ashida}},
  \ and\ \bibinfo {author} {\bibfnamefont {Masahito}\ \bibnamefont {Ueda}},\
  }\bibfield  {title} {\enquote {\bibinfo {title} {Information retrieval and
  criticality in parity-time-symmetric systems},}\ }\href {\doibase
  10.1103/PhysRevLett.119.190401} {\bibfield  {journal} {\bibinfo  {journal}
  {Phys. Rev. Lett.}\ }\textbf {\bibinfo {volume} {119}},\ \bibinfo {pages}
  {190401} (\bibinfo {year} {2017})}\BibitemShut {NoStop}%
\bibitem [{\citenamefont {Chen}\ \emph {et~al.}(2017)\citenamefont {Chen},
  \citenamefont {{\"O}zdemir}, \citenamefont {Zhao}, \citenamefont {Wiersig},\
  and\ \citenamefont {Yang}}]{chen2017exceptional}%
  \BibitemOpen
  \bibfield  {author} {\bibinfo {author} {\bibfnamefont {Weijian}\ \bibnamefont
  {Chen}}, \bibinfo {author} {\bibfnamefont {{\c{S}}ahin~Kaya}\ \bibnamefont
  {{\"O}zdemir}}, \bibinfo {author} {\bibfnamefont {Guangming}\ \bibnamefont
  {Zhao}}, \bibinfo {author} {\bibfnamefont {Jan}\ \bibnamefont {Wiersig}}, \
  and\ \bibinfo {author} {\bibfnamefont {Lan}\ \bibnamefont {Yang}},\
  }\bibfield  {title} {\enquote {\bibinfo {title} {Exceptional points enhance
  sensing in an optical microcavity},}\ }\href@noop {} {\bibfield  {journal}
  {\bibinfo  {journal} {Nature}\ }\textbf {\bibinfo {volume} {548}},\ \bibinfo
  {pages} {192} (\bibinfo {year} {2017})}\BibitemShut {NoStop}%
\bibitem [{\citenamefont {Ding}\ \emph {et~al.}(2016)\citenamefont {Ding},
  \citenamefont {Ma}, \citenamefont {Xiao}, \citenamefont {Zhang},\ and\
  \citenamefont {Chan}}]{ding2016multiple}%
  \BibitemOpen
  \bibfield  {author} {\bibinfo {author} {\bibfnamefont {Kun}\ \bibnamefont
  {Ding}}, \bibinfo {author} {\bibfnamefont {Guancong}\ \bibnamefont {Ma}},
  \bibinfo {author} {\bibfnamefont {Meng}\ \bibnamefont {Xiao}}, \bibinfo
  {author} {\bibfnamefont {Z.~Q.}\ \bibnamefont {Zhang}}, \ and\ \bibinfo
  {author} {\bibfnamefont {C.~T.}\ \bibnamefont {Chan}},\ }\bibfield  {title}
  {\enquote {\bibinfo {title} {Emergence, coalescence, and topological
  properties of multiple exceptional points and their experimental
  realization},}\ }\href {\doibase 10.1103/PhysRevX.6.021007} {\bibfield
  {journal} {\bibinfo  {journal} {Phys. Rev. X}\ }\textbf {\bibinfo {volume}
  {6}},\ \bibinfo {pages} {021007} (\bibinfo {year} {2016})}\BibitemShut
  {NoStop}%
\bibitem [{\citenamefont {Downing}\ and\ \citenamefont
  {Weick}(2017)}]{downing2017}%
  \BibitemOpen
  \bibfield  {author} {\bibinfo {author} {\bibfnamefont {Charles~A.}\
  \bibnamefont {Downing}}\ and\ \bibinfo {author} {\bibfnamefont {Guillaume}\
  \bibnamefont {Weick}},\ }\bibfield  {title} {\enquote {\bibinfo {title}
  {Topological collective plasmons in bipartite chains of metallic
  nanoparticles},}\ }\href {\doibase 10.1103/PhysRevB.95.125426} {\bibfield
  {journal} {\bibinfo  {journal} {Phys. Rev. B}\ }\textbf {\bibinfo {volume}
  {95}},\ \bibinfo {pages} {125426} (\bibinfo {year} {2017})}\BibitemShut
  {NoStop}%
\bibitem [{\citenamefont {{Ozawa}}\ \emph {et~al.}(2018)\citenamefont
  {{Ozawa}}, \citenamefont {{Price}}, \citenamefont {{Amo}}, \citenamefont
  {{Goldman}}, \citenamefont {{Hafezi}}, \citenamefont {{Lu}}, \citenamefont
  {{Rechtsman}}, \citenamefont {{Schuster}}, \citenamefont {{Simon}},
  \citenamefont {{Zilberberg}},\ and\ \citenamefont
  {{Carusotto}}}]{ozawa2018rmp}%
  \BibitemOpen
  \bibfield  {author} {\bibinfo {author} {\bibfnamefont {T.}~\bibnamefont
  {{Ozawa}}}, \bibinfo {author} {\bibfnamefont {H.~M.}\ \bibnamefont
  {{Price}}}, \bibinfo {author} {\bibfnamefont {A.}~\bibnamefont {{Amo}}},
  \bibinfo {author} {\bibfnamefont {N.}~\bibnamefont {{Goldman}}}, \bibinfo
  {author} {\bibfnamefont {M.}~\bibnamefont {{Hafezi}}}, \bibinfo {author}
  {\bibfnamefont {L.}~\bibnamefont {{Lu}}}, \bibinfo {author} {\bibfnamefont
  {M.}~\bibnamefont {{Rechtsman}}}, \bibinfo {author} {\bibfnamefont
  {D.}~\bibnamefont {{Schuster}}}, \bibinfo {author} {\bibfnamefont
  {J.}~\bibnamefont {{Simon}}}, \bibinfo {author} {\bibfnamefont
  {O.}~\bibnamefont {{Zilberberg}}}, \ and\ \bibinfo {author} {\bibfnamefont
  {I.}~\bibnamefont {{Carusotto}}},\ }\bibfield  {title} {\enquote {\bibinfo
  {title} {{Topological Photonics}},}\ }\href@noop {} {\bibfield  {journal}
  {\bibinfo  {journal} {ArXiv e-prints}\ } (\bibinfo {year} {2018})},\ \Eprint
  {http://arxiv.org/abs/1802.04173} {arXiv:1802.04173 [physics.optics]}
  \BibitemShut {NoStop}%
\bibitem [{\citenamefont {Lu}\ \emph {et~al.}(2014)\citenamefont {Lu},
  \citenamefont {Joannopoulos},\ and\ \citenamefont {Soljacic}}]{Lu2014review}%
  \BibitemOpen
  \bibfield  {author} {\bibinfo {author} {\bibfnamefont {Ling}\ \bibnamefont
  {Lu}}, \bibinfo {author} {\bibfnamefont {John~D.}\ \bibnamefont
  {Joannopoulos}}, \ and\ \bibinfo {author} {\bibfnamefont {Marin}\
  \bibnamefont {Soljacic}},\ }\bibfield  {title} {\enquote {\bibinfo {title}
  {Topological photonics},}\ }\href@noop {} {\bibfield  {journal} {\bibinfo
  {journal} {Nat Photon}\ }\textbf {\bibinfo {volume} {8}},\ \bibinfo {pages}
  {821--829} (\bibinfo {year} {2014})}\BibitemShut {NoStop}%
\bibitem [{\citenamefont {El-Ganainy}\ \emph {et~al.}(2018)\citenamefont
  {El-Ganainy}, \citenamefont {Makris}, \citenamefont {Khajavikhan},
  \citenamefont {Musslimani}, \citenamefont {Rotter},\ and\ \citenamefont
  {Christodoulides}}]{el2018non}%
  \BibitemOpen
  \bibfield  {author} {\bibinfo {author} {\bibfnamefont {Ramy}\ \bibnamefont
  {El-Ganainy}}, \bibinfo {author} {\bibfnamefont {Konstantinos~G}\
  \bibnamefont {Makris}}, \bibinfo {author} {\bibfnamefont {Mercedeh}\
  \bibnamefont {Khajavikhan}}, \bibinfo {author} {\bibfnamefont {Ziad~H}\
  \bibnamefont {Musslimani}}, \bibinfo {author} {\bibfnamefont {Stefan}\
  \bibnamefont {Rotter}}, \ and\ \bibinfo {author} {\bibfnamefont
  {Demetrios~N}\ \bibnamefont {Christodoulides}},\ }\bibfield  {title}
  {\enquote {\bibinfo {title} {Non-hermitian physics and pt symmetry},}\
  }\href@noop {} {\bibfield  {journal} {\bibinfo  {journal} {Nature Physics}\
  }\textbf {\bibinfo {volume} {14}},\ \bibinfo {pages} {11} (\bibinfo {year}
  {2018})}\BibitemShut {NoStop}%
\bibitem [{\citenamefont {{Longhi}}(2018)}]{longhi2018}%
  \BibitemOpen
  \bibfield  {author} {\bibinfo {author} {\bibfnamefont {S.}~\bibnamefont
  {{Longhi}}},\ }\bibfield  {title} {\enquote {\bibinfo {title} {{Parity-Time
  Symmetry meets Photonics: A New Twist in non-Hermitian Optics}},}\
  }\href@noop {} {\bibfield  {journal} {\bibinfo  {journal} {ArXiv e-prints}\ }
  (\bibinfo {year} {2018})},\ \Eprint {http://arxiv.org/abs/1802.05025}
  {arXiv:1802.05025 [physics.optics]} \BibitemShut {NoStop}%
\bibitem [{\citenamefont {{Kozii}}\ and\ \citenamefont
  {{Fu}}(2017)}]{kozii2017}%
  \BibitemOpen
  \bibfield  {author} {\bibinfo {author} {\bibfnamefont {V.}~\bibnamefont
  {{Kozii}}}\ and\ \bibinfo {author} {\bibfnamefont {L.}~\bibnamefont {{Fu}}},\
  }\bibfield  {title} {\enquote {\bibinfo {title} {{Non-Hermitian Topological
  Theory of Finite-Lifetime Quasiparticles: Prediction of Bulk Fermi Arc Due to
  Exceptional Point}},}\ }\href@noop {} {\bibfield  {journal} {\bibinfo
  {journal} {ArXiv e-prints}\ } (\bibinfo {year} {2017})},\ \Eprint
  {http://arxiv.org/abs/1708.05841} {arXiv:1708.05841 [cond-mat.mes-hall]}
  \BibitemShut {NoStop}%
\bibitem [{\citenamefont {{Papaj}}\ \emph {et~al.}(2018)\citenamefont
  {{Papaj}}, \citenamefont {{Isobe}},\ and\ \citenamefont
  {{Fu}}}]{papa2018bulk}%
  \BibitemOpen
  \bibfield  {author} {\bibinfo {author} {\bibfnamefont {M.}~\bibnamefont
  {{Papaj}}}, \bibinfo {author} {\bibfnamefont {H.}~\bibnamefont {{Isobe}}}, \
  and\ \bibinfo {author} {\bibfnamefont {L.}~\bibnamefont {{Fu}}},\ }\bibfield
  {title} {\enquote {\bibinfo {title} {{Bulk Fermi arc of disordered Dirac
  fermions in two dimensions}},}\ }\href@noop {} {\bibfield  {journal}
  {\bibinfo  {journal} {ArXiv e-prints}\ } (\bibinfo {year} {2018})},\ \Eprint
  {http://arxiv.org/abs/1802.00443} {arXiv:1802.00443 [cond-mat.dis-nn]}
  \BibitemShut {NoStop}%
\bibitem [{\citenamefont {{Shen}}\ and\ \citenamefont
  {{Fu}}(2018)}]{shen2018quantum}%
  \BibitemOpen
  \bibfield  {author} {\bibinfo {author} {\bibfnamefont {H.}~\bibnamefont
  {{Shen}}}\ and\ \bibinfo {author} {\bibfnamefont {L.}~\bibnamefont {{Fu}}},\
  }\bibfield  {title} {\enquote {\bibinfo {title} {{Quantum Oscillation from
  In-gap States and non-Hermitian Landau Level Problem}},}\ }\href@noop {}
  {\bibfield  {journal} {\bibinfo  {journal} {ArXiv e-prints}\ } (\bibinfo
  {year} {2018})},\ \Eprint {http://arxiv.org/abs/1802.03023} {arXiv:1802.03023
  [cond-mat.str-el]} \BibitemShut {NoStop}%
\bibitem [{\citenamefont {Esaki}\ \emph {et~al.}(2011)\citenamefont {Esaki},
  \citenamefont {Sato}, \citenamefont {Hasebe},\ and\ \citenamefont
  {Kohmoto}}]{esaki2011}%
  \BibitemOpen
  \bibfield  {author} {\bibinfo {author} {\bibfnamefont {Kenta}\ \bibnamefont
  {Esaki}}, \bibinfo {author} {\bibfnamefont {Masatoshi}\ \bibnamefont {Sato}},
  \bibinfo {author} {\bibfnamefont {Kazuki}\ \bibnamefont {Hasebe}}, \ and\
  \bibinfo {author} {\bibfnamefont {Mahito}\ \bibnamefont {Kohmoto}},\
  }\bibfield  {title} {\enquote {\bibinfo {title} {Edge states and topological
  phases in non-hermitian systems},}\ }\href {\doibase
  10.1103/PhysRevB.84.205128} {\bibfield  {journal} {\bibinfo  {journal} {Phys.
  Rev. B}\ }\textbf {\bibinfo {volume} {84}},\ \bibinfo {pages} {205128}
  (\bibinfo {year} {2011})}\BibitemShut {NoStop}%
\bibitem [{\citenamefont {Lee}(2016)}]{lee2016anomalous}%
  \BibitemOpen
  \bibfield  {author} {\bibinfo {author} {\bibfnamefont {Tony~E.}\ \bibnamefont
  {Lee}},\ }\bibfield  {title} {\enquote {\bibinfo {title} {Anomalous edge
  state in a non-hermitian lattice},}\ }\href {\doibase
  10.1103/PhysRevLett.116.133903} {\bibfield  {journal} {\bibinfo  {journal}
  {Phys. Rev. Lett.}\ }\textbf {\bibinfo {volume} {116}},\ \bibinfo {pages}
  {133903} (\bibinfo {year} {2016})}\BibitemShut {NoStop}%
\bibitem [{\citenamefont {Leykam}\ \emph {et~al.}(2017)\citenamefont {Leykam},
  \citenamefont {Bliokh}, \citenamefont {Huang}, \citenamefont {Chong},\ and\
  \citenamefont {Nori}}]{leykam2017}%
  \BibitemOpen
  \bibfield  {author} {\bibinfo {author} {\bibfnamefont {Daniel}\ \bibnamefont
  {Leykam}}, \bibinfo {author} {\bibfnamefont {Konstantin~Y.}\ \bibnamefont
  {Bliokh}}, \bibinfo {author} {\bibfnamefont {Chunli}\ \bibnamefont {Huang}},
  \bibinfo {author} {\bibfnamefont {Y.~D.}\ \bibnamefont {Chong}}, \ and\
  \bibinfo {author} {\bibfnamefont {Franco}\ \bibnamefont {Nori}},\ }\bibfield
  {title} {\enquote {\bibinfo {title} {Edge modes, degeneracies, and
  topological numbers in non-hermitian systems},}\ }\href {\doibase
  10.1103/PhysRevLett.118.040401} {\bibfield  {journal} {\bibinfo  {journal}
  {Phys. Rev. Lett.}\ }\textbf {\bibinfo {volume} {118}},\ \bibinfo {pages}
  {040401} (\bibinfo {year} {2017})}\BibitemShut {NoStop}%
\bibitem [{\citenamefont {Menke}\ and\ \citenamefont
  {Hirschmann}(2017)}]{menke2017}%
  \BibitemOpen
  \bibfield  {author} {\bibinfo {author} {\bibfnamefont {Henri}\ \bibnamefont
  {Menke}}\ and\ \bibinfo {author} {\bibfnamefont {Moritz~M.}\ \bibnamefont
  {Hirschmann}},\ }\bibfield  {title} {\enquote {\bibinfo {title} {Topological
  quantum wires with balanced gain and loss},}\ }\href {\doibase
  10.1103/PhysRevB.95.174506} {\bibfield  {journal} {\bibinfo  {journal} {Phys.
  Rev. B}\ }\textbf {\bibinfo {volume} {95}},\ \bibinfo {pages} {174506}
  (\bibinfo {year} {2017})}\BibitemShut {NoStop}%
\bibitem [{\citenamefont {Lieu}(2018)}]{lieu2018ssh}%
  \BibitemOpen
  \bibfield  {author} {\bibinfo {author} {\bibfnamefont {Simon}\ \bibnamefont
  {Lieu}},\ }\bibfield  {title} {\enquote {\bibinfo {title} {Topological phases
  in the non-hermitian su-schrieffer-heeger model},}\ }\href {\doibase
  10.1103/PhysRevB.97.045106} {\bibfield  {journal} {\bibinfo  {journal} {Phys.
  Rev. B}\ }\textbf {\bibinfo {volume} {97}},\ \bibinfo {pages} {045106}
  (\bibinfo {year} {2018})}\BibitemShut {NoStop}%
\bibitem [{\citenamefont {Shen}\ \emph {et~al.}(2018)\citenamefont {Shen},
  \citenamefont {Zhen},\ and\ \citenamefont {Fu}}]{shen2017topological}%
  \BibitemOpen
  \bibfield  {author} {\bibinfo {author} {\bibfnamefont {Huitao}\ \bibnamefont
  {Shen}}, \bibinfo {author} {\bibfnamefont {Bo}~\bibnamefont {Zhen}}, \ and\
  \bibinfo {author} {\bibfnamefont {Liang}\ \bibnamefont {Fu}},\ }\bibfield
  {title} {\enquote {\bibinfo {title} {Topological band theory for
  non-hermitian hamiltonians},}\ }\href {\doibase
  10.1103/PhysRevLett.120.146402} {\bibfield  {journal} {\bibinfo  {journal}
  {Phys. Rev. Lett.}\ }\textbf {\bibinfo {volume} {120}},\ \bibinfo {pages}
  {146402} (\bibinfo {year} {2018})}\BibitemShut {NoStop}%
\bibitem [{\citenamefont {Yin}\ \emph {et~al.}(2018)\citenamefont {Yin},
  \citenamefont {Jiang}, \citenamefont {Li}, \citenamefont {L\"u},\ and\
  \citenamefont {Chen}}]{yin2018ssh}%
  \BibitemOpen
  \bibfield  {author} {\bibinfo {author} {\bibfnamefont {Chuanhao}\
  \bibnamefont {Yin}}, \bibinfo {author} {\bibfnamefont {Hui}\ \bibnamefont
  {Jiang}}, \bibinfo {author} {\bibfnamefont {Linhu}\ \bibnamefont {Li}},
  \bibinfo {author} {\bibfnamefont {Rong}\ \bibnamefont {L\"u}}, \ and\
  \bibinfo {author} {\bibfnamefont {Shu}\ \bibnamefont {Chen}},\ }\bibfield
  {title} {\enquote {\bibinfo {title} {Geometrical meaning of winding number
  and its characterization of topological phases in one-dimensional chiral
  non-hermitian systems},}\ }\href {\doibase 10.1103/PhysRevA.97.052115}
  {\bibfield  {journal} {\bibinfo  {journal} {Phys. Rev. A}\ }\textbf {\bibinfo
  {volume} {97}},\ \bibinfo {pages} {052115} (\bibinfo {year}
  {2018})}\BibitemShut {NoStop}%
\bibitem [{\citenamefont {Li}\ \emph {et~al.}(2018)\citenamefont {Li},
  \citenamefont {Zhang}, \citenamefont {Zhang},\ and\ \citenamefont
  {Song}}]{li2017kitaev}%
  \BibitemOpen
  \bibfield  {author} {\bibinfo {author} {\bibfnamefont {C.}~\bibnamefont
  {Li}}, \bibinfo {author} {\bibfnamefont {X.~Z.}\ \bibnamefont {Zhang}},
  \bibinfo {author} {\bibfnamefont {G.}~\bibnamefont {Zhang}}, \ and\ \bibinfo
  {author} {\bibfnamefont {Z.}~\bibnamefont {Song}},\ }\bibfield  {title}
  {\enquote {\bibinfo {title} {Topological phases in a kitaev chain with
  imbalanced pairing},}\ }\href {\doibase 10.1103/PhysRevB.97.115436}
  {\bibfield  {journal} {\bibinfo  {journal} {Phys. Rev. B}\ }\textbf {\bibinfo
  {volume} {97}},\ \bibinfo {pages} {115436} (\bibinfo {year}
  {2018})}\BibitemShut {NoStop}%
\bibitem [{\citenamefont {Rudner}\ and\ \citenamefont
  {Levitov}(2009)}]{rudner2009topological}%
  \BibitemOpen
  \bibfield  {author} {\bibinfo {author} {\bibfnamefont {M.~S.}\ \bibnamefont
  {Rudner}}\ and\ \bibinfo {author} {\bibfnamefont {L.~S.}\ \bibnamefont
  {Levitov}},\ }\bibfield  {title} {\enquote {\bibinfo {title} {Topological
  transition in a non-hermitian quantum walk},}\ }\href {\doibase
  10.1103/PhysRevLett.102.065703} {\bibfield  {journal} {\bibinfo  {journal}
  {Phys. Rev. Lett.}\ }\textbf {\bibinfo {volume} {102}},\ \bibinfo {pages}
  {065703} (\bibinfo {year} {2009})}\BibitemShut {NoStop}%
\bibitem [{\citenamefont {Liang}\ and\ \citenamefont
  {Huang}(2013)}]{liang2013topological}%
  \BibitemOpen
  \bibfield  {author} {\bibinfo {author} {\bibfnamefont {Shi-Dong}\
  \bibnamefont {Liang}}\ and\ \bibinfo {author} {\bibfnamefont {Guang-Yao}\
  \bibnamefont {Huang}},\ }\bibfield  {title} {\enquote {\bibinfo {title}
  {Topological invariance and global berry phase in non-hermitian systems},}\
  }\href {\doibase 10.1103/PhysRevA.87.012118} {\bibfield  {journal} {\bibinfo
  {journal} {Phys. Rev. A}\ }\textbf {\bibinfo {volume} {87}},\ \bibinfo
  {pages} {012118} (\bibinfo {year} {2013})}\BibitemShut {NoStop}%
\bibitem [{\citenamefont {Hu}\ and\ \citenamefont
  {Hughes}(2011)}]{hu2011absence}%
  \BibitemOpen
  \bibfield  {author} {\bibinfo {author} {\bibfnamefont {Yi~Chen}\ \bibnamefont
  {Hu}}\ and\ \bibinfo {author} {\bibfnamefont {Taylor~L.}\ \bibnamefont
  {Hughes}},\ }\bibfield  {title} {\enquote {\bibinfo {title} {Absence of
  topological insulator phases in non-hermitian $pt$-symmetric hamiltonians},}\
  }\href {\doibase 10.1103/PhysRevB.84.153101} {\bibfield  {journal} {\bibinfo
  {journal} {Phys. Rev. B}\ }\textbf {\bibinfo {volume} {84}},\ \bibinfo
  {pages} {153101} (\bibinfo {year} {2011})}\BibitemShut {NoStop}%
\bibitem [{\citenamefont {Gong}\ \emph {et~al.}(2017)\citenamefont {Gong},
  \citenamefont {Higashikawa},\ and\ \citenamefont {Ueda}}]{gong2017zeno}%
  \BibitemOpen
  \bibfield  {author} {\bibinfo {author} {\bibfnamefont {Zongping}\
  \bibnamefont {Gong}}, \bibinfo {author} {\bibfnamefont {Sho}\ \bibnamefont
  {Higashikawa}}, \ and\ \bibinfo {author} {\bibfnamefont {Masahito}\
  \bibnamefont {Ueda}},\ }\bibfield  {title} {\enquote {\bibinfo {title} {Zeno
  hall effect},}\ }\href {\doibase 10.1103/PhysRevLett.118.200401} {\bibfield
  {journal} {\bibinfo  {journal} {Phys. Rev. Lett.}\ }\textbf {\bibinfo
  {volume} {118}},\ \bibinfo {pages} {200401} (\bibinfo {year}
  {2017})}\BibitemShut {NoStop}%
\bibitem [{\citenamefont {Gong}\ and\ \citenamefont
  {Wang}(2010)}]{gong2010geometrical}%
  \BibitemOpen
  \bibfield  {author} {\bibinfo {author} {\bibfnamefont {Jiangbin}\
  \bibnamefont {Gong}}\ and\ \bibinfo {author} {\bibfnamefont {Qing-hai}\
  \bibnamefont {Wang}},\ }\bibfield  {title} {\enquote {\bibinfo {title}
  {Geometric phase in $\mathcal{PT}$-symmetric quantum mechanics},}\ }\href
  {\doibase 10.1103/PhysRevA.82.012103} {\bibfield  {journal} {\bibinfo
  {journal} {Phys. Rev. A}\ }\textbf {\bibinfo {volume} {82}},\ \bibinfo
  {pages} {012103} (\bibinfo {year} {2010})}\BibitemShut {NoStop}%
\bibitem [{\citenamefont {{Rudner}}\ \emph {et~al.}(2016)\citenamefont
  {{Rudner}}, \citenamefont {{Levin}},\ and\ \citenamefont
  {{Levitov}}}]{rudner2016survival}%
  \BibitemOpen
  \bibfield  {author} {\bibinfo {author} {\bibfnamefont {M.~S.}\ \bibnamefont
  {{Rudner}}}, \bibinfo {author} {\bibfnamefont {M.}~\bibnamefont {{Levin}}}, \
  and\ \bibinfo {author} {\bibfnamefont {L.~S.}\ \bibnamefont {{Levitov}}},\
  }\bibfield  {title} {\enquote {\bibinfo {title} {{Survival, decay, and
  topological protection in non-Hermitian quantum transport}},}\ }\href@noop {}
  {\bibfield  {journal} {\bibinfo  {journal} {ArXiv e-prints}\ } (\bibinfo
  {year} {2016})},\ \Eprint {http://arxiv.org/abs/1605.07652} {arXiv:1605.07652
  [cond-mat.mes-hall]} \BibitemShut {NoStop}%
\bibitem [{\citenamefont {Kawabata}\ \emph {et~al.}(2018)\citenamefont
  {Kawabata}, \citenamefont {Ashida}, \citenamefont {Katsura},\ and\
  \citenamefont {Ueda}}]{kawabata2018PT}%
  \BibitemOpen
  \bibfield  {author} {\bibinfo {author} {\bibfnamefont {Kohei}\ \bibnamefont
  {Kawabata}}, \bibinfo {author} {\bibfnamefont {Yuto}\ \bibnamefont {Ashida}},
  \bibinfo {author} {\bibfnamefont {Hosho}\ \bibnamefont {Katsura}}, \ and\
  \bibinfo {author} {\bibfnamefont {Masahito}\ \bibnamefont {Ueda}},\
  }\bibfield  {title} {\enquote {\bibinfo {title} {Parity-time-symmetric
  topological superconductor},}\ }\href@noop {} {\bibfield  {journal} {\bibinfo
   {journal} {arXiv preprint arXiv:1801.00499}\ } (\bibinfo {year}
  {2018})}\BibitemShut {NoStop}%
\bibitem [{\citenamefont {{Ni}}\ \emph {et~al.}(2018)\citenamefont {{Ni}},
  \citenamefont {{Smirnova}}, \citenamefont {{Poddubny}}, \citenamefont
  {{Leykam}}, \citenamefont {{Chong}},\ and\ \citenamefont
  {{Khanikaev}}}]{ni2018exceptional}%
  \BibitemOpen
  \bibfield  {author} {\bibinfo {author} {\bibfnamefont {X.}~\bibnamefont
  {{Ni}}}, \bibinfo {author} {\bibfnamefont {D.}~\bibnamefont {{Smirnova}}},
  \bibinfo {author} {\bibfnamefont {A.}~\bibnamefont {{Poddubny}}}, \bibinfo
  {author} {\bibfnamefont {D.}~\bibnamefont {{Leykam}}}, \bibinfo {author}
  {\bibfnamefont {Y.}~\bibnamefont {{Chong}}}, \ and\ \bibinfo {author}
  {\bibfnamefont {A.~B.}\ \bibnamefont {{Khanikaev}}},\ }\bibfield  {title}
  {\enquote {\bibinfo {title} {{Exceptional points in topological edge spectrum
  of PT symmetric domain walls}},}\ }\href@noop {} {\bibfield  {journal}
  {\bibinfo  {journal} {ArXiv e-prints}\ } (\bibinfo {year} {2018})},\ \Eprint
  {http://arxiv.org/abs/1801.04689} {arXiv:1801.04689 [cond-mat.mes-hall]}
  \BibitemShut {NoStop}%
\bibitem [{\citenamefont {Zyuzin}\ and\ \citenamefont
  {Zyuzin}(2018)}]{zyuzin2018flat}%
  \BibitemOpen
  \bibfield  {author} {\bibinfo {author} {\bibfnamefont {A.~A.}\ \bibnamefont
  {Zyuzin}}\ and\ \bibinfo {author} {\bibfnamefont {A.~Yu.}\ \bibnamefont
  {Zyuzin}},\ }\bibfield  {title} {\enquote {\bibinfo {title} {Flat band in
  disorder-driven non-hermitian weyl semimetals},}\ }\href {\doibase
  10.1103/PhysRevB.97.041203} {\bibfield  {journal} {\bibinfo  {journal} {Phys.
  Rev. B}\ }\textbf {\bibinfo {volume} {97}},\ \bibinfo {pages} {041203}
  (\bibinfo {year} {2018})}\BibitemShut {NoStop}%
\bibitem [{\citenamefont {Cerjan}\ \emph {et~al.}(2018)\citenamefont {Cerjan},
  \citenamefont {Xiao}, \citenamefont {Yuan},\ and\ \citenamefont
  {Fan}}]{cerjan2018weyl}%
  \BibitemOpen
  \bibfield  {author} {\bibinfo {author} {\bibfnamefont {Alexander}\
  \bibnamefont {Cerjan}}, \bibinfo {author} {\bibfnamefont {Meng}\ \bibnamefont
  {Xiao}}, \bibinfo {author} {\bibfnamefont {Luqi}\ \bibnamefont {Yuan}}, \
  and\ \bibinfo {author} {\bibfnamefont {Shanhui}\ \bibnamefont {Fan}},\
  }\bibfield  {title} {\enquote {\bibinfo {title} {Effects of non-hermitian
  perturbations on weyl hamiltonians with arbitrary topological charges},}\
  }\href {\doibase 10.1103/PhysRevB.97.075128} {\bibfield  {journal} {\bibinfo
  {journal} {Phys. Rev. B}\ }\textbf {\bibinfo {volume} {97}},\ \bibinfo
  {pages} {075128} (\bibinfo {year} {2018})}\BibitemShut {NoStop}%
\bibitem [{\citenamefont {{Zhou}}\ \emph
  {et~al.}(2017{\natexlab{a}})\citenamefont {{Zhou}}, \citenamefont {{Wang}},
  \citenamefont {{Wang}},\ and\ \citenamefont {{Gong}}}]{zhou2017dynamical}%
  \BibitemOpen
  \bibfield  {author} {\bibinfo {author} {\bibfnamefont {L.}~\bibnamefont
  {{Zhou}}}, \bibinfo {author} {\bibfnamefont {Q.-h.}\ \bibnamefont {{Wang}}},
  \bibinfo {author} {\bibfnamefont {H.}~\bibnamefont {{Wang}}}, \ and\ \bibinfo
  {author} {\bibfnamefont {J.}~\bibnamefont {{Gong}}},\ }\bibfield  {title}
  {\enquote {\bibinfo {title} {{Dynamical quantum phase transitions in
  non-Hermitian lattices}},}\ }\href@noop {} {\bibfield  {journal} {\bibinfo
  {journal} {ArXiv e-prints}\ } (\bibinfo {year} {2017}{\natexlab{a}})},\
  \Eprint {http://arxiv.org/abs/1711.10741} {arXiv:1711.10741
  [cond-mat.stat-mech]} \BibitemShut {NoStop}%
\bibitem [{\citenamefont {Gonz\'alez}\ and\ \citenamefont
  {Molina}(2017)}]{gonzalez2017}%
  \BibitemOpen
  \bibfield  {author} {\bibinfo {author} {\bibfnamefont {J.}~\bibnamefont
  {Gonz\'alez}}\ and\ \bibinfo {author} {\bibfnamefont {R.~A.}\ \bibnamefont
  {Molina}},\ }\bibfield  {title} {\enquote {\bibinfo {title} {Topological
  protection from exceptional points in weyl and nodal-line semimetals},}\
  }\href {\doibase 10.1103/PhysRevB.96.045437} {\bibfield  {journal} {\bibinfo
  {journal} {Phys. Rev. B}\ }\textbf {\bibinfo {volume} {96}},\ \bibinfo
  {pages} {045437} (\bibinfo {year} {2017})}\BibitemShut {NoStop}%
\bibitem [{\citenamefont {Klett}\ \emph {et~al.}(2017)\citenamefont {Klett},
  \citenamefont {Cartarius}, \citenamefont {Dast}, \citenamefont {Main},\ and\
  \citenamefont {Wunner}}]{klett2017sshkitaev}%
  \BibitemOpen
  \bibfield  {author} {\bibinfo {author} {\bibfnamefont {Marcel}\ \bibnamefont
  {Klett}}, \bibinfo {author} {\bibfnamefont {Holger}\ \bibnamefont
  {Cartarius}}, \bibinfo {author} {\bibfnamefont {Dennis}\ \bibnamefont
  {Dast}}, \bibinfo {author} {\bibfnamefont {J\"org}\ \bibnamefont {Main}}, \
  and\ \bibinfo {author} {\bibfnamefont {G\"unter}\ \bibnamefont {Wunner}},\
  }\bibfield  {title} {\enquote {\bibinfo {title} {Relation between
  $\mathcal{PT}$-symmetry breaking and topologically nontrivial phases in the
  su-schrieffer-heeger and kitaev models},}\ }\href {\doibase
  10.1103/PhysRevA.95.053626} {\bibfield  {journal} {\bibinfo  {journal} {Phys.
  Rev. A}\ }\textbf {\bibinfo {volume} {95}},\ \bibinfo {pages} {053626}
  (\bibinfo {year} {2017})}\BibitemShut {NoStop}%
\bibitem [{\citenamefont {{Klett}}\ \emph {et~al.}(2018)\citenamefont
  {{Klett}}, \citenamefont {{Cartarius}}, \citenamefont {{Dast}}, \citenamefont
  {{Main}},\ and\ \citenamefont {{Wunner}}}]{klett2018ssh}%
  \BibitemOpen
  \bibfield  {author} {\bibinfo {author} {\bibfnamefont {M.}~\bibnamefont
  {{Klett}}}, \bibinfo {author} {\bibfnamefont {H.}~\bibnamefont
  {{Cartarius}}}, \bibinfo {author} {\bibfnamefont {D.}~\bibnamefont {{Dast}}},
  \bibinfo {author} {\bibfnamefont {J.}~\bibnamefont {{Main}}}, \ and\ \bibinfo
  {author} {\bibfnamefont {G.}~\bibnamefont {{Wunner}}},\ }\bibfield  {title}
  {\enquote {\bibinfo {title} {{Topological edge states in the
  Su-Schrieffer-Heeger model subject to balanced particle gain and loss}},}\
  }\href@noop {} {\bibfield  {journal} {\bibinfo  {journal} {ArXiv e-prints}\ }
  (\bibinfo {year} {2018})},\ \Eprint {http://arxiv.org/abs/1802.06128}
  {arXiv:1802.06128 [quant-ph]} \BibitemShut {NoStop}%
\bibitem [{\citenamefont {Yuce}(2016)}]{yuce2016majorana}%
  \BibitemOpen
  \bibfield  {author} {\bibinfo {author} {\bibfnamefont {C.}~\bibnamefont
  {Yuce}},\ }\bibfield  {title} {\enquote {\bibinfo {title} {Majorana edge
  modes with gain and loss},}\ }\href {\doibase 10.1103/PhysRevA.93.062130}
  {\bibfield  {journal} {\bibinfo  {journal} {Phys. Rev. A}\ }\textbf {\bibinfo
  {volume} {93}},\ \bibinfo {pages} {062130} (\bibinfo {year}
  {2016})}\BibitemShut {NoStop}%
\bibitem [{\citenamefont {Yuce}(2015)}]{yuce2015topological}%
  \BibitemOpen
  \bibfield  {author} {\bibinfo {author} {\bibfnamefont {Cem}\ \bibnamefont
  {Yuce}},\ }\bibfield  {title} {\enquote {\bibinfo {title} {Topological phase
  in a non-hermitian pt symmetric system},}\ }\href@noop {} {\bibfield
  {journal} {\bibinfo  {journal} {Physics Letters A}\ }\textbf {\bibinfo
  {volume} {379}},\ \bibinfo {pages} {1213--1218} (\bibinfo {year}
  {2015})}\BibitemShut {NoStop}%
\bibitem [{\citenamefont {Xu}\ \emph {et~al.}(2017)\citenamefont {Xu},
  \citenamefont {Wang},\ and\ \citenamefont {Duan}}]{xu2017weyl}%
  \BibitemOpen
  \bibfield  {author} {\bibinfo {author} {\bibfnamefont {Yong}\ \bibnamefont
  {Xu}}, \bibinfo {author} {\bibfnamefont {Sheng-Tao}\ \bibnamefont {Wang}}, \
  and\ \bibinfo {author} {\bibfnamefont {L.-M.}\ \bibnamefont {Duan}},\
  }\bibfield  {title} {\enquote {\bibinfo {title} {Weyl exceptional rings in a
  three-dimensional dissipative cold atomic gas},}\ }\href {\doibase
  10.1103/PhysRevLett.118.045701} {\bibfield  {journal} {\bibinfo  {journal}
  {Phys. Rev. Lett.}\ }\textbf {\bibinfo {volume} {118}},\ \bibinfo {pages}
  {045701} (\bibinfo {year} {2017})}\BibitemShut {NoStop}%
\bibitem [{\citenamefont {Hu}\ \emph {et~al.}(2017)\citenamefont {Hu},
  \citenamefont {Wang}, \citenamefont {Shum},\ and\ \citenamefont
  {Chong}}]{hu2017exceptional}%
  \BibitemOpen
  \bibfield  {author} {\bibinfo {author} {\bibfnamefont {Wenchao}\ \bibnamefont
  {Hu}}, \bibinfo {author} {\bibfnamefont {Hailong}\ \bibnamefont {Wang}},
  \bibinfo {author} {\bibfnamefont {Perry~Ping}\ \bibnamefont {Shum}}, \ and\
  \bibinfo {author} {\bibfnamefont {Y.~D.}\ \bibnamefont {Chong}},\ }\bibfield
  {title} {\enquote {\bibinfo {title} {Exceptional points in a non-hermitian
  topological pump},}\ }\href {\doibase 10.1103/PhysRevB.95.184306} {\bibfield
  {journal} {\bibinfo  {journal} {Phys. Rev. B}\ }\textbf {\bibinfo {volume}
  {95}},\ \bibinfo {pages} {184306} (\bibinfo {year} {2017})}\BibitemShut
  {NoStop}%
\bibitem [{\citenamefont {Wang}\ \emph {et~al.}(2015)\citenamefont {Wang},
  \citenamefont {Liu}, \citenamefont {Xiong},\ and\ \citenamefont
  {Tong}}]{wang2015spontaneous}%
  \BibitemOpen
  \bibfield  {author} {\bibinfo {author} {\bibfnamefont {Xiaohui}\ \bibnamefont
  {Wang}}, \bibinfo {author} {\bibfnamefont {Tingting}\ \bibnamefont {Liu}},
  \bibinfo {author} {\bibfnamefont {Ye}~\bibnamefont {Xiong}}, \ and\ \bibinfo
  {author} {\bibfnamefont {Peiqing}\ \bibnamefont {Tong}},\ }\bibfield  {title}
  {\enquote {\bibinfo {title} {Spontaneous $\mathcal{PT}$-symmetry breaking in
  non-hermitian kitaev and extended kitaev models},}\ }\href {\doibase
  10.1103/PhysRevA.92.012116} {\bibfield  {journal} {\bibinfo  {journal} {Phys.
  Rev. A}\ }\textbf {\bibinfo {volume} {92}},\ \bibinfo {pages} {012116}
  (\bibinfo {year} {2015})}\BibitemShut {NoStop}%
\bibitem [{\citenamefont {Ke}\ \emph {et~al.}(2017)\citenamefont {Ke},
  \citenamefont {Wang}, \citenamefont {Long}, \citenamefont {Wang},\ and\
  \citenamefont {Lu}}]{ke2017topological}%
  \BibitemOpen
  \bibfield  {author} {\bibinfo {author} {\bibfnamefont {Shaolin}\ \bibnamefont
  {Ke}}, \bibinfo {author} {\bibfnamefont {Bing}\ \bibnamefont {Wang}},
  \bibinfo {author} {\bibfnamefont {Hua}\ \bibnamefont {Long}}, \bibinfo
  {author} {\bibfnamefont {Kai}\ \bibnamefont {Wang}}, \ and\ \bibinfo {author}
  {\bibfnamefont {Peixiang}\ \bibnamefont {Lu}},\ }\bibfield  {title} {\enquote
  {\bibinfo {title} {Topological edge modes in non-hermitian plasmonic
  waveguide arrays},}\ }\href@noop {} {\bibfield  {journal} {\bibinfo
  {journal} {Optics Express}\ }\textbf {\bibinfo {volume} {25}},\ \bibinfo
  {pages} {11132--11143} (\bibinfo {year} {2017})}\BibitemShut {NoStop}%
\bibitem [{\citenamefont {Rivolta}\ \emph {et~al.}(2017)\citenamefont
  {Rivolta}, \citenamefont {Benisty},\ and\ \citenamefont
  {Maes}}]{rivolta2017}%
  \BibitemOpen
  \bibfield  {author} {\bibinfo {author} {\bibfnamefont {Nicolas X.~A.}\
  \bibnamefont {Rivolta}}, \bibinfo {author} {\bibfnamefont {Henri}\
  \bibnamefont {Benisty}}, \ and\ \bibinfo {author} {\bibfnamefont {Bjorn}\
  \bibnamefont {Maes}},\ }\bibfield  {title} {\enquote {\bibinfo {title}
  {Topological edge modes with $\mathcal{PT}$ symmetry in a quasiperiodic
  structure},}\ }\href {\doibase 10.1103/PhysRevA.96.023864} {\bibfield
  {journal} {\bibinfo  {journal} {Phys. Rev. A}\ }\textbf {\bibinfo {volume}
  {96}},\ \bibinfo {pages} {023864} (\bibinfo {year} {2017})}\BibitemShut
  {NoStop}%
\bibitem [{\citenamefont {{Gong}}\ \emph {et~al.}(2018)\citenamefont {{Gong}},
  \citenamefont {{Ashida}}, \citenamefont {{Kawabata}}, \citenamefont
  {{Takasan}}, \citenamefont {{Higashikawa}},\ and\ \citenamefont
  {{Ueda}}}]{gong2018nonhermitian}%
  \BibitemOpen
  \bibfield  {author} {\bibinfo {author} {\bibfnamefont {Z.}~\bibnamefont
  {{Gong}}}, \bibinfo {author} {\bibfnamefont {Y.}~\bibnamefont {{Ashida}}},
  \bibinfo {author} {\bibfnamefont {K.}~\bibnamefont {{Kawabata}}}, \bibinfo
  {author} {\bibfnamefont {K.}~\bibnamefont {{Takasan}}}, \bibinfo {author}
  {\bibfnamefont {S.}~\bibnamefont {{Higashikawa}}}, \ and\ \bibinfo {author}
  {\bibfnamefont {M.}~\bibnamefont {{Ueda}}},\ }\bibfield  {title} {\enquote
  {\bibinfo {title} {{Topological phases of non-Hermitian systems}},}\
  }\href@noop {} {\bibfield  {journal} {\bibinfo  {journal} {ArXiv e-prints}\ }
  (\bibinfo {year} {2018})},\ \Eprint {http://arxiv.org/abs/1802.07964v1}
  {arXiv:1802.07964v1 [cond-mat.mes-hall]} \BibitemShut {NoStop}%
\bibitem [{\citenamefont {Harari}\ \emph {et~al.}(2018)\citenamefont {Harari},
  \citenamefont {Bandres}, \citenamefont {Lumer}, \citenamefont {Rechtsman},
  \citenamefont {Chong}, \citenamefont {Khajavikhan}, \citenamefont
  {Christodoulides},\ and\ \citenamefont {Segev}}]{harari2018topological}%
  \BibitemOpen
  \bibfield  {author} {\bibinfo {author} {\bibfnamefont {Gal}\ \bibnamefont
  {Harari}}, \bibinfo {author} {\bibfnamefont {Miguel~A}\ \bibnamefont
  {Bandres}}, \bibinfo {author} {\bibfnamefont {Yaakov}\ \bibnamefont {Lumer}},
  \bibinfo {author} {\bibfnamefont {Mikael~C}\ \bibnamefont {Rechtsman}},
  \bibinfo {author} {\bibfnamefont {YD}~\bibnamefont {Chong}}, \bibinfo
  {author} {\bibfnamefont {Mercedeh}\ \bibnamefont {Khajavikhan}}, \bibinfo
  {author} {\bibfnamefont {Demetrios~N}\ \bibnamefont {Christodoulides}}, \
  and\ \bibinfo {author} {\bibfnamefont {Mordechai}\ \bibnamefont {Segev}},\
  }\bibfield  {title} {\enquote {\bibinfo {title} {Topological insulator laser:
  Theory},}\ }\href@noop {} {\bibfield  {journal} {\bibinfo  {journal}
  {Science}\ ,\ \bibinfo {pages} {eaar4003}} (\bibinfo {year}
  {2018})}\BibitemShut {NoStop}%
\bibitem [{\citenamefont {Zeuner}\ \emph {et~al.}(2015)\citenamefont {Zeuner},
  \citenamefont {Rechtsman}, \citenamefont {Plotnik}, \citenamefont {Lumer},
  \citenamefont {Nolte}, \citenamefont {Rudner}, \citenamefont {Segev},\ and\
  \citenamefont {Szameit}}]{zeuner2015bulk}%
  \BibitemOpen
  \bibfield  {author} {\bibinfo {author} {\bibfnamefont {Julia~M.}\
  \bibnamefont {Zeuner}}, \bibinfo {author} {\bibfnamefont {Mikael~C.}\
  \bibnamefont {Rechtsman}}, \bibinfo {author} {\bibfnamefont {Yonatan}\
  \bibnamefont {Plotnik}}, \bibinfo {author} {\bibfnamefont {Yaakov}\
  \bibnamefont {Lumer}}, \bibinfo {author} {\bibfnamefont {Stefan}\
  \bibnamefont {Nolte}}, \bibinfo {author} {\bibfnamefont {Mark~S.}\
  \bibnamefont {Rudner}}, \bibinfo {author} {\bibfnamefont {Mordechai}\
  \bibnamefont {Segev}}, \ and\ \bibinfo {author} {\bibfnamefont {Alexander}\
  \bibnamefont {Szameit}},\ }\bibfield  {title} {\enquote {\bibinfo {title}
  {Observation of a topological transition in the bulk of a non-hermitian
  system},}\ }\href {\doibase 10.1103/PhysRevLett.115.040402} {\bibfield
  {journal} {\bibinfo  {journal} {Phys. Rev. Lett.}\ }\textbf {\bibinfo
  {volume} {115}},\ \bibinfo {pages} {040402} (\bibinfo {year}
  {2015})}\BibitemShut {NoStop}%
\bibitem [{\citenamefont {Zhan}\ \emph {et~al.}(2017)\citenamefont {Zhan},
  \citenamefont {Xiao}, \citenamefont {Bian}, \citenamefont {Wang},
  \citenamefont {Qiu}, \citenamefont {Sanders}, \citenamefont {Yi},\ and\
  \citenamefont {Xue}}]{zhan2017detecting}%
  \BibitemOpen
  \bibfield  {author} {\bibinfo {author} {\bibfnamefont {Xiang}\ \bibnamefont
  {Zhan}}, \bibinfo {author} {\bibfnamefont {Lei}\ \bibnamefont {Xiao}},
  \bibinfo {author} {\bibfnamefont {Zhihao}\ \bibnamefont {Bian}}, \bibinfo
  {author} {\bibfnamefont {Kunkun}\ \bibnamefont {Wang}}, \bibinfo {author}
  {\bibfnamefont {Xingze}\ \bibnamefont {Qiu}}, \bibinfo {author}
  {\bibfnamefont {Barry~C.}\ \bibnamefont {Sanders}}, \bibinfo {author}
  {\bibfnamefont {Wei}\ \bibnamefont {Yi}}, \ and\ \bibinfo {author}
  {\bibfnamefont {Peng}\ \bibnamefont {Xue}},\ }\bibfield  {title} {\enquote
  {\bibinfo {title} {Detecting topological invariants in nonunitary
  discrete-time quantum walks},}\ }\href {\doibase
  10.1103/PhysRevLett.119.130501} {\bibfield  {journal} {\bibinfo  {journal}
  {Phys. Rev. Lett.}\ }\textbf {\bibinfo {volume} {119}},\ \bibinfo {pages}
  {130501} (\bibinfo {year} {2017})}\BibitemShut {NoStop}%
\bibitem [{\citenamefont {Xiao}\ \emph {et~al.}(2017)\citenamefont {Xiao},
  \citenamefont {Zhan}, \citenamefont {Bian}, \citenamefont {Wang},
  \citenamefont {Zhang}, \citenamefont {Wang}, \citenamefont {Li},
  \citenamefont {Mochizuki}, \citenamefont {Kim}, \citenamefont {Kawakami},
  \citenamefont {Yi}, \citenamefont {Obuse}, \citenamefont {Sanders},\ and\
  \citenamefont {Xue}}]{xiao2017observation}%
  \BibitemOpen
  \bibfield  {author} {\bibinfo {author} {\bibfnamefont {L.}~\bibnamefont
  {Xiao}}, \bibinfo {author} {\bibfnamefont {X.}~\bibnamefont {Zhan}}, \bibinfo
  {author} {\bibfnamefont {Z.~H.}\ \bibnamefont {Bian}}, \bibinfo {author}
  {\bibfnamefont {K.~K.}\ \bibnamefont {Wang}}, \bibinfo {author}
  {\bibfnamefont {X.}~\bibnamefont {Zhang}}, \bibinfo {author} {\bibfnamefont
  {X.~P.}\ \bibnamefont {Wang}}, \bibinfo {author} {\bibfnamefont
  {J.}~\bibnamefont {Li}}, \bibinfo {author} {\bibfnamefont {K.}~\bibnamefont
  {Mochizuki}}, \bibinfo {author} {\bibfnamefont {D.}~\bibnamefont {Kim}},
  \bibinfo {author} {\bibfnamefont {N.}~\bibnamefont {Kawakami}}, \bibinfo
  {author} {\bibfnamefont {W.}~\bibnamefont {Yi}}, \bibinfo {author}
  {\bibfnamefont {H.}~\bibnamefont {Obuse}}, \bibinfo {author} {\bibfnamefont
  {B.~C.}\ \bibnamefont {Sanders}}, \ and\ \bibinfo {author} {\bibfnamefont
  {P.}~\bibnamefont {Xue}},\ }\bibfield  {title} {\enquote {\bibinfo {title}
  {Observation of topological edge states in parity--time-symmetric quantum
  walks},}\ }\href@noop {} {\bibfield  {journal} {\bibinfo  {journal} {Nature
  Physics}\ }\textbf {\bibinfo {volume} {13}},\ \bibinfo {pages} {1117}
  (\bibinfo {year} {2017})}\BibitemShut {NoStop}%
\bibitem [{\citenamefont {Weimann}\ \emph {et~al.}(2017)\citenamefont
  {Weimann}, \citenamefont {Kremer}, \citenamefont {Plotnik}, \citenamefont
  {Lumer}, \citenamefont {Nolte}, \citenamefont {Makris}, \citenamefont
  {Segev}, \citenamefont {Rechtsman},\ and\ \citenamefont
  {Szameit}}]{weimann2017topologically}%
  \BibitemOpen
  \bibfield  {author} {\bibinfo {author} {\bibfnamefont {S}~\bibnamefont
  {Weimann}}, \bibinfo {author} {\bibfnamefont {M}~\bibnamefont {Kremer}},
  \bibinfo {author} {\bibfnamefont {Y}~\bibnamefont {Plotnik}}, \bibinfo
  {author} {\bibfnamefont {Y}~\bibnamefont {Lumer}}, \bibinfo {author}
  {\bibfnamefont {S}~\bibnamefont {Nolte}}, \bibinfo {author} {\bibfnamefont
  {KG}~\bibnamefont {Makris}}, \bibinfo {author} {\bibfnamefont
  {M}~\bibnamefont {Segev}}, \bibinfo {author} {\bibfnamefont {MC}~\bibnamefont
  {Rechtsman}}, \ and\ \bibinfo {author} {\bibfnamefont {A}~\bibnamefont
  {Szameit}},\ }\bibfield  {title} {\enquote {\bibinfo {title} {Topologically
  protected bound states in photonic parity--time-symmetric crystals},}\
  }\href@noop {} {\bibfield  {journal} {\bibinfo  {journal} {Nature materials}\
  }\textbf {\bibinfo {volume} {16}},\ \bibinfo {pages} {433} (\bibinfo {year}
  {2017})}\BibitemShut {NoStop}%
\bibitem [{\citenamefont {{Parto}}\ \emph {et~al.}(2017)\citenamefont
  {{Parto}}, \citenamefont {{Wittek}}, \citenamefont {{Hodaei}}, \citenamefont
  {{Harari}}, \citenamefont {{Bandres}}, \citenamefont {{Ren}}, \citenamefont
  {{Rechtsman}}, \citenamefont {{Segev}}, \citenamefont {{Christodoulides}},\
  and\ \citenamefont {{Khajavikhan}}}]{parto2017SSHexperiment}%
  \BibitemOpen
  \bibfield  {author} {\bibinfo {author} {\bibfnamefont {M.}~\bibnamefont
  {{Parto}}}, \bibinfo {author} {\bibfnamefont {S.}~\bibnamefont {{Wittek}}},
  \bibinfo {author} {\bibfnamefont {H.}~\bibnamefont {{Hodaei}}}, \bibinfo
  {author} {\bibfnamefont {G.}~\bibnamefont {{Harari}}}, \bibinfo {author}
  {\bibfnamefont {M.~A.}\ \bibnamefont {{Bandres}}}, \bibinfo {author}
  {\bibfnamefont {J.}~\bibnamefont {{Ren}}}, \bibinfo {author} {\bibfnamefont
  {M.~C.}\ \bibnamefont {{Rechtsman}}}, \bibinfo {author} {\bibfnamefont
  {M.}~\bibnamefont {{Segev}}}, \bibinfo {author} {\bibfnamefont {D.~N.}\
  \bibnamefont {{Christodoulides}}}, \ and\ \bibinfo {author} {\bibfnamefont
  {M.}~\bibnamefont {{Khajavikhan}}},\ }\bibfield  {title} {\enquote {\bibinfo
  {title} {{Complex Edge-State Phase Transitions in 1D Topological Laser
  Arrays}},}\ }\href@noop {} {\bibfield  {journal} {\bibinfo  {journal} {ArXiv
  e-prints}\ } (\bibinfo {year} {2017})},\ \Eprint
  {http://arxiv.org/abs/1709.00523} {arXiv:1709.00523 [physics.optics]}
  \BibitemShut {NoStop}%
\bibitem [{\citenamefont {{Zhao}}\ \emph {et~al.}(2017)\citenamefont {{Zhao}},
  \citenamefont {{Miao}}, \citenamefont {{Teimourpour}}, \citenamefont
  {{Malzard}}, \citenamefont {{El-Ganainy}}, \citenamefont {{Schomerus}},\ and\
  \citenamefont {{Feng}}}]{zhao2017Topological}%
  \BibitemOpen
  \bibfield  {author} {\bibinfo {author} {\bibfnamefont {H.}~\bibnamefont
  {{Zhao}}}, \bibinfo {author} {\bibfnamefont {P.}~\bibnamefont {{Miao}}},
  \bibinfo {author} {\bibfnamefont {M.~H.}\ \bibnamefont {{Teimourpour}}},
  \bibinfo {author} {\bibfnamefont {S.}~\bibnamefont {{Malzard}}}, \bibinfo
  {author} {\bibfnamefont {R.}~\bibnamefont {{El-Ganainy}}}, \bibinfo {author}
  {\bibfnamefont {H.}~\bibnamefont {{Schomerus}}}, \ and\ \bibinfo {author}
  {\bibfnamefont {L.}~\bibnamefont {{Feng}}},\ }\bibfield  {title} {\enquote
  {\bibinfo {title} {{Topological Hybrid Silicon Microlasers}},}\ }\href@noop
  {} {\bibfield  {journal} {\bibinfo  {journal} {ArXiv e-prints}\ } (\bibinfo
  {year} {2017})},\ \Eprint {http://arxiv.org/abs/1709.02747} {arXiv:1709.02747
  [physics.optics]} \BibitemShut {NoStop}%
\bibitem [{\citenamefont {{Zhou}}\ \emph
  {et~al.}(2017{\natexlab{b}})\citenamefont {{Zhou}}, \citenamefont {{Peng}},
  \citenamefont {{Yoon}}, \citenamefont {{Hsu}}, \citenamefont {{Nelson}},
  \citenamefont {{Fu}}, \citenamefont {{Joannopoulos}}, \citenamefont
  {{Soljacic}},\ and\ \citenamefont {{Zhen}}}]{zhou2017observation}%
  \BibitemOpen
  \bibfield  {author} {\bibinfo {author} {\bibfnamefont {H.}~\bibnamefont
  {{Zhou}}}, \bibinfo {author} {\bibfnamefont {C.}~\bibnamefont {{Peng}}},
  \bibinfo {author} {\bibfnamefont {Y.}~\bibnamefont {{Yoon}}}, \bibinfo
  {author} {\bibfnamefont {C.~W.}\ \bibnamefont {{Hsu}}}, \bibinfo {author}
  {\bibfnamefont {K.~A.}\ \bibnamefont {{Nelson}}}, \bibinfo {author}
  {\bibfnamefont {L.}~\bibnamefont {{Fu}}}, \bibinfo {author} {\bibfnamefont
  {J.~D.}\ \bibnamefont {{Joannopoulos}}}, \bibinfo {author} {\bibfnamefont
  {M.}~\bibnamefont {{Soljacic}}}, \ and\ \bibinfo {author} {\bibfnamefont
  {B.}~\bibnamefont {{Zhen}}},\ }\bibfield  {title} {\enquote {\bibinfo {title}
  {{Observation of Bulk Fermi Arc and Polarization Half Charge from Paired
  Exceptional Points}},}\ }\href@noop {} {\bibfield  {journal} {\bibinfo
  {journal} {ArXiv e-prints}\ } (\bibinfo {year} {2017}{\natexlab{b}})},\
  \Eprint {http://arxiv.org/abs/1709.03044} {arXiv:1709.03044 [physics.optics]}
  \BibitemShut {NoStop}%
\bibitem [{\citenamefont {{Xiong}}(2017)}]{xiong2017}%
  \BibitemOpen
  \bibfield  {author} {\bibinfo {author} {\bibfnamefont {Y.}~\bibnamefont
  {{Xiong}}},\ }\bibfield  {title} {\enquote {\bibinfo {title} {{Why does bulk
  boundary correspondence fail in some non-hermitian topological models}},}\
  }\href@noop {} {\bibfield  {journal} {\bibinfo  {journal} {ArXiv e-prints}\ }
  (\bibinfo {year} {2017})},\ \Eprint {http://arxiv.org/abs/1705.06039v1}
  {arXiv:1705.06039v1 [cond-mat.mes-hall]} \BibitemShut {NoStop}%
\bibitem [{\citenamefont {Longhi}(2017)}]{Longhi2017}%
  \BibitemOpen
  \bibfield  {author} {\bibinfo {author} {\bibfnamefont {Stefano}\ \bibnamefont
  {Longhi}},\ }\bibfield  {title} {\enquote {\bibinfo {title} {Nonadiabatic
  robust excitation transfer assisted by an imaginary gauge field},}\ }\href
  {\doibase 10.1103/PhysRevA.95.062122} {\bibfield  {journal} {\bibinfo
  {journal} {Phys. Rev. A}\ }\textbf {\bibinfo {volume} {95}},\ \bibinfo
  {pages} {062122} (\bibinfo {year} {2017})}\BibitemShut {NoStop}%
\bibitem [{\citenamefont {Hatano}\ and\ \citenamefont
  {Nelson}(1996)}]{Hatano1996}%
  \BibitemOpen
  \bibfield  {author} {\bibinfo {author} {\bibfnamefont {Naomichi}\
  \bibnamefont {Hatano}}\ and\ \bibinfo {author} {\bibfnamefont {David~R.}\
  \bibnamefont {Nelson}},\ }\bibfield  {title} {\enquote {\bibinfo {title}
  {Localization transitions in non-hermitian quantum mechanics},}\ }\href
  {\doibase 10.1103/PhysRevLett.77.570} {\bibfield  {journal} {\bibinfo
  {journal} {Phys. Rev. Lett.}\ }\textbf {\bibinfo {volume} {77}},\ \bibinfo
  {pages} {570--573} (\bibinfo {year} {1996})}\BibitemShut {NoStop}%
\bibitem [{\citenamefont {Su}\ \emph {et~al.}(1980)\citenamefont {Su},
  \citenamefont {Schrieffer},\ and\ \citenamefont {Heeger}}]{su1980}%
  \BibitemOpen
  \bibfield  {author} {\bibinfo {author} {\bibfnamefont {W\_P}\ \bibnamefont
  {Su}}, \bibinfo {author} {\bibfnamefont {JR}~\bibnamefont {Schrieffer}}, \
  and\ \bibinfo {author} {\bibfnamefont {AJ}~\bibnamefont {Heeger}},\
  }\bibfield  {title} {\enquote {\bibinfo {title} {Soliton excitations in
  polyacetylene},}\ }\href@noop {} {\bibfield  {journal} {\bibinfo  {journal}
  {Physical Review B}\ }\textbf {\bibinfo {volume} {22}},\ \bibinfo {pages}
  {2099} (\bibinfo {year} {1980})}\BibitemShut {NoStop}%
\bibitem [{Note1()}]{Note1}%
  \BibitemOpen
  \bibinfo {note} {Related models have been studied, for example, in Refs.
  \cite {zhu2014PT,yin2018ssh,lieu2018ssh}.}\BibitemShut {Stop}%
\bibitem [{\citenamefont {Poli}\ \emph {et~al.}(2015)\citenamefont {Poli},
  \citenamefont {Bellec}, \citenamefont {Kuhl}, \citenamefont {Mortessagne},\
  and\ \citenamefont {Schomerus}}]{poli2015selective}%
  \BibitemOpen
  \bibfield  {author} {\bibinfo {author} {\bibfnamefont {Charles}\ \bibnamefont
  {Poli}}, \bibinfo {author} {\bibfnamefont {Matthieu}\ \bibnamefont {Bellec}},
  \bibinfo {author} {\bibfnamefont {Ulrich}\ \bibnamefont {Kuhl}}, \bibinfo
  {author} {\bibfnamefont {Fabrice}\ \bibnamefont {Mortessagne}}, \ and\
  \bibinfo {author} {\bibfnamefont {Henning}\ \bibnamefont {Schomerus}},\
  }\bibfield  {title} {\enquote {\bibinfo {title} {Selective enhancement of
  topologically induced interface states in a dielectric resonator chain},}\
  }\href@noop {} {\bibfield  {journal} {\bibinfo  {journal} {Nature
  communications}\ }\textbf {\bibinfo {volume} {6}},\ \bibinfo {pages} {6710}
  (\bibinfo {year} {2015})}\BibitemShut {NoStop}%
\bibitem [{Note2()}]{Note2}%
  \BibitemOpen
  \bibinfo {note} {We note that the numerical precision of Ref.\cite
  {lee2016anomalous} is improvable. According to our exact results, the
  zero-mode line in their Fig. 3(a) should span the entire $[-1/\protect \sqrt
  {2},1/\protect \sqrt {2}]$ interval, instead of the two disconnected lines
  there.}\BibitemShut {Stop}%
\bibitem [{sup()}]{supplemental}%
  \BibitemOpen
  \href@noop {} {}\bibinfo {howpublished} {Supplemental Material.}\BibitemShut
  {Stop}%
\bibitem [{Note3()}]{Note3}%
  \BibitemOpen
  \bibinfo {note} {Recently we noticed Ref.\cite {alvarez2017}, in which
  similar localization is found numerically; however, in contrast to our
  viewpoint, it is suggested there that the localization lessens the relevance
  of zero modes and destroys bulk-boundary correspondence. Also note that the
  zero-mode interval in their Fig.1 differs from our exact
  solutions.}\BibitemShut {Stop}%
\bibitem [{Note4()}]{Note4}%
  \BibitemOpen
  \bibinfo {note} {We emphasize that the bulk energy spectra remain insensitive
  to a small perturbation at the ends of a long chain. In fact, the eigenstates
  of $H^\protect \dag $ (namely left eigenstates) have opposite exponential
  decay and the outcome of a perturbation depends on the product of right and
  left eigenstates.}\BibitemShut {Stop}%
\bibitem [{Note5()}]{Note5}%
  \BibitemOpen
  \bibinfo {note} {For example, we find that it is the case for the model
  numerically studied in Ref. \cite {lieu2018ssh}.}\BibitemShut {Stop}%
\bibitem [{\citenamefont {Zhu}\ \emph {et~al.}(2014)\citenamefont {Zhu},
  \citenamefont {L\"u},\ and\ \citenamefont {Chen}}]{zhu2014PT}%
  \BibitemOpen
  \bibfield  {author} {\bibinfo {author} {\bibfnamefont {Baogang}\ \bibnamefont
  {Zhu}}, \bibinfo {author} {\bibfnamefont {Rong}\ \bibnamefont {L\"u}}, \ and\
  \bibinfo {author} {\bibfnamefont {Shu}\ \bibnamefont {Chen}},\ }\bibfield
  {title} {\enquote {\bibinfo {title} {$\mathcal{PT}$ symmetry in the
  non-hermitian su-schrieffer-heeger model with complex boundary potentials},}\
  }\href {\doibase 10.1103/PhysRevA.89.062102} {\bibfield  {journal} {\bibinfo
  {journal} {Phys. Rev. A}\ }\textbf {\bibinfo {volume} {89}},\ \bibinfo
  {pages} {062102} (\bibinfo {year} {2014})}\BibitemShut {NoStop}%
\bibitem [{\citenamefont {Martinez~Alvarez}\ \emph {et~al.}(2018)\citenamefont
  {Martinez~Alvarez}, \citenamefont {Barrios~Vargas},\ and\ \citenamefont
  {Foa~Torres}}]{alvarez2017}%
  \BibitemOpen
  \bibfield  {author} {\bibinfo {author} {\bibfnamefont {V.~M.}\ \bibnamefont
  {Martinez~Alvarez}}, \bibinfo {author} {\bibfnamefont {J.~E.}\ \bibnamefont
  {Barrios~Vargas}}, \ and\ \bibinfo {author} {\bibfnamefont {L.~E.~F.}\
  \bibnamefont {Foa~Torres}},\ }\bibfield  {title} {\enquote {\bibinfo {title}
  {Non-hermitian robust edge states in one dimension: Anomalous localization
  and eigenspace condensation at exceptional points},}\ }\href {\doibase
  10.1103/PhysRevB.97.121401} {\bibfield  {journal} {\bibinfo  {journal} {Phys.
  Rev. B}\ }\textbf {\bibinfo {volume} {97}},\ \bibinfo {pages} {121401}
  (\bibinfo {year} {2018})}\BibitemShut {NoStop}%
\end{thebibliography}%


\clearpage

{\bf Supplemental Material}

\vspace{7mm}

\begin{figure}[htb]
\subfigure{\includegraphics[width=3.7cm, height=3.5cm]{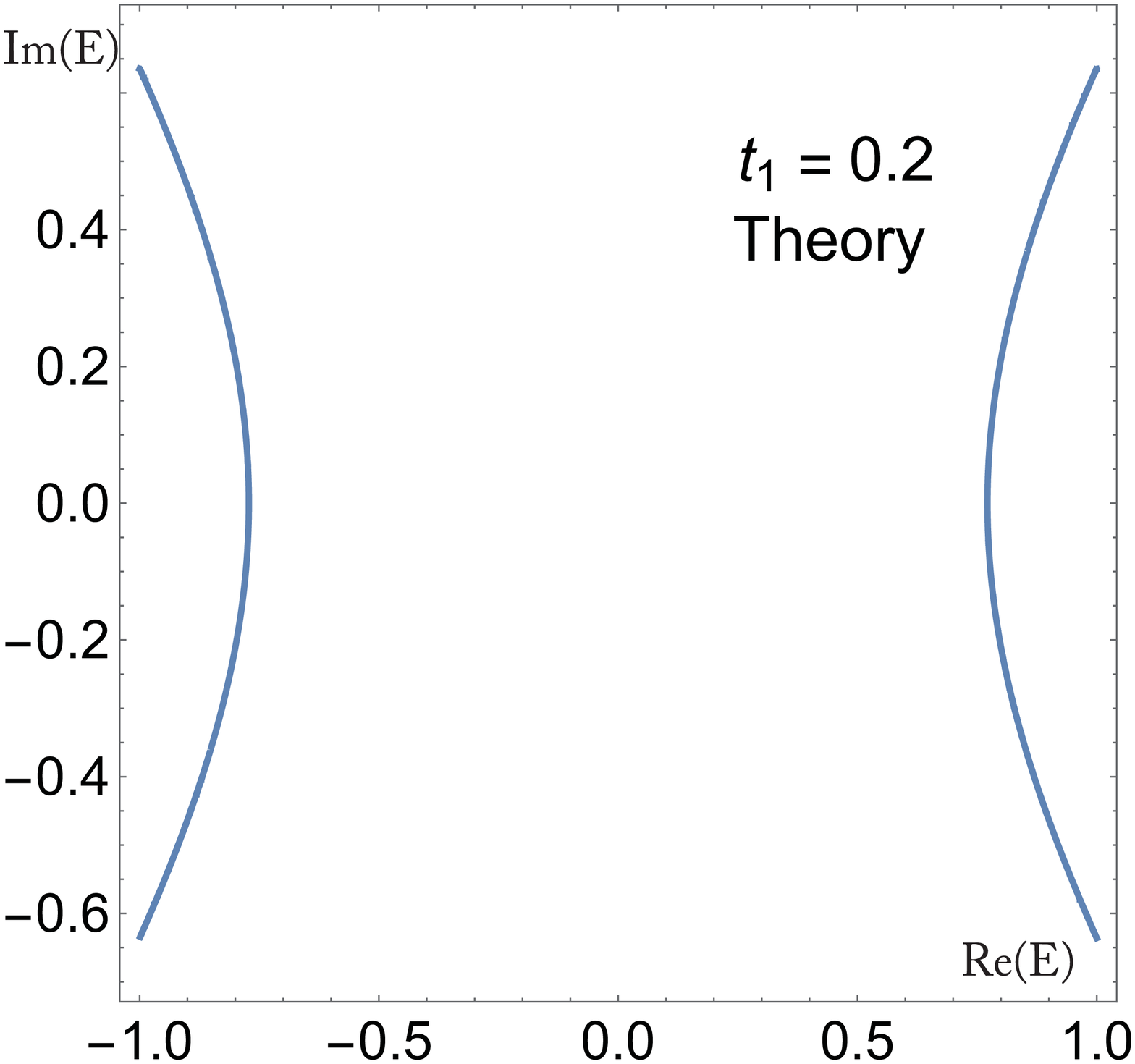}}
\subfigure{\includegraphics[width=3.7cm, height=3.5cm]{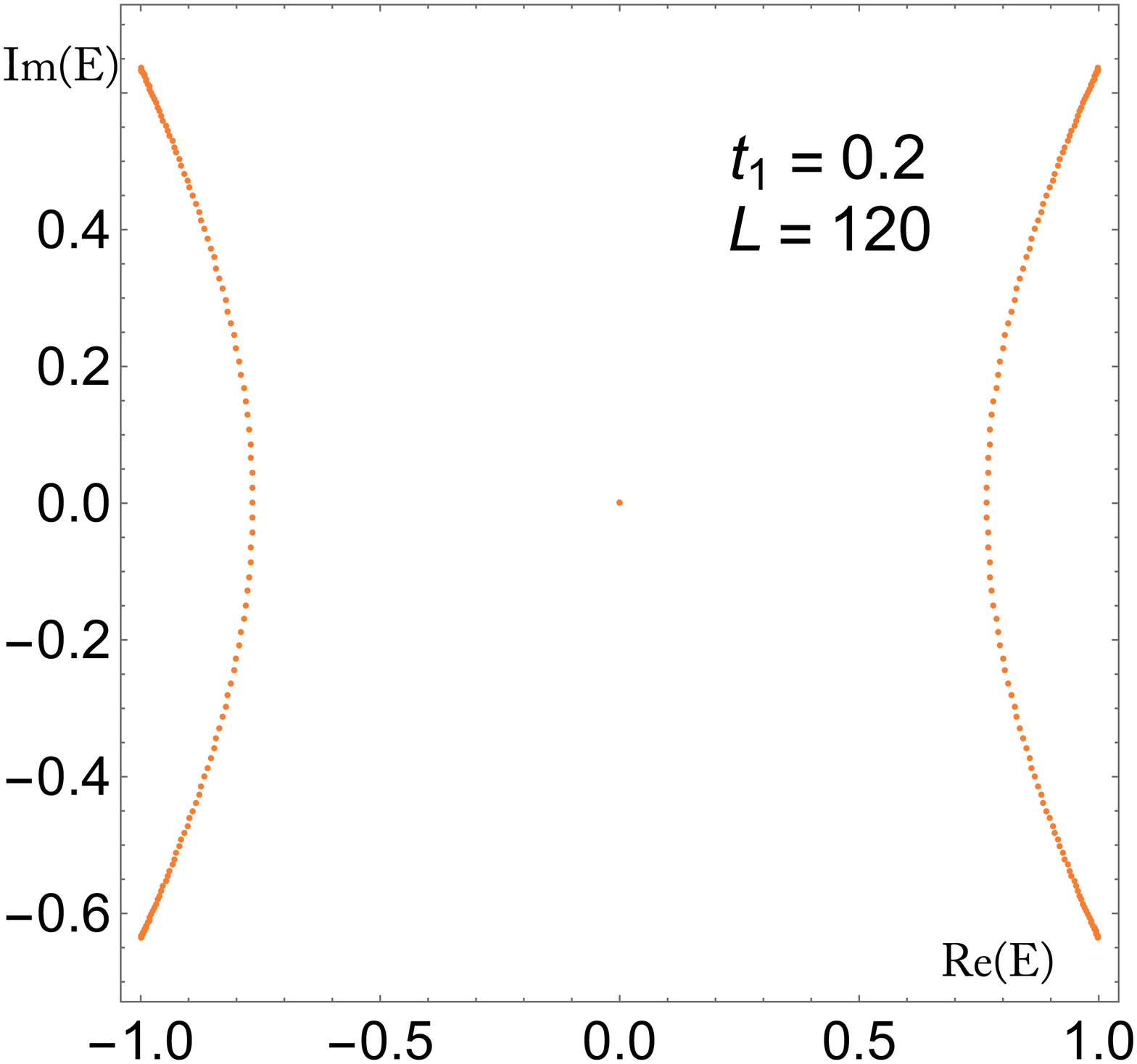}}
\subfigure{\includegraphics[width=3.7cm, height=3.5cm]{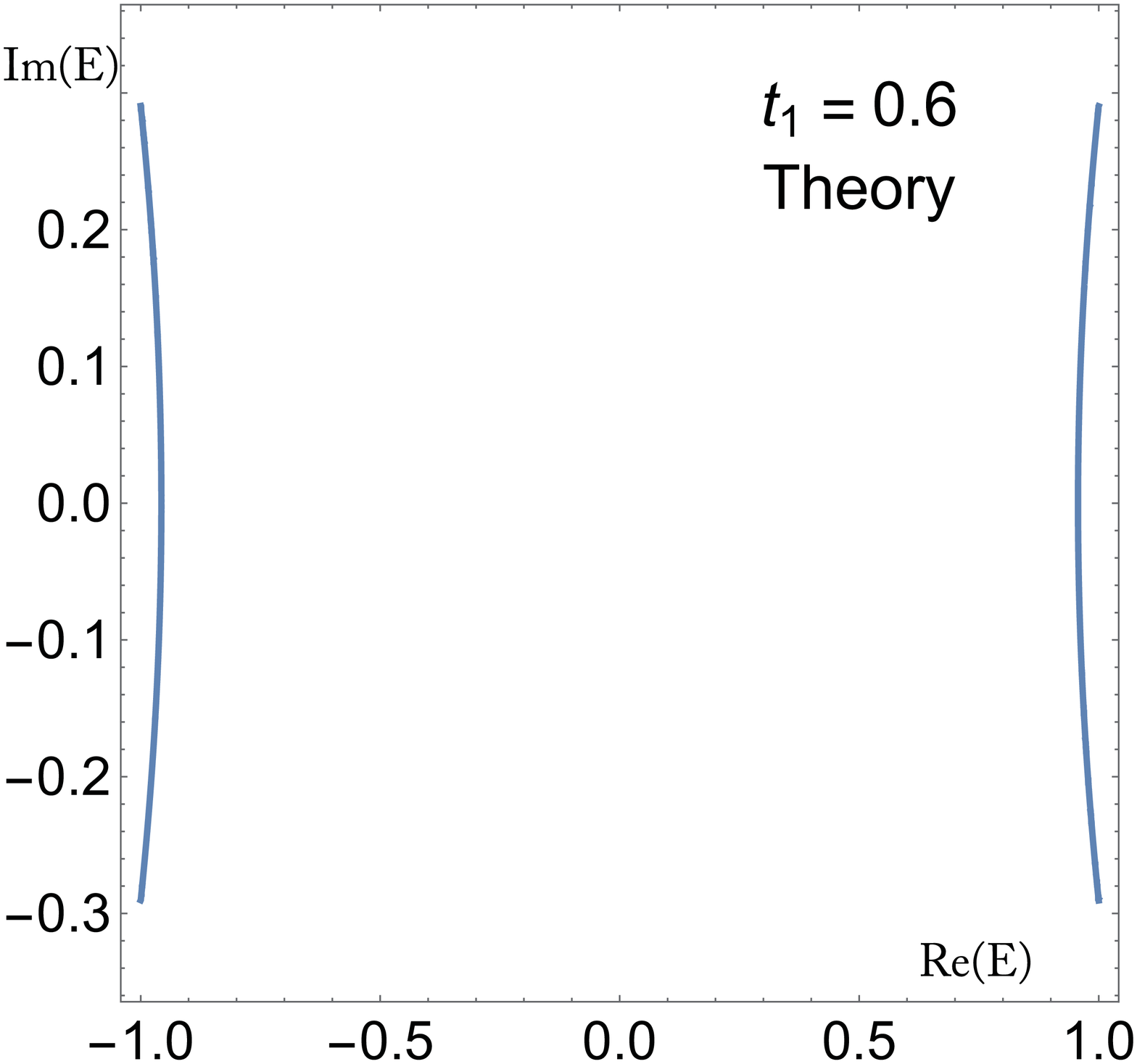}}
\subfigure{\includegraphics[width=3.7cm, height=3.5cm]{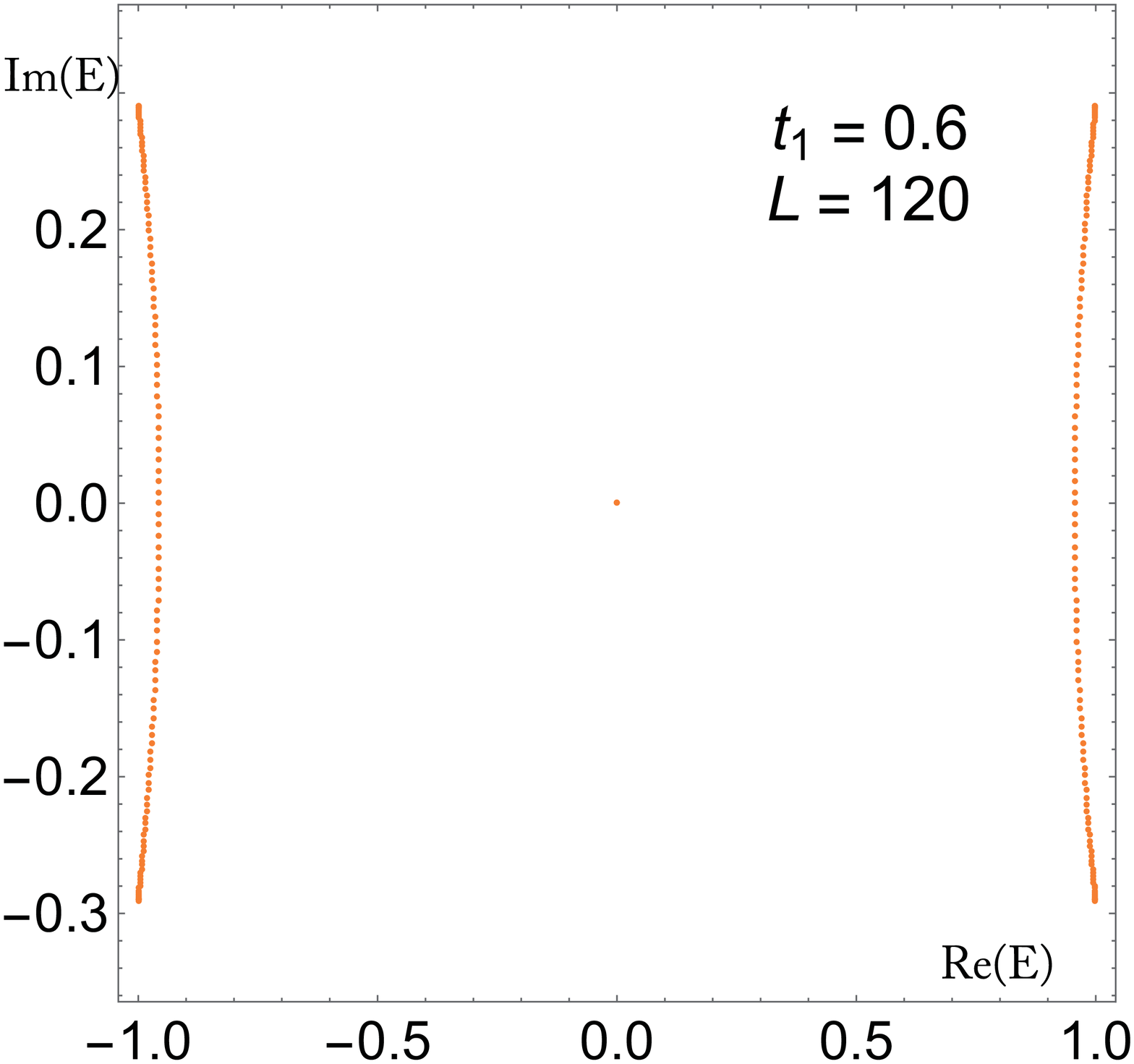}}
\subfigure{\includegraphics[width=3.7cm, height=3.5cm]{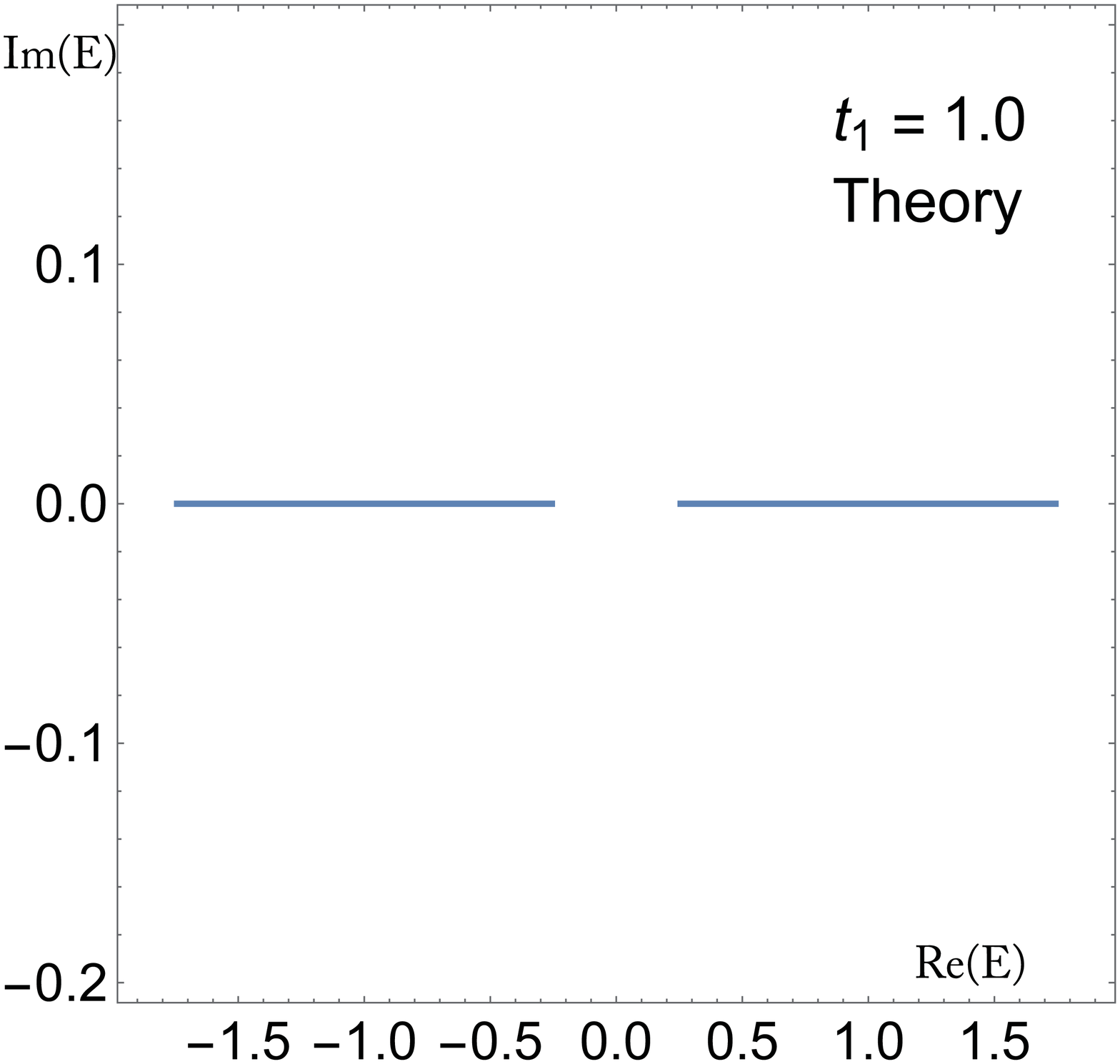}}
\subfigure{\includegraphics[width=3.7cm, height=3.5cm]{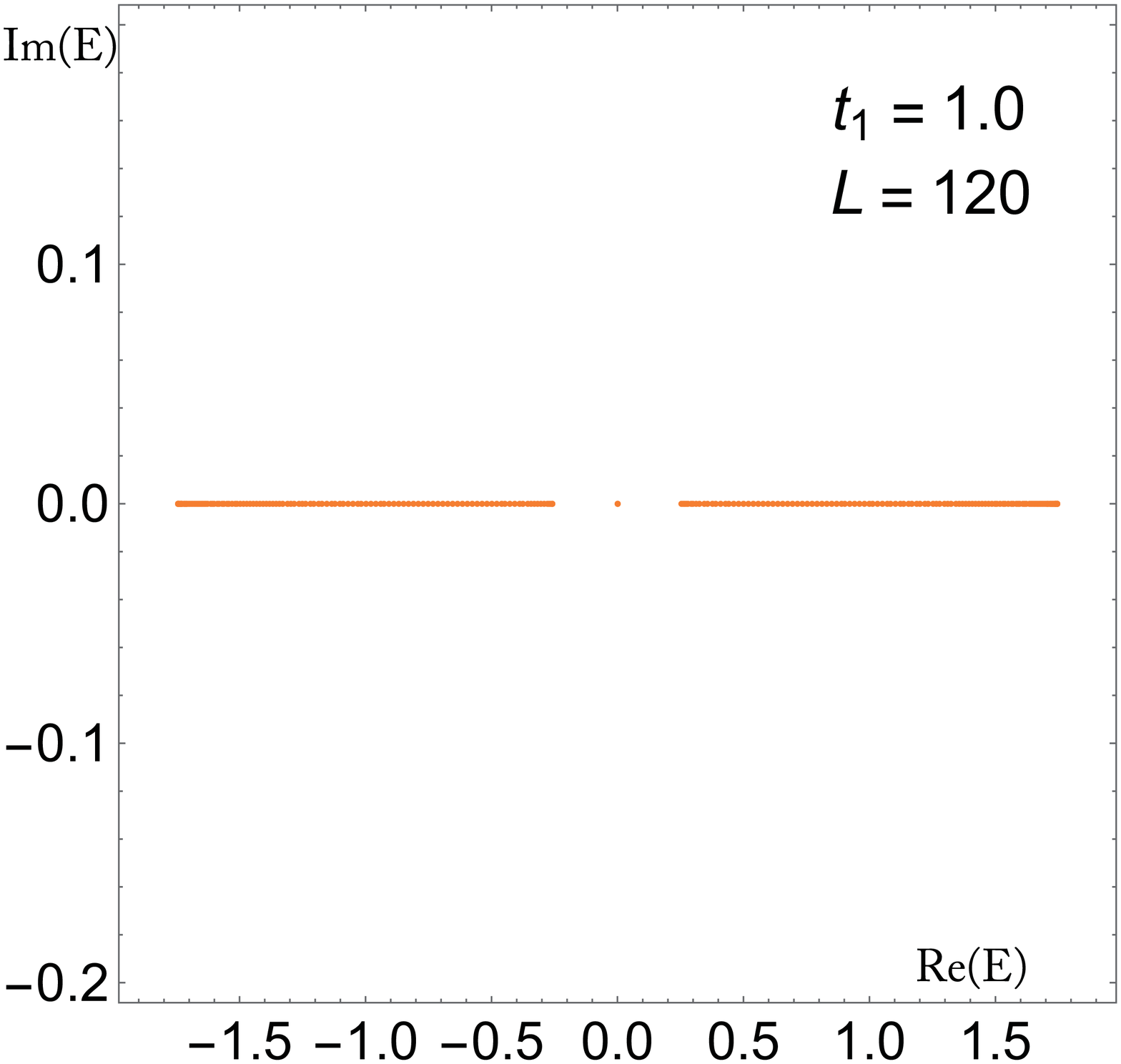}}
\caption{ Left panels: Energies ($E$) solved from $|\beta_1(E)|=|\beta_2(E)|$ [see the discussion below Eq.(12) in the main article]; Right panels: Numerical eigenenergies of open chains with length $L=120$. Common parameters are $t_2=1, \gamma=4/3$. }\label{supplemental-1}
\end{figure}

\emph{Two supplemental figures.--}As explained in the main article, the equation $|\beta_1(E)|=|\beta_2(E)|$ determines the bulk-band energies [see the discussion below Eq. (12) in the main article]. In fact, in the complex $E$ plane, $|\beta_1(E)|=|\beta_2(E)|$ determines one-dimensional curves containing the bulk-band energies. Fig.\ref{supplemental-1} illustrates calculating bulk-band energies by solving $|\beta_1(E)|=|\beta_2(E)|$ for three values of $t_1$.

\begin{figure}[htb]
\includegraphics[width=8.0cm, height=5.0cm]{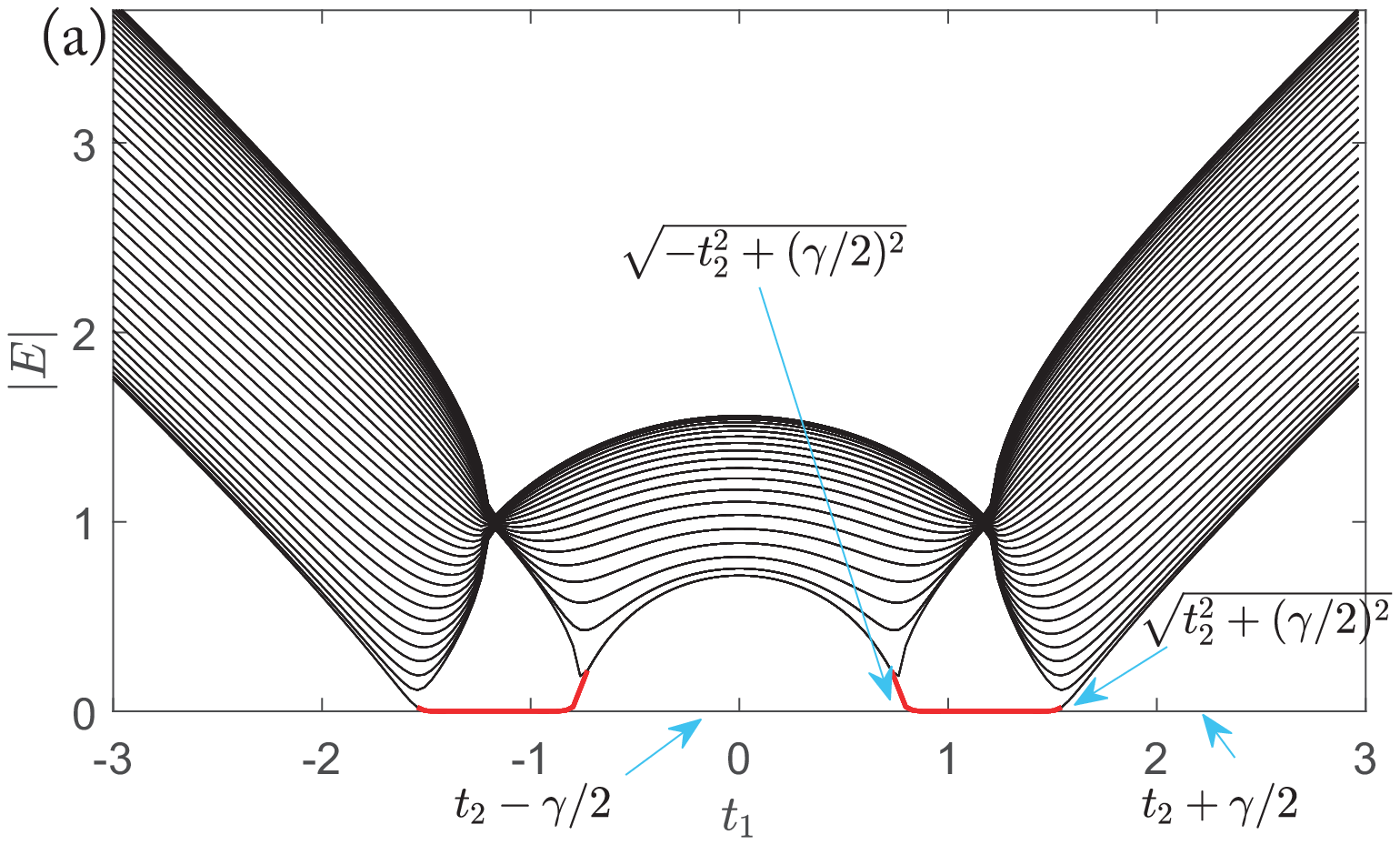}
\includegraphics[width=7.8cm, height=5.0cm]{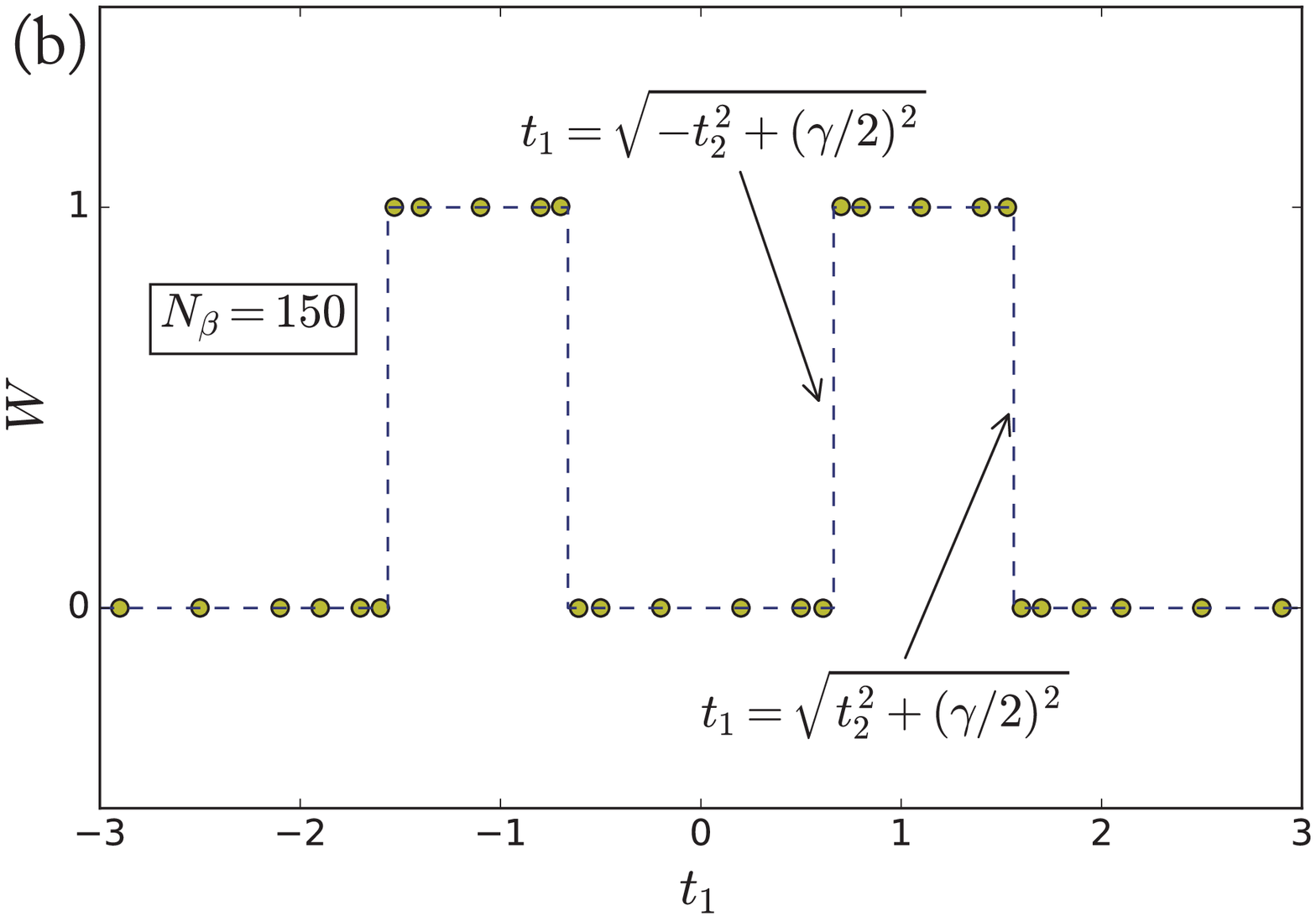}
\caption{ (a) The modulus of energy for an open chain with length $L=40$. (b) Numerical results of the topological invariant.  $t_2=1.0, \gamma=2.4$. According to the analytical solution, in the regime $|t_2|<|\gamma|/2$, there are four transition points $t_1=\pm\sqrt{\pm t_2^2+(\gamma/2)^2}$. The theory is consistent with the numerical results. The topological invariant correctly predicts the number of zero modes.  }\label{supplemental-2}
\end{figure}

Fig.\ref{supplemental-2} shows the energies and topological invariant for the parameter regime $|t_2|<|\gamma/2|$ (In the main article, we have focused on $|t_2|>|\gamma/2|$).

\vspace{10mm}

\emph{Nonzero $t_3$.--}Let us outline the calculation of generalized Brillouin zone $C_\beta$ for nonzero $t_3$. We consider an open-boundary chain with length $L$.  In the bulk, the real-space eigenequations are  $t_2\psi_{n-1,B}+(t_1+\frac{\gamma}{2})\psi_{n,B}+t_3\psi_{n+1,B}=E\psi_{n,A}$ and
$t_3\psi_{n-1,A}+(t_1-\frac{\gamma}{2})\psi_{n,A}+t_2\psi_{n+1,A}=E\psi_{n,B}$.  Similar to Eq. (6) of the main article, we now have
\begin{equation} \label{}
\begin{aligned}
&[t_2\beta^{-1}+(t_1+\frac{\gamma}{2})+t_3\beta]\phi_{B}   =E\phi_{A},\\
&[t_3\beta^{-1}+(t_1-\frac{\gamma}{2})+t_2\beta]\phi_{A}   =E\phi_{B}.
\end{aligned}
\end{equation}
Therefore, $\beta$ and $E$ satisfy
\begin{equation} \label{t3E}
\begin{aligned}
E^2=[t_2\beta^{-1}+(t_1+\gamma/2)+t_3\beta][t_3\beta^{-1}+(t_1-\gamma/2)+t_2\beta].
\end{aligned}
\end{equation}
As a quartic equation of $\beta$, it has four roots $\beta_j(E)$ ($j=1,2,3,4$). As explained in the main article, the bulk-band energies have to satisfy $|\beta_i(E)|=|\beta_j(E)|$ for a pair of $i,j$. In fact, this equation can also be intuitively understood as follows. Suppose that a wave with $\beta_i$ propagates from the left end towards the right. It hits the right end and gets reflected, and the reflected waves with $\beta_j$ propagates back to the left end. To satisfy certain standing-wave conditions for an energy eigenstate, the magnitudes of the initial and the final waves have to be of the same order, therefore, one must have $|\beta_i(E)|^L\sim |\beta_j(E)|^L$ or $|\beta_i(E)|= |\beta_j(E)|$.
Each equation $|\beta_i(E)|= |\beta_j(E)|$ determines a one-dimensional curve in the complex $E$ plane, and the $\beta$ curve follows from the $E$ curves.

There is also a more brute-force approach to find the $C_\beta$ curve. One can numerically solve the eigen-energies of an open chain, and then find $\beta_j(E)$'s from Eq. (\ref{t3E}). In this calculations, one has to discard $\beta_i(E),\beta_j(E)$ that do not satisfy $|\beta_i(E)|=|\beta_j(E)|$, as they should not be regarded as bulk components of the eigenstates. This disposal is similar to the Hermitian case: A typical eigenstate of an open chain is a superposition of right-propagating and left-propagating Bloch waves (both have $|\beta|=1$) and certain decaying components localized at the two ends. The (Hermitian) topological invariants are defined in terms of the bulk components, namely the Bloch waves.

\end{document}